%% file: main.tex
\documentclass[
    reprint,
	amsmath,
    amssymb,
	aps,
	prx,
    floatfix,
    superscriptaddress
]{revtex4-2}

\usepackage[english]{babel}
\usepackage[utf8x]{inputenc}
\usepackage[T1]{fontenc}
\usepackage{braket}
\usepackage{mathtools}
\usepackage{flushend}

\usepackage{amsmath}
\usepackage{siunitx}
\usepackage{graphicx}
\usepackage{dcolumn}
\usepackage{bm}
\usepackage{array}
\usepackage[shortlabels]{enumitem}
\newcolumntype{?}{!{\vrule width 1pt}}
\usepackage[colorinlistoftodos]{todonotes}
\usepackage[colorlinks=true,linkcolor=blue,citecolor=blue,urlcolor=blue]{hyperref}
\usepackage{slashed}

\usepackage{bibunits}
\defaultbibliographystyle{apsrev4-1}
\usepackage{etoolbox}

\makeatletter
\newcommand*{\newbibstartnumber}[1]{%
  \apptocmd{\thebibliography}{%
    \global\c@NAT@ctr #1\relax
    \addtocounter{NAT@ctr}{-1}%
  }{}{}%
}
\makeatother

\newcommand{\blue}[1]{\textcolor{blue}{#1}} 

\newcommand\bb[1]{\mbox{\boldmath{$#1$}}}

\def\apjl{The Astrophysical Journal Letters}

\makeatletter
\let\cat@comma@active\@empty
\makeatother
\begin{document}
\title{Intermittency and electron heating in kinetic-Alfvén-wave turbulence}
\author{Muni Zhou}
\affiliation{Plasma Science and Fusion Center$,$ Massachusetts Institute of Technology$,$ Cambridge$,$ MA 02139$,$ USA}
\author{Zhuo Liu}
\affiliation{Plasma Science and Fusion Center$,$ Massachusetts Institute of Technology$,$ Cambridge$,$ MA 02139$,$ USA}
\author{Nuno F. Loureiro}
\affiliation{Plasma Science and Fusion Center$,$ Massachusetts Institute of Technology$,$ Cambridge$,$ MA 02139$,$ USA}
\thanks{Corresponding author.\\\href{mailto:who@mit.edu}{who@mit.edu} \vspace{0.5cm}}

\date{\today}

\begin{abstract}
We report analytical and numerical investigations of sub-ion-scale turbulence in low-beta plasmas, focusing on the spectral properties of the fluctuations and electron heating.
In the isothermal limit, the numerical results strongly support a description of the turbulence as a critically-balanced Kolmogorov-like cascade of kinetic Alfv\'en wave fluctuations, as amended by Boldyrev \& Perez (Astrophys. J. Lett. \textbf{758}, L44 (2012)) to include intermittent effects. 
When the constraint of isothermality is removed (i.e., with the inclusion of electron kinetic physics), the energy spectrum is found to steepen due to electron Landau damping, which is enabled by the local weakening of advective nonlinearities around current sheets, and yields significant energy dissipation via a velocity-space cascade. The use of a Hermite-polynomial representation to express the velocity-space dependence of the electron distribution function allows us to obtain an analytical, lowest-order solution for the Hermite moments of the distribution, which is borne out by numerical simulations.
\end{abstract}

\maketitle


\section{Introduction} 
\label{sec:intro}
Plasma turbulence is ubiquitous in space and astrophysical systems as diverse as the Earth's magnetosphere~\cite{borovsky2003mhd}, the solar wind~\cite{Goldreich1995,tu1995mhd}, the solar corona~\cite{parker1983}, accretion disks~\cite{balbus1998instability}, and the interstellar and intracluster media~\cite{begelman1990turbulent}.
Many of these environments are sufficiently dilute that particle collisions are rare on the dynamical timescales of interest. As such, kinetic plasma descriptions are needed to understand turbulence in those environments.
Despite decades of intensive research, formulating a predictive theoretical framework for the dynamics of turbulence in the phase space of positions and velocities has been a notoriously difficult problem, especially in the sub-ion scale range (the kinetic range), where various microphysical plasma processes are dynamically important.

Recent high-resolution, {\it in situ} measurements of electromagnetic fluctuations and plasma distribution functions from satellites such as the Magnetospheric Multiscale mission ({\it MMS}), {\it Cluster}, and the Parker Solar Probe ({\it PSP}) have provided unprecedented opportunities to study the rich plasma dynamics in the sub-ion range of the turbulence~\cite{sahraoui2009evidence,alexandrova2009universality,alexandrova2012solar,alexandrova2013solar,chen2019evidence}.
The interpretation and understanding of those observations essentially reduces to answering two questions: (i) What are the spectral properties of the electromagnetic fluctuations, and more specifically, what physical processes determine the energy spectra in the kinetic range? (ii) What is the nature of the dissipative processes, i.e., how are ions and electrons heated?  
Answering these two questions requires a comprehensive understanding of the phase-space dynamics of kinetic plasma turbulence.

Alfv\'en waves (AWs), an essential building block of long-wavelength magnetohydrodynamic (MHD) turbulence, convert into (dispersive) kinetic Alfv\'en waves (KAWs) in the sub-ion range.
The KAW nature of kinetic turbulence has been supported by a substantial amount of analytical and numerical work~\cite{gary2004kinetic,howes2008model,howes2008kinetic,salem2012identification,tenbarge2012interpreting,franci2018solar,cerri2019kinetic,grovselj2019kinetic}, as well as observations~\cite{leamon1998observational, sahraoui2009evidence,podesta2010kinetic,chen2013nature,kiyani2012enhanced,chen2016recent}.
Therefore, theoretical models conjecture a critically balanced~\cite{Goldreich1995} cascade of KAWs, arriving at predictions of a $\sim k_\perp^{-7/3}$ magnetic energy spectrum~\cite{cho2004anisotropy,cho2009simulations,howes2008model,schekochihin2009astrophysical}.
However, a large number of observations~\cite{leamon1998observational,bale2005measurement,sahraoui2009evidence,salem2012identification,chen2013nature} and numerical simulations~\cite{tenbarge2012interpreting,servidio2015kinetic,wan2015intermittent,told2015multiscale,cerri2018dual,arzamasskiy2019hybrid,grovselj2018fully} yield steeper spectra, typically with a spectral index close to $-2.8$ down to electron scales (e.g., ~\cite{alexandrova2009universality,kiyani2009global,sahraoui2013scaling}).

One important attempt to reconcile theory with observations invokes intermittency.
Indeed, several studies suggest a positive correlation between intermittency and energy dissipation~\cite{Sundkvist2007,karimabadi2011flux,Osman2014,wan2015,chasapis2015thin,Camporeale2018} in the kinetic range; and a phenomenological model has been proposed to describe the role of intermittency in the energy cascade ~\cite{boldyrev2012spectrum}: assuming that the fluctuations, and thus their energy cascade, are concentrated in two-dimensional (2D) structures, a $\propto k_\perp^{-8/3}$ magnetic energy spectrum can be derived, which to some extent explains the steepening of the spectrum (without invoking electron kinetic effects) with respect to a standard KAW-cascade model.

A distinct possibility that has been mooted is that of the KAW cascade being mediated by the tearing mode instability. 
This hypothesis is inspired by similar thinking within the scope of MHD turbulence.
There, the assumption of progressive alignment of the turbulent fluctuations towards smaller scales~\cite{boldyrev2006spectrum} has led to the theoretical conjecture (and tentative numerical confirmation) that, at sufficiently small scales, turbulent eddies should possess a large aspect ratio and become susceptible to tearing, yielding a dynamical range where the energy cascade is mediated by that instability~\cite{loureiro2017role,mallet2017statistical,boldyrev2017magnetohydrodynamic,dong2018role,walker2018influence,loureiro2020nonlinear}. 
Extending these ideas to the sub-ion range is non-trivial because of complications caused by various kinetic effects and the lack of analytical models that predict the scale-dependent anisotropy of turbulent fluctuations in this range.
However, if tearing mediation were to exist in the sub-ion range, a $k_\perp^{-8/3}$ (or $k_\perp^{-3}$, depending on the choice of local magnetic configuration) magnetic spectrum is predicted~\cite{loureiro2017collisionless}, which is also consistent with the steepening of the spectrum. It is an unfortunate coincidence that this prediction is numerically the same as that arising from~\citet{boldyrev2012spectrum} intermittency model: the underlying physics is entirely different.

Apart from understanding the spatial properties of the fluctuations, one of the ultimate goals in studying kinetic turbulence is to understand how energy is dissipated in weakly-collisional plasmas and how the electrons and ions are energized~\cite{parashar2015turbulent}. 
In a collisional plasma, energy can only be thermalized through the ``fluid-channel'', in which energy cascades to small spatial scales through nonlinear advection and finally dissipates by viscosity and resistivity~\cite{dmitruk2004test,Matthaeus2011who,servidio2011,osman2011,zhdankin2013statistical}, while the particle distribution remains close to local thermodynamic equilibrium~\cite{chapman1990mathematical}.
In a weakly collisional plasma, however, particle free streaming along magnetic field lines leads to the ``phase mixing'' process, which smooths out the electromagnetic fluctuations and develops complex structures in velocity space. 
This kinetic effect enables an additional ``kinetic-channel'' for dissipation, in which energy transfers to small scales in velocity space (i.e., high velocity moments of the distribution function) and dissipates through collisions~\cite{howes2008model,schekochihin2008gyrokinetic,navarro2016structure,grovselj2017fully,servidio2017magnetospheric}.
Energy dissipation in kinetic turbulence occurs via a combination of these two channels --- their relative importance is determined by the complex phase-space dynamics of kinetic turbulence~\cite{schekochihin2016phase,adkins2018solvable,cerri2018dual,eyink2018cascades}.

The kinetic dissipation channel is, in principle, susceptible to the {\it plasma echo} effect~\cite{Gould1967echo,malmberg1968plasma}, whereby free energy inverse cascades in velocity space, returning from the high to the low moments of the distribution function (i.e., phase unmixing).
The occurrence of a collection of the stochastic plasma echoes can cause kinetic turbulence to resemble fluid turbulence, as the only allowed energy-cascade channel in that case is the fluid-type cascade towards smaller spatial scales~\cite{schekochihin2016phase,adkins2018solvable}.
The significance of this effect has been confirmed by numerical simulations of collisionless plasma turbulence at scales above the ion Larmor radius ($\rho_i$) with isothermal electrons focusing on compressive fluctuations~\cite{meyrand2019fluidization}, and below $\rho_i$ with an electrostatic drift-kinetic model~\cite{parker2016suppression}.
However, how the possible occurrence of plasma echoes can be reconciled with observations of efficient electron and ion heating (e.g.,~\cite{servidio2017magnetospheric,chen2019evidence}) at kinetic scales is an open question.

In this work, we provide answers to the aforementioned two questions --- what the physical mechanism that sets the energy spectrum is, and how electron heating occurs --- in the low-$\beta$ limit of sub-$\rho_i$ turbulence.
The theoretical framework we adopt is described in Sec.~\ref{sec:KREHM}. 
In Sec.~\ref{sec:KAW_turbulence}, we first review three leading models for the KAW spectrum in the isothermal limit, and then, with electron kinetic effects accounted for, derive a (nonlinear) lowest-order solution of the electron distribution function in the velocity space.
We then test these models using numerical simulations, whose details are provided in Sec.~\ref{sec:numerical} and from which the results presented in Sec.~\ref{sec:numerical_results} are obtained.
Sec.~\ref{sec:conclusion} presents our conclusions and discussion of our results.

\section{Theoretical framework}
\label{sec:KREHM}
Our results in this paper are obtained with an analytical framework known as the Kinetic Reduced Electron Heating Model (KREHM)~\cite{zocco2011reduced}.
KREHM is a rigorous asymptotic reduction of gyrokinetics valid in the limit of low electron plasma-beta, $\beta_e \sim m_e/m_i$, with $\beta_e \equiv 8\pi n_{0e}T_{0e}/B_0^2$, where $B_0$ is the background magnetic (guide) field strength (assumed constant and directed along $\bb{\hat{z}}$), and $n_{0e}$, $T_{0e}$ are the background electron density and temperature, respectively; $m_e$ and $m_i$ denote the electron and ion masses.    
Within this limit, the parallel streaming and electromagnetic effects are ordered out in the ion gyrokinetic equation. It follows that the ion flow velocity parallel to the background field is $u_{z i}=0$, and ions become isothermal and electrostatic. 
Ion finite Larmor radius (FLR) effects are retained in the gyrokinetic Poisson's law,
\begin{equation}
    \label{eq:gk_poisson}
    \frac{\delta n_e}{n_{0e}}=\frac{1}{\tau}(\hat{\Gamma}_0-1)\frac{e\varphi}{T_{0e}},
\end{equation}
where $\varphi$ is the electrostatic potential, $\delta n_e/n_{0e}$ is the electron density perturbation normalized to its background value, $\tau \equiv T_{0i}/T_{0e}$ is the temperature ratio, and $\hat{\Gamma}_0$ is a gyroaveraging operator that has the closed-form expression $\Gamma_0(\alpha)=I_0(\alpha)e^{-\alpha}$ in Fourier space;
here, $I_0$ is the zeroth-order modified Bessel function of the first kind and $\alpha = k_\perp^2 \rho_i^2/2$, where $\rho_i = v_{{\rm th}i}/\Omega_i$ is the ion Larmor radius, $v_{{\rm th}i}=\sqrt{2T_{0i}/m_i}$ is the ion thermal velocity, and $\Omega_i=eB_0/m_ic$ is the ion Larmor frequency (we consider single charge ions in this work, $Z=1$).

The electrons are described by a perturbed distribution function which, to order $\sqrt{m_e/m_i} \sim \sqrt{\beta_e}$ in the gyrokinetic expansion, can be written as $\delta f_e = g_e + (\delta n_e/n_{0e} + 2 v_z u_{z e}/v_{{\rm th}e}^2)F_{0e}$, where $F_{0e}$ is the equilibrium Maxwellian defined with the mean electron temperature $T_{0e}$ and its corresponding thermal speed $v_{{\rm th}e}=\sqrt{2T_{0e}/m_e}$, and $\delta n_e$ and $u_{z e}$ (electron flow parallel to the guide field) are the zeroth and first moments of $\delta f_e$, respectively.  
Since $u_{z i}=0$, the parallel component of Amp\`ere's law, $J_z=-(c/4\pi)\nabla_\perp^2 A_z$, where $A_z$ is the parallel component of the vector potential, leads to $u_{ze}=(e/cm_e)d_e^2 \nabla_\perp^2 A_z$, where $d_e=c/\omega_{pe}$ is the electron skin depth.
Information about the second and higher moments of $\delta f_e$ is contained in the (reduced) distribution function $g_e$.

The dynamics of this system are described by fluid equations for the first two moments, coupled to a drift-kinetic equation for $g_e$:
\begin{equation}
    \frac{1}{n_{0e}}\frac{d \delta n_e}{dt} = -\hat{\bb{b}} \cdot \bb{\nabla}  \frac{e}{cm_e} d_e^2 \nabla_\perp^2A_z,
\label{eq:ne}
\end{equation}
\begin{equation}
    \frac{d}{dt}(A_z-d_e^2 \nabla_\perp^2 A_z) = -c\frac{\partial \varphi}{\partial z}+\frac{cT_{e0}}{e}\hat{\bb{b}} \cdot \bb{\nabla} \left( \frac{\delta n_e}{n_{0e}}+\frac{\delta T_{z e}}{T_{0e}} \right),
\label{eq:Az}
\end{equation}
\begin{equation}
\begin{split}
    &\frac{d g_e}{dt}+v_z \hat{\bb{b}} \cdot \bb{\nabla} \left(g_e-\frac{\delta T_{z e}}{T_{0e}}F_{0e} \right) \\ &= C[g_e] + \left(1-\frac{v_z^2}{v_{{\rm th}e}^2} \right) F_{0e} \hat{\bb{b}} \cdot \bb{\nabla} \frac{e}{cm_e} d_e^2 \nabla_\perp^2 A_z.  
\end{split}
\label{eq:ge}
\end{equation}
In these equations, $d/dt \equiv \partial/\partial t+c/B_0\{\varphi,...\}$ denotes the convective time derivative, with the Poisson bracket defined as $\{P,Q\} \equiv \partial_x P \partial_y Q-\partial_y P \partial_x Q$. 
The parallel (to the total field) gradient operator is $\hat{\bb{b}} \cdot \bb{\nabla} P \equiv \partial_z P -\{A_z,P \}/B_0$. 
The last term in the generalized Ohm's law features the (normalized) parallel electron temperature perturbation, defined in terms of the reduced electron distribution function as
\begin{equation}
    \frac{\delta T_{z e}}{T_{0e}}=\frac{1}{n_{0e}}\left(\int d^3{\bb v} \frac{2 v_z ^2}{v_{{\rm th} e}^2}g_e\right);
\end{equation}
 the isothermal limit of these equations corresponds to $\delta T_{z e}=0$, that is, $g_e=0$. 
Lastly, $C[g_e]$ represents the collision operator. 
Since Eq.~\eqref{eq:ge} has no explicit dependence on $v_\perp$, this coordinate can be integrated out of the problem if a collision operator is chosen that bares no such dependence, such as the (modified) Lenard-Bernstein collision operator~\cite{zocco2011reduced}. In that case, the reduced distribution function becomes four-dimensional, $g_e=g_e(\bb{r}, v_z, t)$.

In the absence of collisions,  Eqs.~\eqref{eq:ne}-\eqref{eq:ge} conserve a quadratic invariant that is usually referred to as the total free energy
\begin{equation}
\begin{aligned}
W &\equiv \int \frac{d^3\bb{r}}{V} \left[\frac{1}{\tau}(1-\hat{\Gamma}_0) \frac{e^2 n_{0e} \varphi^2}{2T_{0e}} + \frac{|\nabla_\perp A_z|^2}{8 \pi} \right. \\
&+\left. \frac{1}{\tau^2}(1-\hat{\Gamma}_0)^2 \frac{e^2 n_{0e} \varphi^2}{2T_{0e}} + \frac{d_e^2|\nabla^2_\perp A_z|^2}{8 \pi} + \int d^3\bb{v}\frac{T_{0e}g_e^2}{2F_{0e}} \right],
\label{eq:free_energy}
\end{aligned}
\end{equation}
where the different terms on the right-hand side, in the order in which they appear, correspond to the ion perturbed entropy, the magnetic energy, the electron density variance, the kinetic energy of the parallel electron flow, and the electron free energy, respectively.  
For completeness, we note that the isothermal ($g_e=0$) limit of these equations admits another ideal invariant --- the generalized helicity,
\begin{equation}
    H \equiv \int \frac{(1-\hat{\Gamma}_0)\varphi}{\rho_i^2} (1-d_e^2 \nabla_\perp^2)A_z dV, 
    \label{eq:helicity}
\end{equation}
which reduces to $H = \int n_e A_z dV$ in the range $\rho_i \gg k_\perp^{-1} \gg d_e$, and to the cross-helicity at MHD scales ($k_\perp \rho_i \ll 1$). 

To better describe the velocity-space dynamics, we follow Ref.~\cite{zocco2011reduced} and expand $g_e$ in Hermite polynomials:
\begin{equation}
g_e(\bb{r},v_z,t)=\frac{1}{\sqrt{2^mm!}} \sum_{m=0}^\infty H_m \left(\frac{v_z}{v_{{\rm th}e}}\right)g_m(\bb{r},t)F_{0e}(v_z),
    \label{eq:hemite_expansion}
\end{equation}
where
\begin{equation}
g_m=\frac{1}{n_{0e}}\frac{1}{\sqrt{2^mm!}}\int_\infty^\infty dv_z H_m \left(\frac{v_z}{v_{{\rm th}e}}\right)g_e.
    \label{eq:hemite_expansion_2}
\end{equation}
Since $g_e$ only contains moments higher than $\delta n_e$ and $u_{z e}$, it follows that $g_0=g_1=0$.
Using the Hermite expansion, Eq.~\eqref{eq:ge} can be decomposed into a series of coupled equations for the Hermite coefficients $g_m$ (for $m \ge 2$):
\begin{equation}
\begin{split}
    \frac{d g_m}{dt} &= -v_{{\rm th}e} \hat{\bb{b}} \cdot \bb{\nabla} \left( \sqrt{\frac{m+1}{2}} g_{m+1} + \sqrt{\frac{m}{2}}g_{m-1}-\delta_{m,1} g_2 \right) \\ & -\sqrt{2}\delta_{m,2}\hat{\bb{b}} \cdot \bb{\nabla} \frac{e}{cm_e}d_e^2\nabla_\perp^2A_z-C[g_m].
\label{eq:gm}
\end{split}
\end{equation}
At large $m$, free energy is converted to electron entropy and thence to heat via the collisional operator $C[g_m]$.
In summary, the KREHM framework captures ion FLR effects and electron (drift) kinetic dynamics, including linear Landau damping. It is accurate at all scales above the electron Larmor radius, under the low-beta constraint $\beta_e\sim m_e/m_i$. The only linear mode captured by these equations is the kinetic Alfv\'en wave (KAW); as such, it provides the simplest-possible fully kinetic platform to study turbulence below the ion scales. 
In the long-wavelength limit $k_\perp \rho_i \ll 1$, the KREHM equations reduce to the reduced-magnetohydrodynamic (RMHD) equations~\cite{kadomtsev1973nonlinear,strauss1976nonlinear,zank1992,schekochihin2009astrophysical}.

\section{Theoretical models of Kinetic-Alfv\'en-wave turbulence}
\label{sec:KAW_turbulence}
In this section, we first focus on the $\rho_i \gg k_\perp^{-1} \gg d_e$ scale range and discuss the spectral properties of the fluctuations in position and velocity space.
The position-space spectra is discussed in the limit of isothermal electrons, i.e., the limit when the perturbed electron distribution vanishes, $g_e=0$.
In the case of $g_e\neq 0$, we obtain a lowest-order solution for $g_e$ in velocity space, and discuss its implications for electron heating.
We then extend our discussion to the sub-electron scale range, $k_\perp d_e\gtrsim 1$.

\subsection{Spectra and spectral anisotropy}
\label{sec:theory_spectra}
At sub-$\rho_i$ scales, the turbulent fluctuations are composed of nonlinearly interacting, strongly anisotropic KAWs propagating along the ``local mean'' magnetic field.
The total magnetic field is $\bb{B}=B_0 \bb{\hat{z}}+\delta \bb{B}_\perp$, where the perpendicular (fluctuating) magnetic field is $\delta \bb{B}_\perp= -\hat{\bb{z}} \times \bb{\nabla} A_z$, and $\delta B_\perp/B_0 \ll 1$.
The wave numbers of the fluctuations can be decomposed into parallel and perpendicular components, $k_\parallel$ and $k_\perp$, with respect to $\bb{B}$.
The multi-scale fluctuations can be characterized as turbulent eddies with the correlation length $\ell \sim 1/k_\parallel$ in the parallel (to the local magnetic field) direction, the width $\xi$ in the direction of the local perpendicular fluctuation, and the thickness $\lambda \sim 1/k_\perp$ in the direction perpendicular to the other two.
In the rest of the paper, we use the subscript $\lambda$ to refer to the amplitude of fluctuating quantities at scale $\lambda\sim k_\perp^{-1}$.

First, assuming scale-by-scale equipartition between the density and magnetic energy fluctuations, $(\delta n_e/n_{0e})^2 n_{0e}T_{0e} \sim |\nabla_\perp A_z|^2/8\pi$, leads to the relation
\begin{equation}
    \varphi_\lambda \sim (\rho_i V_A/c) \delta B_{\perp \lambda},
\label{eq:equipartition}
\end{equation}
with $V_A=B_0/\sqrt{4 \pi \rho_0}$ the Alfv\'en speed pertaining to the guide field, where $\rho_0=n_0 m_i$ is the background mass density, and we have used the $k_\perp \rho_i \gg 1$ limit of GK Poisson's law (Eq.~\eqref{eq:gk_poisson}), 
\begin{equation}
    \delta n_{e \lambda}/n_{0e} \sim -e \varphi_\lambda/(\tau T_{0e}).
\label{eq:Gk_poisson_krhoi_gg_1}
\end{equation}

Dynamically, the parallel component of fluctuations at scale $\ell \sim k_\parallel^{-1}$ is assumed to be determined by the  propagation of KAWs, and so the characteristic (parallel) time scale is set by the (linear) frequency of those waves:
\begin{equation}
    \gamma_{l} \sim \omega_{\rm KAW} \propto k_\perp \rho_s k_\parallel V_A \sim \rho_s V_A/(\ell \lambda),
\label{eq:gamma_linear}
\end{equation}
where $\rho_s=\rho_i/\sqrt{2 \tau}$ is the ion sound Larmor radius.
The perpendicular fluctuations at scale $\lambda \sim k_\perp^{-1}$ undergo nonlinear interactions which lead to a cascade of constant energy flux toward smaller scales. 
Dimensionally, this inverse nonlinear time scale, usually referred to as the ``eddy-turnover rate'', is 
\begin{equation}
    \gamma_{nl} \sim \varepsilon / (\rho_0 v_{A \lambda}^2/2) \sim \varepsilon/(\delta B_{\perp \lambda}^2/8\pi),
\label{eq:gamma_nl}
\end{equation}
where $\varepsilon$ is the constant energy flux and $v_{A \lambda}=\delta B_{\perp \lambda}/\sqrt{4 \pi \rho_0}$.
The critical balance~\cite{Goldreich1995} conjecture declares that these two frequencies should be comparable at each scale, $\gamma_l \sim \gamma_{nl}$, setting the energy cascade rate of the turbulence. 

The above arguments are relatively general (not model-specific) as they do not involve any assumption on the nonlinear physics of the fluctuations.
In what follows, we discuss three existing models for describing the spectrum of KAW turbulence in the low-$\beta$ regime. 
Each model makes specific assumptions about the nonlinear turbulent cascade and derives the corresponding spectra and spectral anisotropy.

\paragraph{KAW cascade model.} 
Conventional models~\cite{cho2004anisotropy,howes2008model,schekochihin2009astrophysical} of KAW turbulence assume isotropic fluctuations in the field-perpendicular plane (i.e., $\xi\sim\lambda$) and a Kolmogorov-like energy cascade; the nonlinear eddy-turnover rate (Eq.~\eqref{eq:gamma_nl}) can thus be written as $\gamma_{nl} \sim u_{\lambda}/\lambda$, where $u_{\lambda} \sim (c/B_0) \varphi_\lambda/\lambda$ is the velocity of the $\bb{E}\times \bb{B}$ flow.
Combined with the equipartition relation Eq.~\eqref{eq:equipartition}, a $\propto k_\perp^{-7/3}$ energy spectrum for magnetic and density fluctuations is obtained, with a $k_\parallel \propto k_\perp^{-1/3}$ spectral anisotropy following from imposing critical balance.

\paragraph{Intermittency model.} 
As discussed in the Introduction, observations and numerical simulations yield spectra steeper than the $k_\perp^{-7/3}$ predicted by this simple KAW cascade model. To address this discrepancy, \citet{boldyrev2012spectrum} conjectured that intermittent effects might play an important role in KAW turbulence.
They argued that density and magnetic fluctuations, and thus their energy cascade, are concentrated in highly-intermittent, quasi-two-dimensional (2D) structures elongated in the $\hat{\bb{z}}$-direction.
Such 2D structures can be viewed as ``energy-containing sheets'' (not to be confused with current
sheets, where the gradient of the magnetic fluctuation
is large)~\footnote{In the conventional $\beta$-model~\cite{frisch1995turbulence} for intermittency widely used in hydrodynamics, the dimension of intermittent structures is a fractal dimension and is not necessarily the same as their spatial dimension. However, in~\citet{boldyrev2012spectrum} and other turbulence simulations, it is found that intermittent structures do appear to be 2D in position space.}.
The volume-filling fraction of these ``active'' 2D structures at each scale $\lambda$ should scale as $p_\lambda \propto \lambda$~\cite{frisch1995turbulence}.
It follows that the energy density of the fluctuations should scale as $E_{B_\lambda} \sim \delta B_{\perp \lambda}^2 p_\lambda$, and the energy cascade rate (Eq.~\eqref{eq:gamma_nl}) thus becomes $\gamma_{nl} \sim \varepsilon/ (\delta B_{\perp \lambda}^2 \ p_\lambda)$.
The turbulent eddies, however, are assumed to be isotropic on field-perpendicular planes (consistent with the numerical results reported below in Sec.~\ref{sec:spectra}). 
The intermittent ``energy-containing'' sheets can be viewed as chains of such cylindrical magnetic structures. 
No reduction of nonlinearity is expected for such circular eddies, and thus the nonlinear eddy-turnover rate should remain unchanged: $\gamma_{\rm nl} \sim \varphi_\lambda/\lambda^2$. 
Combined with the equipartition relation (Eq.~\eqref{eq:equipartition}), this leads to the scaling $\delta B_{\perp \lambda} \propto \lambda^{1/3}$ and energy density $E_{B_\lambda} \propto \lambda^{5/3}$.
The perpendicular magnetic energy spectrum is thus
\begin{equation}
\begin{aligned}
    E_B (k_\perp) dk_\perp \propto k_\perp^{-8/3} dk_\perp.
    \label{eq:Bspec_intermittent}
\end{aligned}
\end{equation}
Balancing the frequency of KAWs and the nonlinear turnover rate, $\omega_{\rm KAW} \sim \gamma_{\rm nl}$, yields $\ell \propto  \lambda^{2/3}$ or, equivalently, 
\begin{equation}
\begin{aligned}
    k_\parallel \propto k_\perp^{2/3},
    \label{eq:anisotropy_intermittent}
\end{aligned}
\end{equation}
and the parallel magnetic energy spectrum
\begin{equation}
\begin{aligned}
    E_B (k_\parallel) dk_\parallel \propto k_\parallel^{-7/2} dk_\parallel.
    \label{eq:Bspec_intermittent_para}
\end{aligned}
\end{equation}

We denote by $E_\varphi(k_\perp)$ the spectrum of the ion perturbed entropy $K_i$ (the first term on the {\it r.h.s.} of the total free energy invariant Eq.~\eqref{eq:free_energy}).
For $k_\perp \rho_i \gg 1$, $K_i$ reduces to the energy of density fluctuations, and for $k_\perp \rho_i \ll 1$ (in RMHD), $K_i$ reduces to the kinetic energy of the $\bb{E}\times\bb{B}$ flow.
In both regimes (i.e., for all the scales above $d_e$ ), $K_i$ is expected to be in equipartition with the magnetic energy, and so in the KAW turbulence, $E_\varphi (k_\perp) dk_\perp \propto k_\perp^{-8/3} dk_\perp$.

\paragraph{Tearing-mediation model.} 
Another fundamental process that could modify the physics of the nonlinear cascade is the onset of tearing instability in the turbulent eddies. 
In analogy with the reasoning underlying the conjecture of  tearing-mediation of MHD turbulence ~\cite{loureiro2017role,mallet2017statistical,boldyrev2017magnetohydrodynamic}, if dynamical alignment~\cite{boldyrev2006spectrum} were to be present in the sub-$\rho_i$ range and lead to sufficiently elongated eddies, tearing modes could become unstable in those eddies and modify the spectrum of the turbulence.
A critical assumption of tearing-mediated turbulence models~\cite{loureiro2017role,mallet2017disruption,loureiro2017collisionless} is that the eddy turnover rate, or the nonlinear frequency, is determined by the time scale of the tearing instability, i.e., $\gamma_{nl}\sim \gamma_t$, where $\gamma_t \sim v_{A\lambda} \rho_i d_e^{1/2} \lambda^{-5/2}$ is the growth rate of the fastest-growing tearing mode in a collisionless plasma for scales $\lambda>\rho_i$, assuming sinusoidal (rather than Harris-sheet-like) profiles of magnetic fluctuations~\cite{loureiro2017collisionless}.
This leads to $v_{A\lambda} \propto \lambda^{5/6}$, and the spectrum of $\delta B_{\perp \lambda}^2$ thus scales as
\begin{equation}
\begin{aligned}
    E_B (k_\perp) dk_\perp &\propto k_\perp^{-8/3} dk_\perp.
    \label{eq:Bspectrum_scaling}
\end{aligned}
\end{equation}
Invoking, as before, the critical balance conjecture leads to the spectral anisotropy $k_\parallel \propto k_\perp^{2/3}$.
Note the fact that these are exactly the same scalings with $k_\perp$ as found in the intermittency model of~\citet{boldyrev2012spectrum}, even though the physics underlying each of these models are of a completely different nature. 

The derivation above assumes that the magnetic field profile in the eddies is sinusoidal (i.e., $\delta B_\lambda\sim\sin(x/\lambda$)), and that the frozen flux constraint is broken by electron inertia. 
Different assumptions of spatial profiles of the fluctuations, or different choices of flux unfreezing mechanisms, lead to different scalings of a tearing-mediated spectrum, whereas the intermittency is not expected to be affected.
If a Harris sheet-type configuration (i.e., $\tanh(x/\lambda)$ profile) of magnetic fluctuations is assumed instead, the derived spectrum and anisotropy become $E_B (k_\perp) \propto k_\perp^{-3}$ and $k_\parallel \propto k_\perp$~\footnote{For the Harris sheet-type configuration, the growth rate of the fastest tearing mode becomes $\gamma_t \sim v_{A\lambda} d_e \rho_s/\lambda^3$~\cite{loureiro2017collisionless}, leading to the scale dependence of magnetic fluctuation  $v_{A\lambda} \sim \varepsilon^{1/3}\lambda (d_e \rho_s)^{-1/3}$, the magnetic spectrum $E_B (k_\perp) \sim 4 \pi \rho_0  \varepsilon^{2/3} (d_e \rho_s)^{-2/3} k_\perp^{-3} dk_\perp$, and the spectral anisotropy $k_\parallel \sim \varepsilon^{1/3} (d_e^2/\rho_s)^{1/3} V_A^{-1} k_\perp$.}.
In numerical simulations, electron inertia can be replaced by (hyper-) resistivity as the flux-unfreezing mechanism by taking $d_e\rightarrow 0$ (more precisely, by setting the (hyper) resistivity to be such that the corresponding thickness of the inner layer of the tearing mode is larger than $d_e$).
In this case, if the cascade is indeed tearing-mediated, the expected magnetic spectrum would be
\begin{equation}
    E_B(k_\perp)dk_\perp \propto k_\perp^{-(7n\alpha+2\alpha+2)/(3n\alpha+2)} dk_\perp,
    \label{eq:spec_tearing_hyper}
\end{equation}
where $\alpha$ is the order of the spatial derivative in the (hyper) resistive term and $n=1$ ($n=2$) corresponds to the Harris-sheet (sinusoidal) profile 
[a detailed derivation of this expression is provided in Sec.~\ref{sec:tearing_calcualtion} of the Supplemental Materials (SM)]. The flexibility introduced by the use of different flux-unfreezing methods enables the identification of which of these turbulence models --- intermittency or tearing-mediation --- is a better description of the turbulent cascade at sub-ion scales. This is one of the main goals of our numerical simulations (see Sec.~\ref{sec:spectra}).

\subsection{Phase-space cascade and electron heating} 
\label{sec:theory_mcr}
The above-described models ignore electron kinetic physics. In reality, however, electrons are not isothermal and the kinetic physics of electrons leads to complex dynamics in velocity space, relevant to energy dissipation and electron heating. 
It follows that velocity-space dynamics is as important to the understanding of KAW turbulence as real-space dynamics --- and, indeed, we will show that the two are tightly connected.

In velocity space, the phase-mixing rate can be estimated through terms on the {\it r.h.s.} of Eq.~\eqref{eq:gm} which represent the transfer of free energy from the $m$-th to $(m+1)$-th moment.
In the large-$m$ limit, these terms can be written approximately as the derivative with respect to $m$, from which follows that the phase-mixing rate can be approximated as $\sim |k_\parallel| v_{{\rm th}e}/\sqrt{m}\sim (v_{{\rm th}e}/B_0)\{A_z,...\}/\sqrt{m}$~\cite{zocco2011reduced,schekochihin2016phase}.
In position space, on the other hand, free energy cascades to smaller spatial scales at the nonlinear advection rate $d/dt \sim (c/B_0)\{\varphi, ...\}$ ({\it l.h.s.} of Eq.~\eqref{eq:gm}).
These rates are not, in general, the same. 
At each scale $\lambda$, there is a critical Hermite mode number, denoted as $m_{\rm cr}$, at which these two rates balance:
\begin{equation}
\frac{c}{B_0}\{\varphi, g_{m_{\rm cr}}\} \sim \frac{v_{{\rm th}e}}{B_0} \{A_z,g_{m_{\rm cr}}\}/\sqrt{m_{\rm cr}},
\label{eq:mk_balance}
\end{equation}
resulting in the relation $\sqrt{m_{\rm cr}} \sim  (v_{{\rm th}e}/c) A_{z \lambda}/\varphi_\lambda$.
Combined with the GK Poisson's law (Eq.~\eqref{eq:gk_poisson}) and the equipartition relation (Eq.~\eqref{eq:equipartition}), we obtain the spatial scale dependence of the critical Hermite order:
\begin{equation}
m_{\rm cr} (\lambda) \sim (\lambda/d_e)^2/(2\tau^2).
\label{eq:mcr}
\end{equation}
That is, for $\sqrt{m} k_\perp d_e \gg 1$, the nonlinear advection is expected to dominate, and the rapid coupling between modes with different $k_\parallel$ can result in a new mode with negative $k_\parallel$ and with a returning free-energy flux to lower Hermite moments.
This classic phenomenon is know as the ``plasma echo''~\cite{Gould1967echo,malmberg1968plasma,schekochihin2016phase}. 

In many astrophysical environments, collisions are so infrequent that the collisional cutoff in velocity space will only occur at asymptotically large $m$ ($\gg m_{\rm cr}$). 
Therefore, one may expect the free energy to reach $m_{\rm cr}$ and return to the low Hermite moments due to the restoring stochastic plasma echo. 
Kinetic turbulence would thus resemble fluid turbulence, in the sense that the only available route for energy dissipation would be the real-space cascade.
However, this argument is seemingly at odds with measurements of strong electron heating in the near-Earth solar wind~\cite{chen2019evidence} and in kinetic simulations~\cite{howes2018spatially,klein2016measuring,howes2017diagnosing}. Consistently, in the numerical results reported in Sec.~\ref{sec:numerical_results}, we observe no evidence for plasma echoes.
We argue that this paradox is resolved by the presence of spontaneously formed current sheets in kinetic turbulent systems.
On the one hand, efficient phase mixing is expected (as a result of the local suppression of nonlinearity) around current sheets~\cite{loureiro2013fast,numata2015}. 
On the other hand, it is at the sites with large gradient of current density (around the edge of current sheets) that most of the energy in fluid quantities is pumped into Hermite moments, because the coupling between them is through $\propto \hat{\bb{b}}\cdot \bb{\nabla}(e/cm_e)d_e^2\nabla_\perp^2A_z \propto \hat{\bb{b}}\cdot \bb{\nabla} J_z$ [last term in Eqs.~\eqref{eq:ge} and \eqref{eq:gm}].
The combination of these two effects of current sheets locally regulates the kinetic turbulence so that the free energy can freely transfer to large Hermite moments, unimpeded by echo, and collisionally dissipate, leading to electron heating.
Therefore, the phase-mixing-dominated regime in phase-space can be much wider than the estimation (Eq.~\eqref{eq:mcr}) based on the simple time-scale comparison (Eq.~\eqref{eq:mk_balance}).

\subsection{A lowest-order solution for the Hermite-expansion coefficients of $g_e$}
\label{sec:theory_zeroth_solution}
Based on the time-scale comparison discussed in the previous section, we present a lowest-order solution for the perturbed electron distribution function $g_e$ in velocity space, valid in the phase-mixing-dominated regime.
In Eq.~\eqref{eq:gm}, each term can be ordered based on its inherent frequency.
In the phase-mixing-dominated regime, as discussed in Sec.~\ref{sec:theory_mcr}, the {\it l.h.s.} (proportional to the nonlinear-advection rate) of Eq.~\eqref{eq:gm} is much smaller than its {\it r.h.s.} (proportional to the phase-mixing rate). 
To lowest order, the two terms on the {\it r.h.s.} should therefore balance, and each Hermite moment should satisfy the relation
\begin{equation}
\begin{aligned}
    \label{eq:gm_solution}
    &g_{m+1} = -\sqrt{m/(m+1)} g_{m-1} &\text{for} \ m \ge 3,\  \text{and} \\
    &g_3 = -2/\sqrt{3}\ e/(cm_ev_{{\rm th}e})d_e^2\nabla_\perp^2 A_z.
\end{aligned}
\end{equation}
That is, to lowest order, all the odd Hermite moments share the same spatial configuration as the current, and all the even ones that of the temperature fluctuations, $g_2$.

The Hermite spectrum of $g_e$, $E_m \equiv \braket{|g_m|^2}/2$ where $\braket{...}$ represents the volume average, can be derived from Eq.~\eqref{eq:gm_solution}.
With the approximation $E_m \approx \braket{|g_{m}||g_{m-1}|}/2$ at large $m$, and using the relation $\lvert g_{m+1}/g_{m-1} \rvert=\sqrt{m/(m+1)}$, we obtain $E_{m+1}/E_{m-1} \approx g_{m+1}g_{m}/(g_{m-1}g_{m-2})=\sqrt{(m-1)/(m+1)}$.
This recursive relation leads to the solution $E_m \propto m^{-1/2}$ --- the same as the expected spectrum in a linear system with a Landau-damped kinetic field, corresponding to a constant flux of free energy transferring from small to large velocity moments~\cite{zocco2011reduced,kanekar2015fluctuation}.

\subsection{Turbulence at scales below the electron skin depth, $\lambda\ll d_e$}
\label{sec:sub_de}
Lastly, one may also use the KREHM formalism to investigate the turbulent cascade at scales below $d_e$, i.e., in the range $\lambda \ll d_e \lesssim \rho_i$, where fluctuations become predominantly electrostatic.
In this range, equipartition between the density fluctuations and kinetic energy of the parallel electron flows,
 $(\delta n_{e}/n_{0e})^2 n_{0e}T_{0e} \sim d_e^2 |\nabla_\perp^2 A_{z}|^2/8\pi$ is expected, leading to the relation $\varphi_\lambda \sim (\rho_iV_A/c) d_e \delta B_{\perp \lambda}/\lambda$ (invoking Eq.~\eqref{eq:Gk_poisson_krhoi_gg_1}).
 Since electromagnetic effects are subdominant in this scale range, neither the intermittency corrections nor the tearing mediation that we discussed earlier are expected to be important here. One might thus expect a standard Kolmogorov-type cascade model to be a reasonable description of turbulence at these scales.
 The energy flux of the cascade is $\varepsilon \sim \gamma_{\rm nl} e^2 n_{0e} \varphi^2/T_{0e}$, where the eddy turn-over rate is $\gamma_{\rm nl} \sim \varphi_\lambda/\lambda^2$.
Together with the equipartition relation, we obtain the kinetic and magnetic spectra
\begin{equation}
\begin{aligned}
    E_\varphi (k_\perp)  \propto k_\perp^{-7/3} , \quad  E_B(k_\perp) \propto k_\perp^{-13/3}.
    \label{eq:spec_sub_de}
\end{aligned}
\end{equation}
Imposing critical balance of these fluctuations, that is, $\omega_{\rm KAW}\sim\gamma_{nl}$ (at $k_\perp d_e \gg 1$, $\omega_{\rm KAW}\sim k_\parallel V_A \rho_s /d_e$), we obtain $k_{\parallel} \propto k_\perp^{4/3}$.

As before, let us now estimate the critical Hermite moment, $m_{\rm cr}$, where the rate of free energy transfer between consecutive Hermite moments through linear phase mixing balances the nonlinear advection rate, following Eq.~\eqref{eq:mk_balance}.
Combined with the equipartition relation, we obtain $ m_{\rm cr} \sim (\lambda/d_e)^4/(2\tau^2)$.
That is, at scales $k_\perp d_e \gg 1$, the nonlinear advection of free energy is always faster than its channeling to larger Hermite moments. 
Linear phase mixing should therefore be subdominant as an energy dissipation channel, and the isothermal closure should be a good approximation to the dynamics at these scales.

\section{Numerical setup}
\label{sec:numerical}
We perform direct numerical simulations of the KREHM equations (see Section~\ref{sec:KREHM}) using the pseudo-spectral code {\tt Viriato}~\cite{loureiro2016viriato}. 
In what follows, quantities are given in dimensionless form.
The domain is a triply periodic cubic box with sides of length $L=2\pi$. 
Hyper-diffusion terms of the form $\nu_H\bb{\nabla}_\perp^6$ are
included in the {\it r.h.s.} of Eq.~\eqref{eq:ne} (electron hyper-viscosity), Eq.~\eqref{eq:Az} (hyper-resistivity), and Eq.~\eqref{eq:gm}, with $\nu_H$ dynamically adjusted to absorb energy at the grid scale~\cite{loureiro2016viriato}.
A hyper-collision operator of the form $-\nu_{\rm coll}m^6$ is added to the {\it r.h.s.} of Eq.~\eqref{eq:gm} for $m \geq 2$, with $\nu_{\rm coll}$ likewise set to remove energy at the smallest velocity scale, determined by $M$, the order of the highest Hermite polynomial kept in simulations.
The simulations are driven with a white-noise forcing term added to Eq.~\eqref{eq:ne}, which injects energy into the largest scales in the simulation box (perpendicular and parallel wavenumbers ranging from 1 to 2).
Once a steady state is reached, the energy injection is balanced by its dissipation either through hyper-collisions at large $m$ or through hyper-diffusion at large $k_\perp$, self-consistently determined by the nature of kinetic turbulence. As mentioned in Section~\ref{sec:intro}, it is one of the goals of this study to identify which of these dissipation channels is privileged by the turbulence.

We perform one set of simulations with isothermal electrons (i.e., $g_e=0$), and another set containing electron kinetic effects ($g_e\ne0$, meaning that Eq.~(\ref{eq:gm}) is solved for the Hermite moments of electron perturbed distribution).
The key simulation parameters are summarized in Table~\ref{table:iso_sim} and Table~\ref{table:gm_sim}, respectively.
For the isothermal simulations, we focus on the sub-ion range ($\rho_i>k_\perp^{-1}>d_e$) and vary the dominant mechanisms that unfreeze the magnetic flux (such as to be able to distinguish intermittency from tearing in regards to the physical mechanism that determines the energy spectrum; see discussion in Section~\ref{sec:KAW_turbulence}). 
For the non-isothermal (i.e., kinetic) simulations, we investigate three dynamical ranges of interest: (1) the inertial range $L>k_\perp^{-1}>\rho_i$, (2) the sub-ion range $\rho_i>k_\perp^{-1}>d_e$, and (3) the electrostatic range $d_e>k_\perp^{-1}$.
In run {\tt K1}, we include both the inertial and sub-$\rho_i$ ranges, enabling us to study the transition across the $\rho_i$ scale.
However, the resolution of the $d_e$ scale is sacrificed and the effect of electron inertia on turbulence is not properly taken into account. 
This is compensated by runs {\tt K2(a,b)} where the whole domain is in the sub-ion range and $d_e$ is well resolved.
Run {\tt K3} is designed to focus on the sub-$d_e$ range only.
Runs {\tt K4(a-e)} consist of a scan of the parameter $\rho_i/d_e$ with fixed $d_e$ and varying $\rho_i$ and are used to study the dependence of the energy dissipation on the scale separation between ion and electron scales. 

\begin{table}[h!]
\centering
\begin{tabular}{ |p{0.9cm}|p{0.8cm} p{0.8cm} p{0.8cm} p{1cm}  p{1cm}  p{2.3cm}|  }
 \hline
 Runs & $\rho_i/L$ & $d_e/L$ &  $\eta$ & $\nu_H$ & $N^3$ & flux-unfreezing\\
 \hline \hline
 {\tt Iso-1} & 2 & 0.03  &0 & 0.11e-9  & $512^3$ & electron inertia\\
 {\tt Iso-2}  &2  &0.001  &0.01 & 0.39e-9& $512^3$ &  resistivity\\
 {\tt Iso-3}  &2  &0.001  &0 & 0.39e-9  & $512^3$ &  hyper-resistivity\\
 \hline
\end{tabular}
\caption{Summary of isothermal ($g_e=0$) simulations'  parameters. The last column shows the dominant mechanism that breaks the frozen-flux constraint in the simulation.}
\label{table:iso_sim}
\end{table}
\begin{table}[h!]
\centering
\begin{tabular}{ |p{1cm}|p{2cm} p{1.8cm}  p{1.4cm} p{0.8cm}  p{0.8cm}|  }
 \hline
 Runs & $\rho_i/L$ & $\rho_i/d_e$ & $k_{\perp {\rm max}}d_e$  & $M$ & $N^3$ \\
 \hline \hline
 {\tt K1} & 0.1 &20  &0.9  &30  & $256^3$\\
 {\tt K2a}  &2  &40   &9  &30  & $256^3$\\
 {\tt K2b}  &2 &20   &9  &60  & $128^3$\\
 {\tt K2c}  &2  &66   &5.4  &30  & $256^3$\\
 {\tt K3} &2 &1 & 180 &30  & $128^3$\\
 {\tt K4a-e} &3,2,1,0.3,0.1 &30,20,10,3,1 &9 &30  & $128^3$\\
 \hline
\end{tabular}
\caption{Summary of kinetic ($g_e \ne 0$) simulations'  parameters. $k_{\perp {\rm max}}$ is the maximum value of the perpendicular wavenumber, representing the resolution of simulations. }
\label{table:gm_sim}
\end{table}

\section{Numerical results}
\label{sec:numerical_results}

In Sec.~\ref{sec:spectra}, we present the results from runs {\tt Iso-(1-3)} (with isothermal electrons) on the spatial energy spectra, structure functions of the fluctuations, and intermittency. 
With such understanding of the isothermal limit of sub-$\rho_i$ turbulence, in Sec.~\ref{sec:gm_space} we proceed to take into account the electron kinetic effects and study the spectra in position and velocity space, the energy dissipation and electron heating, and the correlation of Hermite moments.

\subsection{Energy spectra set by intermittency}
\label{sec:spectra}
\begin{figure}[htp]
    \centering
    \includegraphics[width=0.5\textwidth]{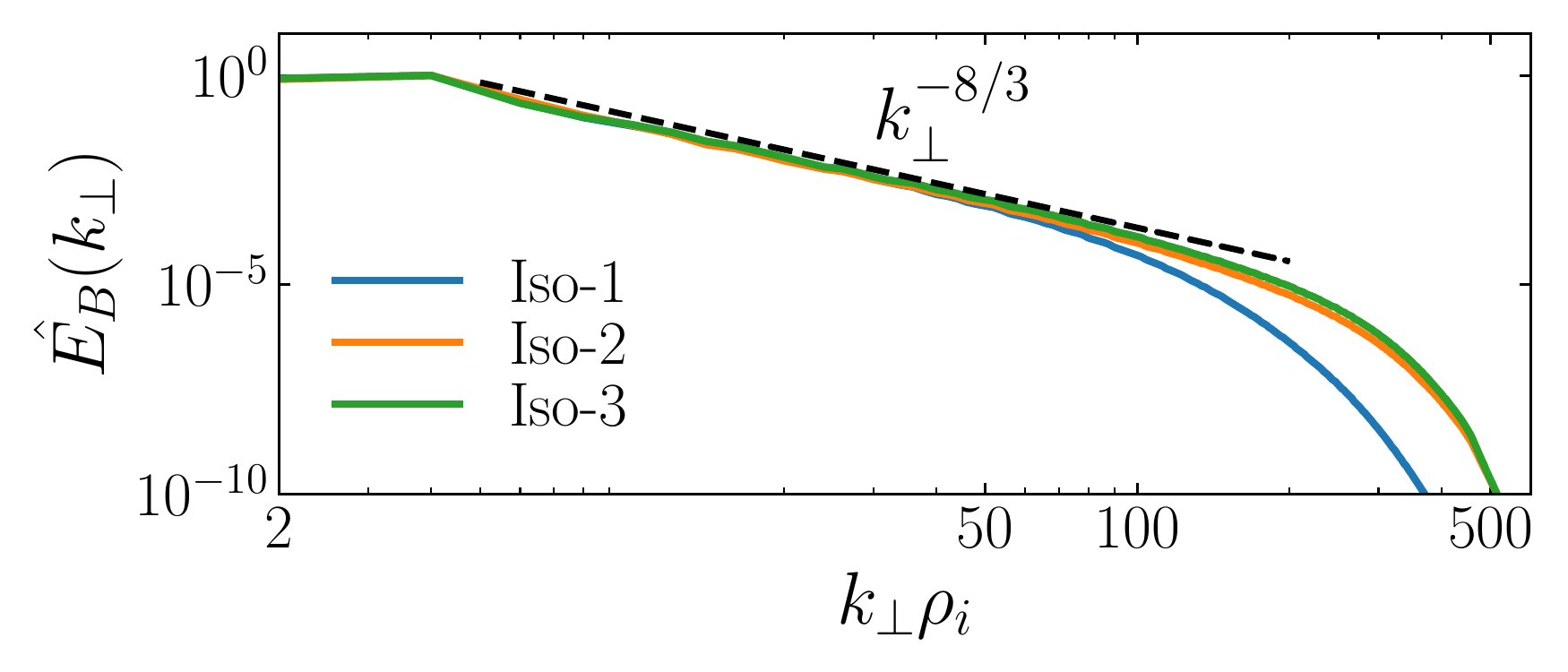}
    \includegraphics[width=0.45\textwidth]{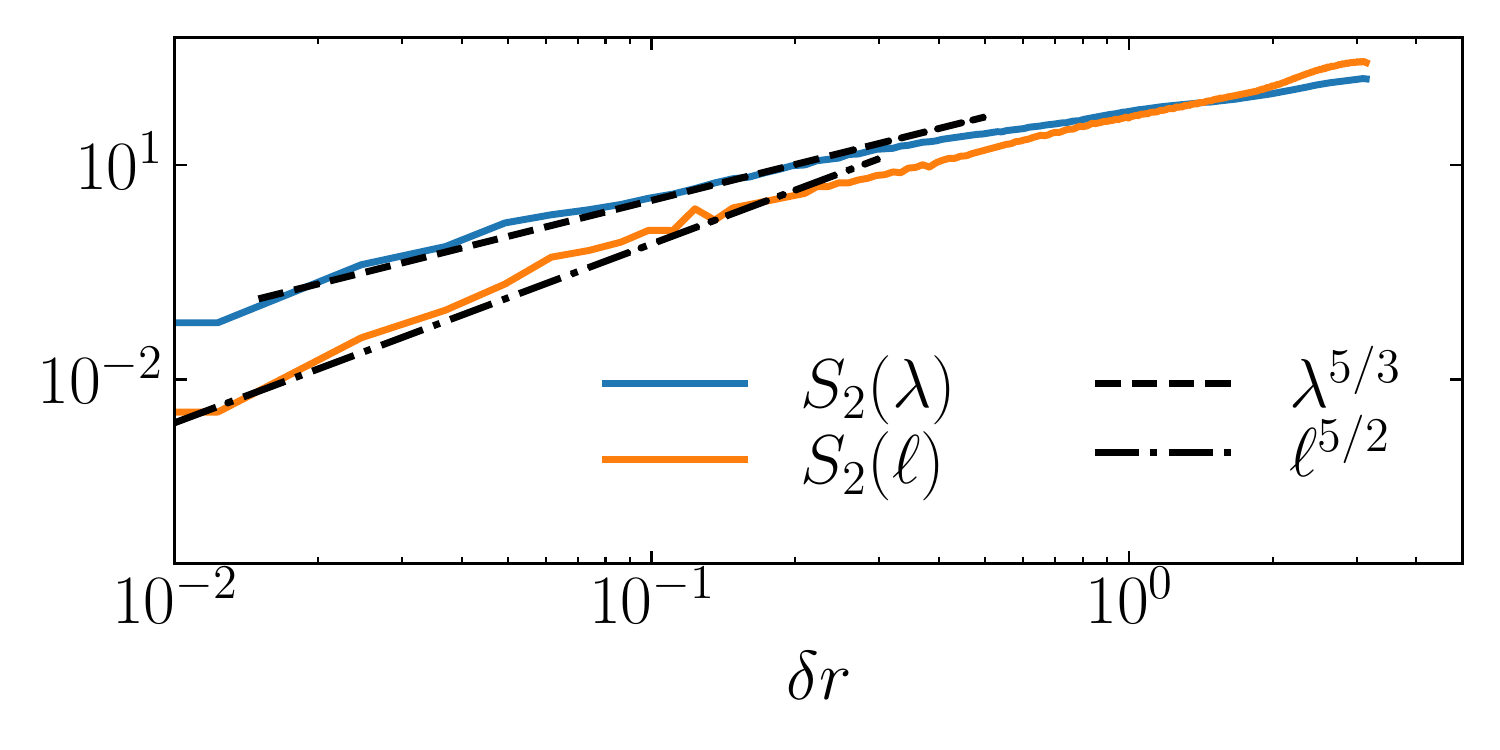}
    \includegraphics[width=0.45\textwidth]{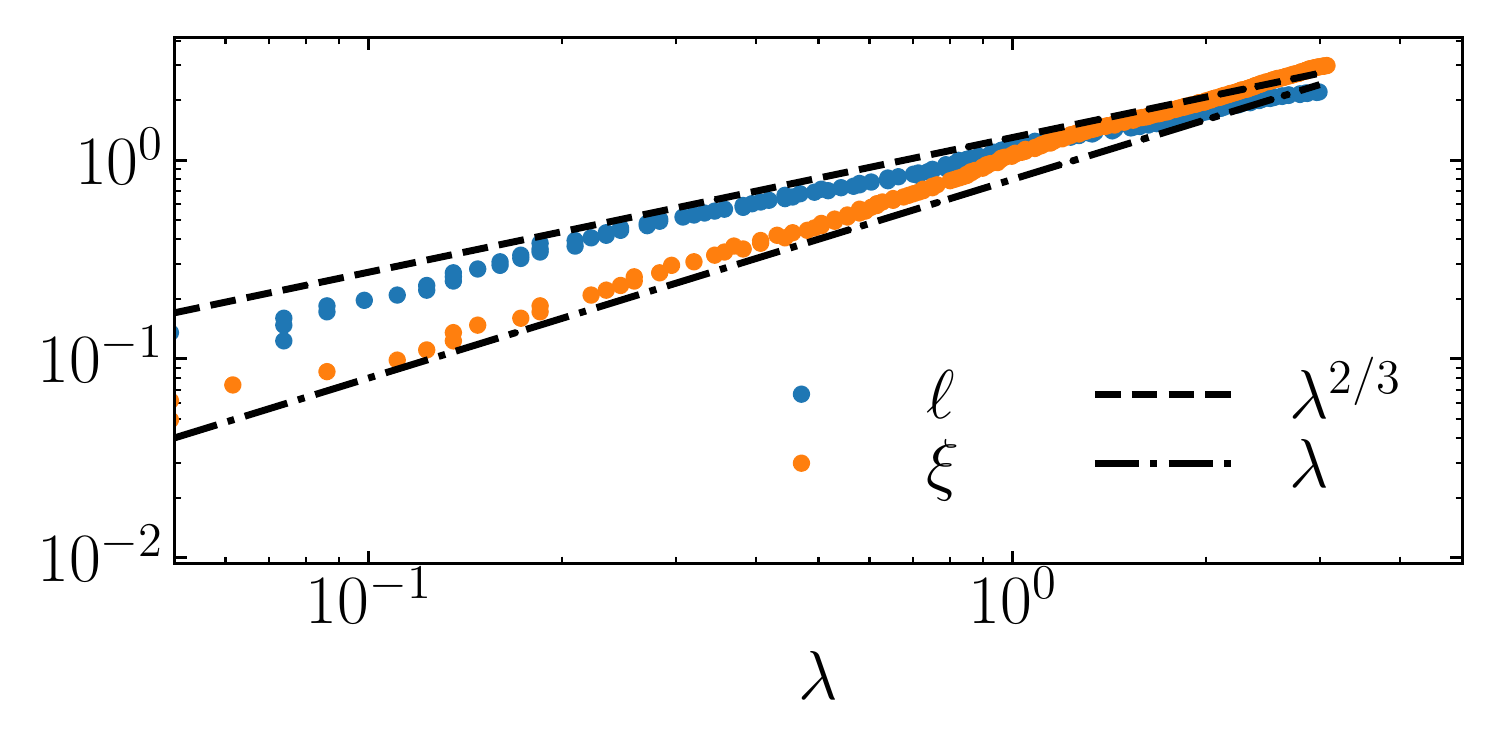}
    \caption{Comparison of magnetic spectra from isothermal runs (see Table~\ref{table:iso_sim}) with different flux-freezing-breaking mechanisms (top). Structure functions (middle) and spectral anisotropy (bottom) from the run {\tt Iso-1} (the other Iso runs show the similar results).}
    \label{fig:spec_compare_Iso}
\end{figure}

In Sec.~\ref{sec:theory_spectra} we discussed three models for energy spectra based on distinct physics that might govern the nonlinear cascade.
As we noted, in spite of the different underlying physical mechanisms, the spectra and spectral anisotropy predicted by the intermittency model and the tearing-mediation model (with electron inertia effects) are the same:
$E_B(k_\perp) \propto k_\perp^{-8/3}$, $E_B(k_\parallel) \propto k_\parallel^{-7/2}$, and $k_\parallel \propto k_\perp^{2/3}$.
This prediction is confirmed by run {\tt Iso-1} in which $d_e$ is resolved (and thus with electron inertia effects accounted for). 
The normalized (to its maximum value) magnetic spectrum exhibits a $\propto k_\perp^{-8/3}$ scaling, as shown in Fig.~\ref{fig:spec_compare_Iso}, top panel.

To obtain a more detailed description of the turbulent fluctuations in these simulations we compute the three-dimensional structure functions of the magnetic fluctuations $S_q(\delta r) \equiv \braket{|\Delta B(\bb{r},\delta \bb{r})|^q}_{\bb{r}}$, where $\delta r$ is the incremental scale, $\Delta B(\bb{r},\delta \bb{r})$ is the local-field increment calculated with a five-point stencil, and $q$ is the order of the structure function. 
The detailed description of this diagnostic and of its numerical implementation can be found in Sec. 3 in SM.
We use the second-order structure function $S_2(\delta r)$ to study the parallel spectrum and spectral anisotropy by projecting it into local coordinates with respect to the local magnetic field.
We denote by $S_2(\lambda)$ and $S_2(\ell)$ the field-perpendicular and field-parallel structure functions, which are mapped to the perpendicular and parallel magnetic energy spectrum through Parseval's theorem: a $\sim k^{-\alpha}$ energy spectrum corresponds to the structure function $S_2(\delta r)\propto \delta r^{\alpha-1}$ with $\delta r \sim k^{-1}$.  
Therefore, the $E_B(k_\perp)\propto k_\perp^{-8/3}$ (Eq.~\eqref{eq:Bspec_intermittent}) and $E_B(k_\parallel) \propto k_\parallel^{-7/2}$ (Eq.~\eqref{eq:Bspec_intermittent_para}) spectra should correspond to the scalings $S_2(\lambda)\propto \lambda^{5/3}$ and $S_2(\ell)\propto \ell^{5/2}$, respectively.
These predicted scalings are consistent with the measurements shown in the middle panel of Fig.~\ref{fig:spec_compare_Iso}.
The predicted spectral anisotropy based on the critical balance of the turbulent fluctuations, $k_\parallel \propto k_\perp^{2/3}$, is also confirmed by the measured $\ell \propto \lambda^{2/3}$ scaling shown in the bottom panel. 
Finally, we find that the fluctuations exhibit a $\xi \propto \lambda$ scaling (where $\xi$ is the coherence length in the fluctuation direction, as explained in section~\ref{sec:theory_spectra}) on perpendicular (to the local magnetic field) planes.
This scaling indicates the isotropic morphology of eddies in perpendicular planes.
That is, the electromagnetic fluctuations in our simulations do not become progressively more aligned with decreasing scale (in contrast with what is conjectured to happen in MHD turbulence~\cite{boldyrev2006spectrum, mallet2016,schekochihin2020mhd}), which is the necessary premise for the hypothesis of tearing-mediation of the energy cascade.

It is worth noting that the three-dimensional eddy anisotropies at sub-ion scales have recently been investigated through {\it in situ} measurements by the {\it MMS} in the Earth's magnetosheath~\cite{wang2020observational} and by the {\it PSP} in the inner heliosphere~\cite{zhang2022three}; the scalings between $\ell$, $\xi$, and $\lambda$ that these investigations report are quantitatively similar to our measurements shown in the bottom panel of Fig.~\ref{fig:spec_compare_Iso}.

As explained earlier, the above results for the magnetic spectrum, by themselves, do not clarify whether it is intermittency or tearing-mediation that determines the spectrum.
To shed light on this issue, we perform two additional runs whose only difference from the one we just discussed ({\tt Iso-1}) is the flux-unfreezing mechanism that is used: instead of electron inertia, it is Laplacian resistivity for run {\tt Iso-2}, and hyper-resistivity for run {\tt Iso-3}. 
In the top panel of Fig.~\ref{fig:spec_compare_Iso}, we compare the perpendicular magnetic spectra from all three isothermal runs.
If the spectra were to be set by tearing-mediation, we would expect the scalings $E_B(k_\perp)\propto k_\perp^{-2.4}$ and $E_B(k_\perp)\propto k_\perp^{-2.6}$ for runs {\tt Iso-2} and {\tt Iso-3}, respectively, according to Eq.~\eqref{eq:spec_tearing_hyper} for the $n=2$ sinusoidal profile.
Instead, we observe a clear overlap of the spectra for all three simulations (in the inertial range). 
This finding, combined with the measured $\xi \sim \lambda$ scaling (Fig.~\ref{fig:spec_compare_Iso}, bottom panel), rules out tearing-mediation as the physical mechanism underpinning the KAW turbulent cascade in our simulations, and leaves intermittency as the only known explanation for the $k_\perp^{-8/3}$ magnetic spectrum in the low-$\beta$, sub-$\rho_i$ regime.

\begin{figure}[htp]
    \centering
    \includegraphics[width=0.5\textwidth]{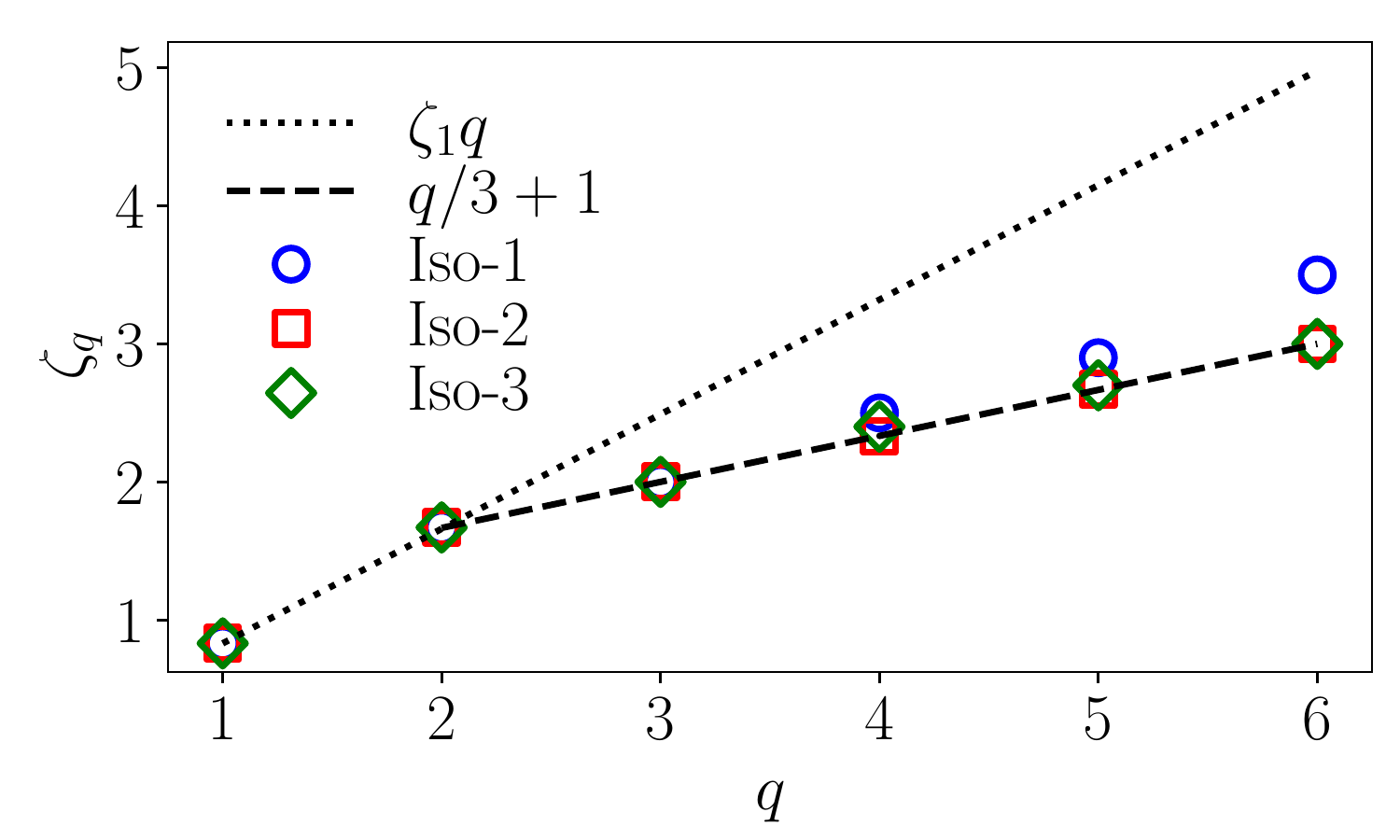}
    \caption{The exponents of $q$-th-order structure functions in the perpendicular plane for isothermal runs. The deviation of high-order exponents from the linear scaling $\zeta_1 q$ (black dashed line) indicates the presence of intermittency.}
    \label{fig:intermittency}
\end{figure}

To characterize the intermittency of the turbulent fluctuations in our simulations we investigate the higher order ($q>2$) structure functions.
In Fig.~\ref{fig:intermittency}, we present the dependence on the order $q$ of the measured index $\zeta_q$ for the field-perpendicular structure function $S_q(\lambda) \propto \lambda^{\zeta_q}$ from the three isothermal runs.
In a turbulent system with purely self-similar fluctuations (i.e., no intermittency), a linear scaling $\zeta_q =\zeta_1 q$ is expected~\cite{frisch1995turbulence}.
The deviation from this linear scaling as $q$ increases confirms the presence of intermittency in our system. 
Furthermore, the measured scaling for $\zeta_q$ is consistent with the $\beta$-model~\cite{frisch1995turbulence} and the main assumption made in ~\citet{boldyrev2012spectrum} that the intermittent structures are essentially 2D.
There are two contributions to $S_q(\lambda)$: the fluctuations $\delta B_{\perp \lambda}$ from the ``active'' intermittent regions, and their volume-filling fraction $p_\lambda$, leading to the relation
 $S_q(\lambda) \equiv \braket{|\Delta B(\bb{r},\lambda)|^q}_{\bb{r}} \sim \delta B_{\perp \lambda}^q p_\lambda \sim \lambda^{q/3+1}$. 
 This $\zeta_q=q/3+1$ scaling for intermittency is indeed measured in the our simulations (with the small deviation of run {\tt Iso-1} caused by the relatively narrow inertial range compared to the other runs.).

\subsection{Electron heating and the absence of plasma echo}
\label{sec:gm_space}

\begin{figure*}[t!]
    \centering
    \includegraphics[width=1.0\textwidth]{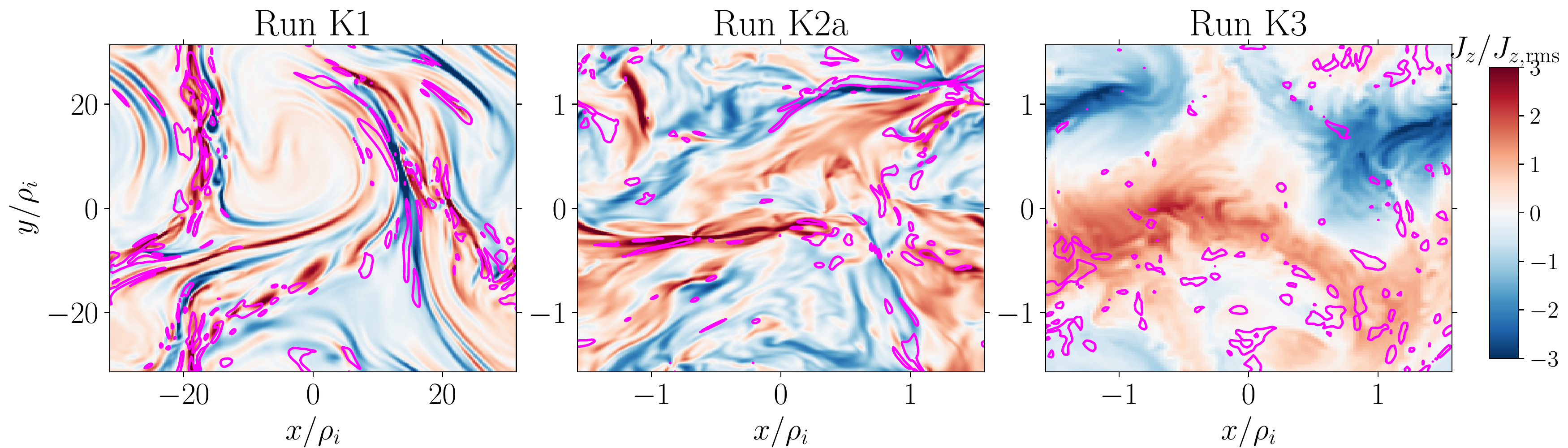}
    \includegraphics[width=0.98\textwidth]{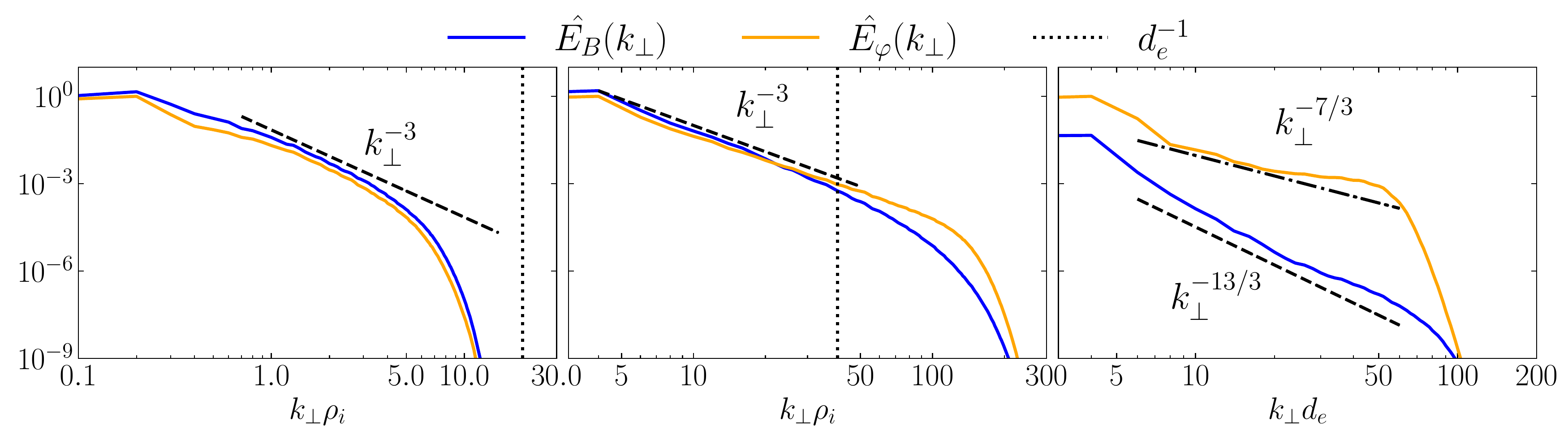}
    \includegraphics[width=1.0\textwidth]{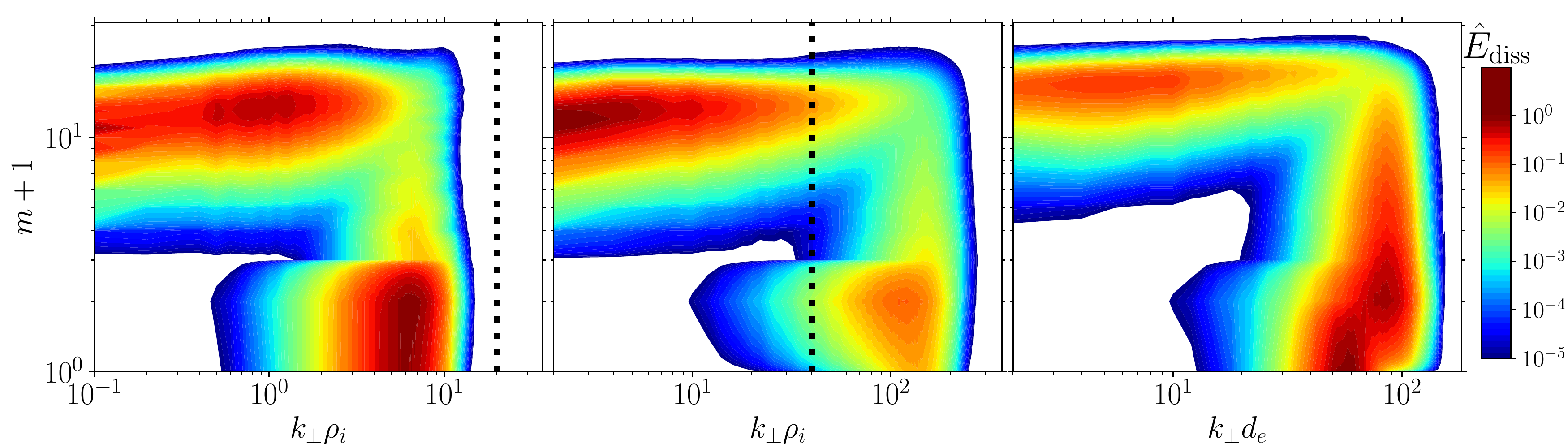}
    \caption {Runs {\tt K1} (left column), {\tt K2a} (middle column), and {\tt K3} (right column). Top: visualization of $J_z$ (color map) and contours of the electron heating rate that is twice its volume-averaged value (magenta lines). Middle: Magnetic energy and density spectra. Bottom: Energy dissipation spectra in $m$-$k_\perp \rho_i$ phase space. The vertical dotted lines indicate the $d_e$ scale.}
    \label{fig:compare_Hruns}
\end{figure*}

With the understanding of the isothermal limit in place, we proceed to investigate the more realistic case where electron kinetic physics --- and, in particular, electron Landau damping --- is included in the simulations (i.e., $g_e\neq0$). As discussed above, this enables the cascade of energy in velocity (or Hermite), as well as in real (or Fourier), space.

In Fig.~\ref{fig:compare_Hruns} we present results from the three main non-isothermal runs ({\tt K1}, {\tt K2a}, and {\tt K3}, corresponding to the left, middle, and right panels) that altogether cover the spatial scales ranging from MHD to sub-$d_e$. 
Run {\tt K1} captures the transition at $\rho_i$ and run {\tt K2} captures that at $d_e$. For each run, the results are plotted at an arbitrarily chosen time during the saturated stage of the turbulence.

\paragraph{Visuals.}
The top row shows the current densities normalized to their root-mean-square values on arbitrarily chosen $xy$ planes. 
The visualization of run {\tt K1} (left panel) is dominated by the features of long-wavelength MHD turbulence, with indications of plasmoid instability~\cite{loureiro2007instability} in some of the large-scale current sheets.
At scales below $\rho_i$ (middle panel), the current still exhibits sheet-type structures, but there are no visual signs of plasmoid generation.
At scales below $d_e$ (right panel), the current profile becomes less sharp and fewer finer structures appear in the system, consistent with the predicted steep magnetic spectrum (Eq.~\eqref{eq:spec_sub_de}) and with the fact that turbulence at these scales should be predominantly electrostatic. 

\paragraph{Energy spectra.}
The middle row shows the normalized perpendicular energy spectra of the magnetic and ion-entropy fluctuations, $\hat{E}_B(k_\perp)$ and $\hat{E}_\varphi(k_\perp)$.
The main purpose of run {\tt K1} (left panel) is to study the transition at around $k_\perp \rho_i \sim 1$, and so both the MHD range $k_\perp \rho_i < 1$ and the kinetic range $k_\perp \rho_i >1$ are included in the simulation, with the compromise that neither of these two ranges are sufficiently wide to form a clear inertial range.
In spite of this limitation, there is a clear steepening of the spectra at $k_\perp \rho_i\sim 1$.
We observe $\hat{E}_B(k_\perp) \approx \hat{E}_\varphi(k_\perp)$ in the sub-$\rho_i$ range, confirming the assumption of equipartition between magnetic and density fluctuations (Eq.~\eqref{eq:equipartition}).

The sub-$\rho_i$ range is explored in run {\tt K2a} (middle panel) where the transition at $k_\perp d_e\sim 1$ is well resolved. 
Both $\hat{E}_B(k_\perp)$ and $\hat{E}_\varphi(k_\perp)$ closely follow a $k_\perp^{-3}$ scaling until they deviate at scales below $d_e$.
This is a steeper spectrum than the predicted (and numerically confirmed) $k_\perp^{-8/3}$ spectrum (Eq.~\eqref{eq:Bspec_intermittent}) in the isothermal limit.
We attribute this steepening of the spectra to electron Landau damping, which enables the dissipation of a non-negligible energy fraction via its cascade to high order Hermite moments through phase-mixing. 
This argument will be elaborated and tested below. 
At scales below $d_e$ (run {\tt K3}, right panel), $\hat{E}_B(k_\perp) \propto k_\perp^{-13/3}$ and $\hat{E}_\varphi(k_\perp) \propto k_\perp^{-7/3}$ are measured, confirming the predictions for the electrostatic limit (Eq.~\eqref{eq:spec_sub_de}).

\paragraph{Energy dissipation.}
The bottom row shows contour maps of the normalized (to its maximum value) dissipation rate of the total free energy (defined in Eq.~\eqref{eq:free_energy}) in the $m$-$k_\perp$ phase space.
Here the $m=0$ and $m=1$ moments correspond to $\delta n_e/n_{0e}$ and $A_z$, respectively.
These results directly address one of the main questions in this paper --- the competition between the kinetic and fluid channels for energy dissipation.
In run {\tt K1} (left panel), we focus on the MHD range and the transition into sub-$\rho_i$ scales. 
The strong energy dissipation at large ($m\gg1$) Hermite moments turns on at around the $\rho_i$ scale and extends towards smaller scales (larger $k_\perp$), suggesting the domination of phase mixing over nonlinear advection at those scales~\footnote{\blue{The large-$m$ dissipation at the system scale is artificial, caused by the forcing.}}.
Because $d_e$ is unresolved in this run, the energy flux in position space reaches the (hyper-)viscous scale without going through the full kinetic range.
This truncation of electron kinetic effects at large $k_\perp$ causes an artificially strong dissipation by the hyper-viscosity (and resistivity), preventing us from making a direct comparison of dissipation at large $m$ and large $k_\perp$ in this run.
This issue is resolved in run {\tt K2a} (middle panel).
With the understanding that the large-$m$ dissipation turns on at $k_\perp \rho_i \sim 1$, we can study the sub-$\rho_i$ range scale and focus on the range between $\rho_i$ and $d_e$, with $d_e$ properly resolved. 
With electron kinetic physics fully accounted for, we find that most of the energy is dissipated at large $m$ (rather than large $k_\perp$).
This result is direct evidence of the absence of plasma echo in the range between $\rho_i$ and $d_e$, and clearly demonstrates that the kinetic channel is dominant for energy dissipation: the free energy transfers to small scales in velocity space through phase mixing, where it dissipates via collisions and heats the electrons.

The large-$m$ dissipation starting at the lowest $k_\perp$ turns off at around the $d_e$ scale. To explore the sub-$d_e$ range, we analyze run {\tt K3} (right panel), and find that the energy dissipation at large $k_\perp$ is clearly dominant. 
This is consistent with our prediction (in Sec.~\ref{sec:sub_de}) that although the Landau damping rate of linear KAWs is large in the sub-$d_e$ range, the nonlinear advection is faster than the phase-mixing process, causing the energy flux to mainly cascade to, and dissipate at, large $k_\perp$ rather than large $m$. 

\begin{figure}[htp]
    \centering
    \includegraphics[width=0.5\textwidth]{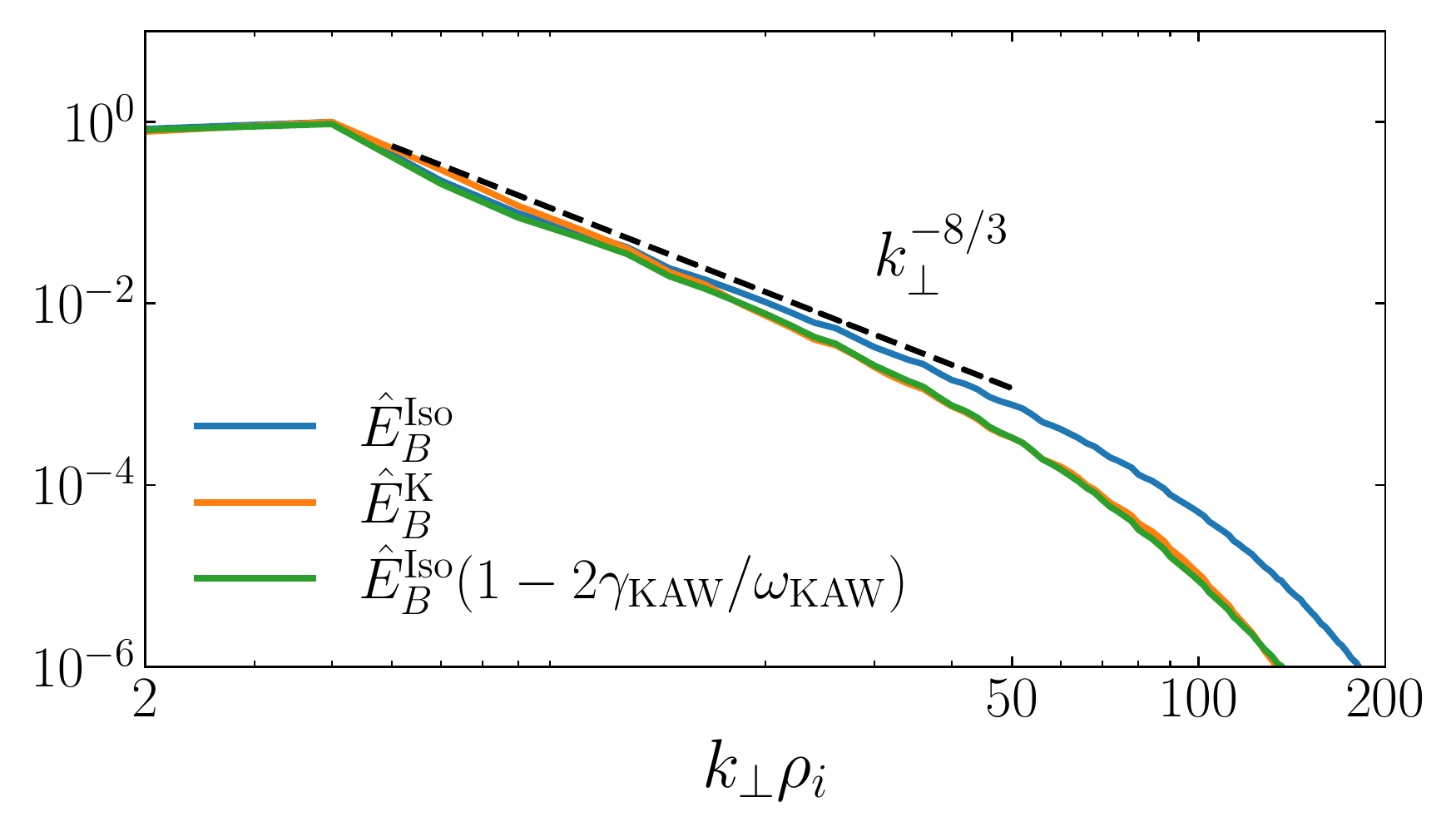}
    \caption{Comparison between the magnetic spectrum from run {\tt K2c} ($\hat{E}_B^{\rm K}$) and that from run {\tt Iso-1} ($\hat{E}_B^{\rm Iso}$) multiplied by a scale-dependent factor $(1-2\gamma_{\rm KAW}/\omega_{\rm KAW})$ that represents the subtraction of the Landau-damped energy within a cascade time. }
    \label{fig:spec_compare}
\end{figure}

\paragraph{Landau damping of KAWs.}
In order to confirm that the strong energy dissipation at high $m$ is indeed caused by the Landau damping of KAWs, we test the following argument with our numerical results.
Let us assume that, at each scale $k_\perp^{-1}$, the electromagnetic fluctuations are damped at the KAW linear damping rate pertaining to that scale, $\gamma_{\rm KAW}(k_\perp)$.
Then, within a nonlinear cascade time $1/\gamma_{nl}$ --- which, by assumption of critical balance, is comparable to the linear time scale, $1/\gamma_{nl}\sim 1/\omega_{\rm KAW}$ --- a fraction of $\sim 2 \gamma_{\rm KAW}/\omega_{\rm KAW}$ of the magnetic energy is damped and becomes the free energy in $g_e$.
All the mentioned quantities are scale-dependent.
With this assumption, if we consider a magnetic spectrum from a simulation with isothermal electrons, $\hat{E}_B^{\rm iso}(k_\perp)$, and subtract from it the energy that would be Landau damped within a cascade time, it should match the spectrum $\hat{E}_B^{\rm K}$ from the corresponding kinetic simulation (i.e., a simulation that differs from the isothermal one only in allowing for kinetic electrons, $g_e\neq0$):
\begin{equation}
   \hat{E}_B^{\rm K}(k_\perp) \approx \hat{E}_B^{\rm iso}(k_\perp)(1-2\gamma_{\rm KAW}/\omega_{\rm KAW}).
    \label{eq:spec_landau_damped_iso}
\end{equation}
This conjecture is confirmed in the comparison of spectra from run {\tt K2a} and its corresponding isothermal run ({\tt Iso-1}), shown in Fig.~\ref{fig:spec_compare}, where the scale-dependent $\gamma_{\rm KAW}$ and $\omega_{\rm KAW}$ are calculated from the linear dispersion relation of KAWs Eq.~\eqref{eq:KAW_DR} (see Sec.~\ref{sec:kaw_dispersion} in SM for more details).
This agreement suggests that the magnetic spectra from kinetic runs, shown in this figure and in Fig.~\ref{fig:compare_Hruns}, which we tentatively fitted with a $k_\perp^{-3}$ scaling, is, in fact, not a simple power law, but rather a more complicated function well approximated by Eq.~(\ref{eq:spec_landau_damped_iso}).
Incidentally, we note that it should not be an artifact of numerical simulations that the sub-$\rho_i$ spectra do not exhibit power-law scalings, since the scale separation between $\rho_i$ and $d_e$ in our simulation is not far-off from that in realistic systems such as the solar wind.
This result is a direct numerical confirmation of the weakened cascade model originally proposed by Howes \textit{et al.}~\cite{howes2008kinetic,howes2011weakened} and numerically tested by TenBarge and co-workers~\cite{tenbarge2013collisionless,tenbarge2013current}.
In a broader context, this result provides a justification for a model~\cite{podesta2010kinetic,howes2010prescription,passot2015model,kunz2018astrophysical} that is widely used in studies of turbulent astrophysical systems, which is based on the assumption that we just confirmed: that the effect of phase mixing is merely to damp the turbulent cascade of the electromagnetic fluctuations at a scale-dependent rate set by linear electron Landau damping.

\begin{figure}[htp]
    \centering
    \includegraphics[width=0.5\textwidth]{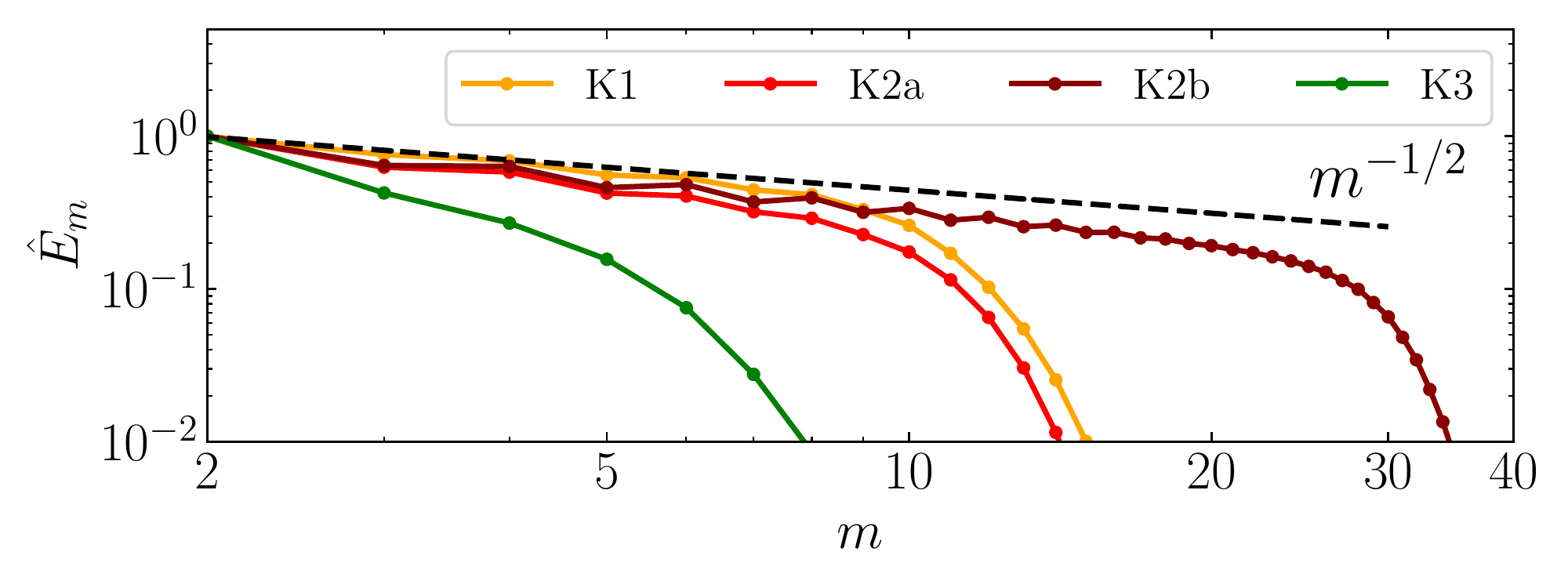}
    \includegraphics[width=0.5\textwidth]{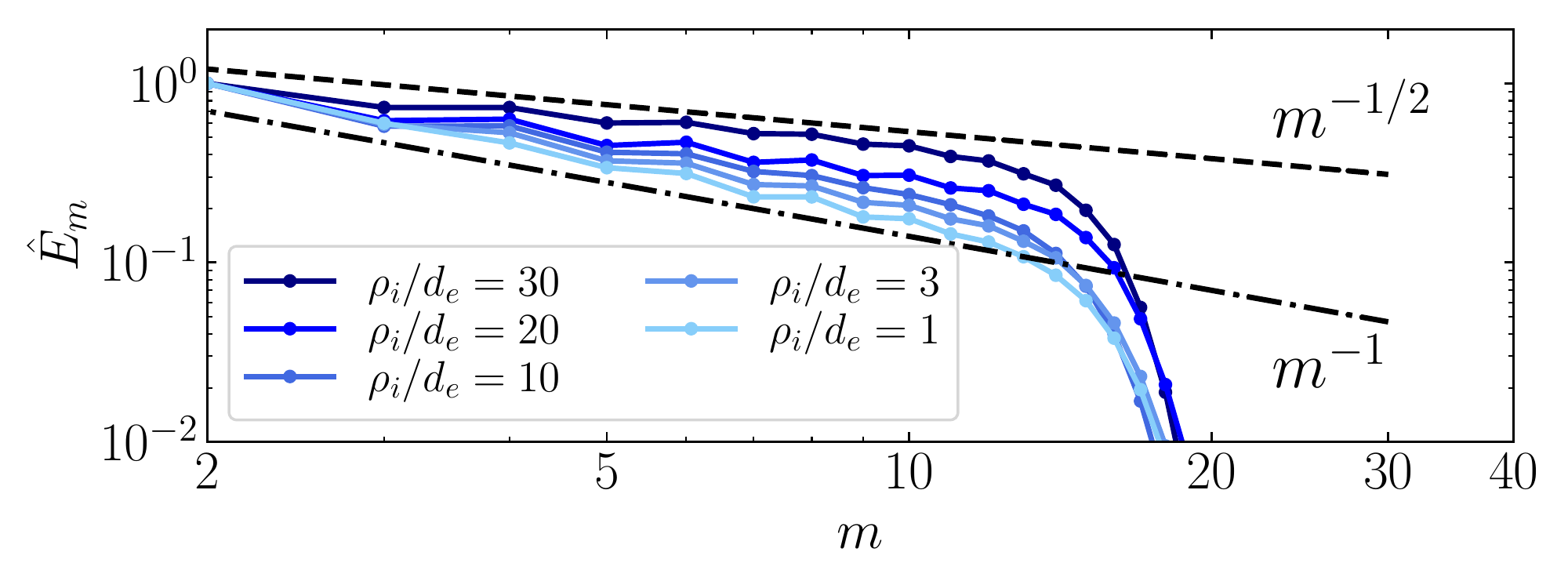}
    \caption{Normalized Hermite spectra from simulations {\tt K1-K3} (top) and {\tt K4a-e} (bottom).  }
    \label{fig:mspec}
\end{figure}

\paragraph{An analytical solution for the Hermite-expansion coefficients of the reduced electron distribution function $g_e$.}
Strong energy dissipation at high Hermite mode numbers $m\gg1$ requires a significant energy cascade towards small scales in the velocity space and should thus correspond to a shallow Hermite spectrum. 
This is consistent with our numerical results shown in the top panel of Fig.~\ref{fig:mspec}.
For runs {\tt K1} and {\tt K2a} with strong large-$m$ dissipation, we measure a shallow $m^{-1/2}$ spectrum.
In contrast, for run {\tt K3} where large-$m$ dissipation is weak and large-$k_\perp$ dissipation dominates, a much steeper Hermite spectrum is measured. 
The $m^{-1/2}$ spectrum in run {\tt K1} is consistent with our earlier explanation that although the dissipation at large-$k_\perp$ is significant in this case, it is due to the truncation of electron kinetic physics at the hyper-viscous scale, and a significant phase-mixing-dominated energy cascade in velocity space does exist at spatial scales larger than the hyper-viscous scale. 

In the bottom panel of Fig.~\ref{fig:mspec} we show the Hermite spectra from runs with different values of $\rho_i/d_e$. 
In these runs, $d_e$ is fixed at small scales (but resolved). A large value of $\rho_i/d_e$ corresponds to a wider electromagnetic sub-$\rho_i$ range where KAWs exist and are subject to Landau damping, whereas runs with small $\rho_i/d_e$ are mostly in the MHD range.
As shown, decreasing values of $\rho_i/d_e$, i.e., a narrower dynamical range for Landau-damped KAWs results in a progressively steeper Hermite spectrum. 
This result further confirms the significant contribution of the Landau damping of KAWs to velocity-space dissipation. 

\begin{figure}[htp]
    \centering
    \includegraphics[width=0.5\textwidth]{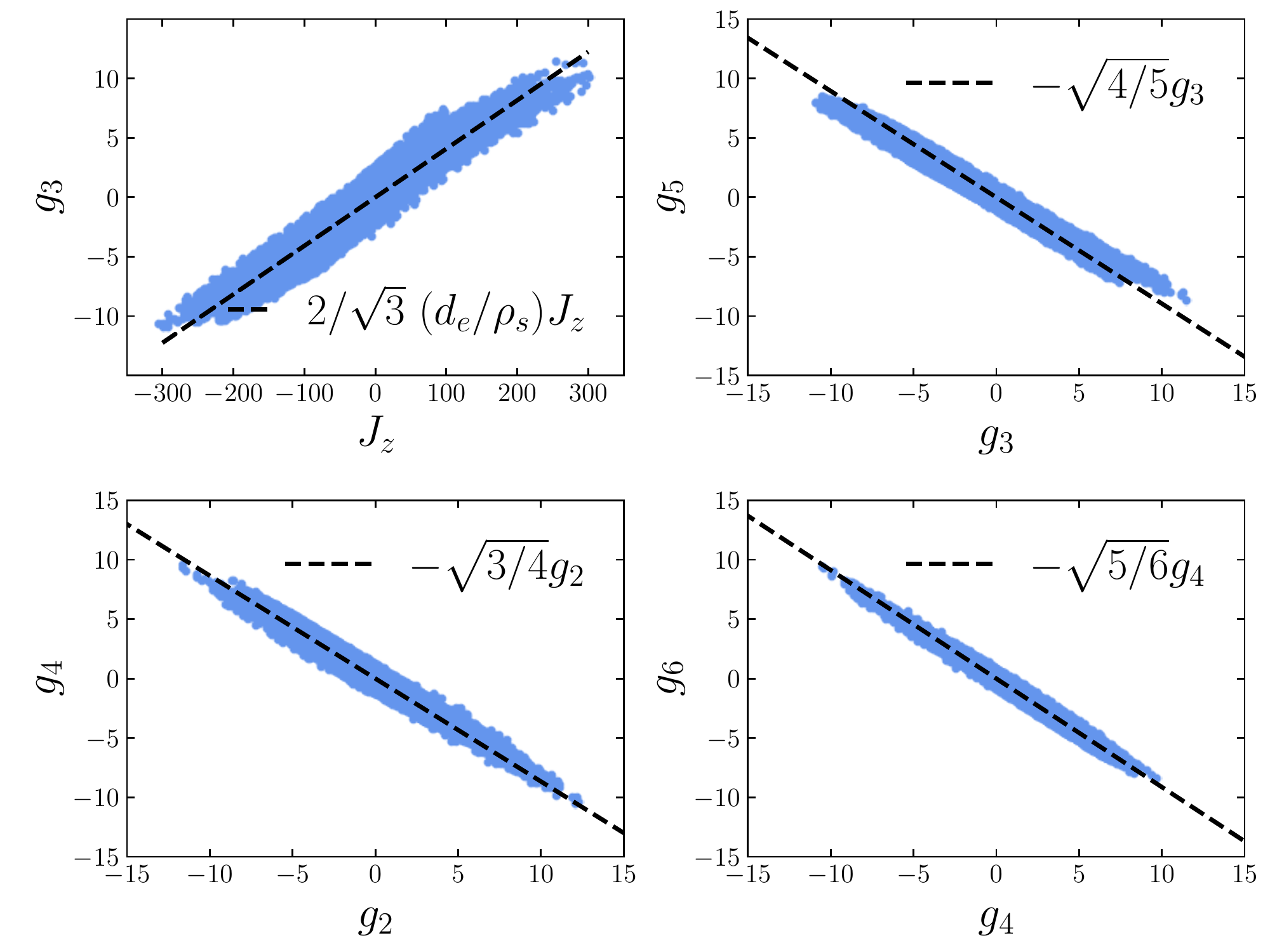}
    \caption{Scatter plots from simulation {\tt K1} showing the correlation between $J_z$ and $g_3$ (top left), $g_5$ and $g_3$ (top right), $g_4$ and $g_2$ (bottom left), and between $g_6$ and $g_4$ (bottom right). }
    \label{fig:scatter_plot}
\end{figure}

The $m^{-1/2}$ scaling is predicted by our analytical model (in Sec.~\ref{sec:theory_zeroth_solution}) based on the timescale ordering between the KAW frequency (which is also the cascade rate) and the phase mixing rate: $\omega_{\rm KAW} \ll k_\parallel v_{{\rm th}e}/\sqrt{m}$; this leads to a lowest-order solution relating different Hermite moments, Eq.~\eqref{eq:gm_solution}.
To further test our theory, we measure the correlation between Hermite moments $g_{m-1}$ and $g_{m+1}$, a few examples of which are shown in Fig.~\ref{fig:scatter_plot}, top and middle panels.
The measured correlations between $J_z$ and $g_3$, $g_5$ and $g_3$, $g_4$ and $g_2$, and between $g_6$ and $g_4$ exhibit remarkable agreement with our lowest-order solution $g_3=2/\sqrt{3}(d_e/\rho_s) J_z$ (in code units) and $g_{m+1}=-\sqrt{m/(m+1)}g_{m-1}$ for $m \geq 3$ (Eq.~\eqref{eq:gm_solution}).

\paragraph{Absence of plasma echoes.}

The unimpeded Hermite cascade of energy that we observe implies that the stochastic echo effect that we discussed earlier is negligible in our simulations. It is revealing to understand why that should be so.
The time-scaling ordering leads to a critical Hermite mode number $m_{\rm cr} \sim (\lambda/d_e)^2/(2\tau^2)$ (Eq.~\eqref{eq:mcr}), below which the phase mixing is faster than the nonlinear advection, and above which the advection can in principle dominate the phase mixing to establish the plasma echo.
That seems to indicate that, for a system with asymptotically large cutoff $M$, the plasma echo should always occur, causing a returning energy flux to low Hermite moments and thus strong energy dissipation at large $k_\perp$.
Even for a system with limited $M$ (as that in our simulations), at sufficiently small scale $\lambda$, $m_{\rm cr}$ will be resolved ($m_{\rm cr}<M$), and phase mixing should be suppressed at those scales. 
However, these arguments are contradictory to our numerical results of the $m^{-1/2}$ Hermite spectrum and efficient electron heating.

\begin{figure}[htp]
    \centering
    \includegraphics[width=0.5\textwidth]{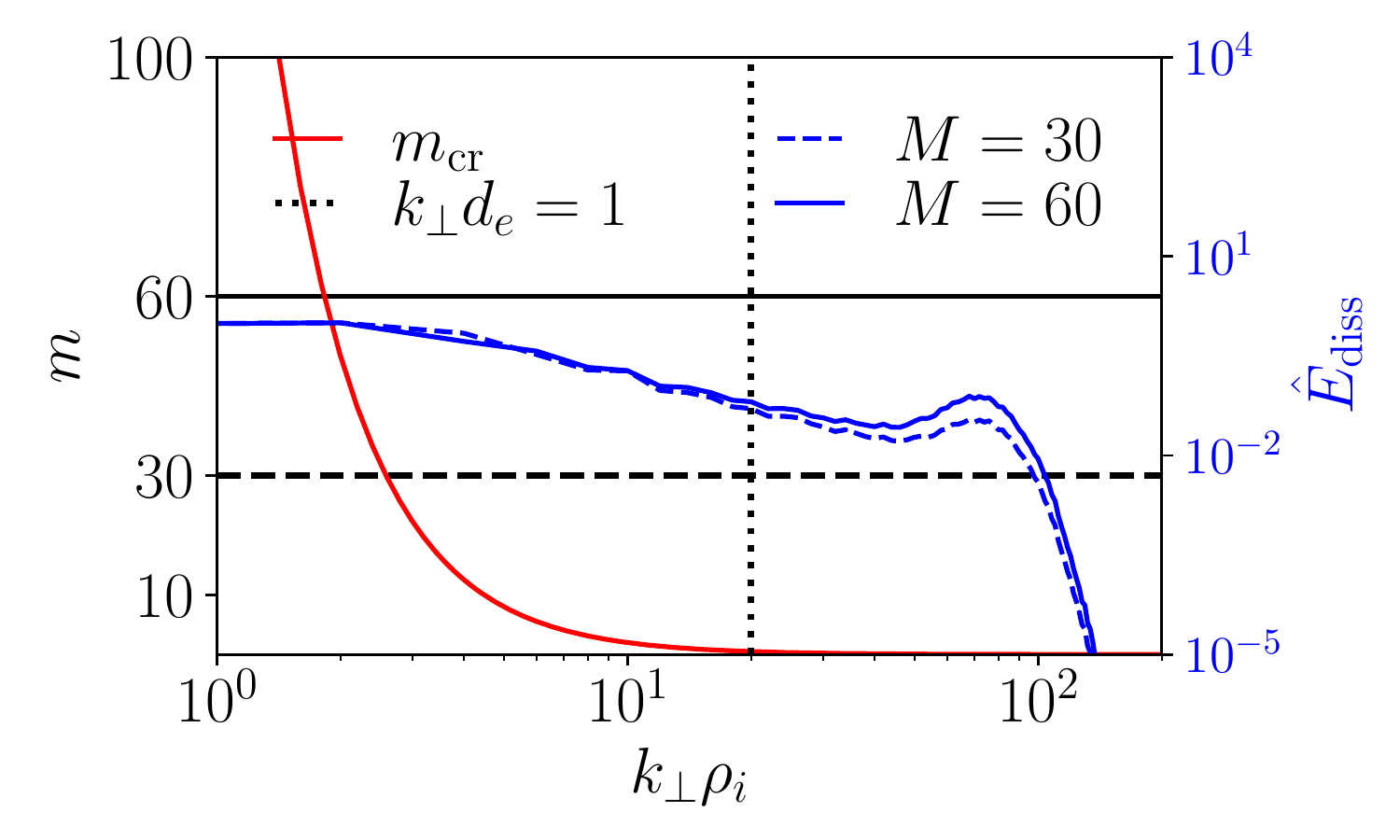}
    \caption{Comparison between normalized total energy dissipation spectra (integrated over $m$) between run {\tt K2b} (with $M=60$; blue solid curve) and run {\tt K4b} (with $M=30$; blue dashed line). Also in the figure is the dependence of the critical Hermite moment order, $m_{\rm cr}$ (Eq.~\eqref{eq:mcr}), as a function of $k_\perp \rho_i$. 
    The horizontal lines indicate the cutoff of $m$ at 30 and 60.
    For both $M=30$ and $M=60$ runs, a wide range of wavenumbers exists that in principle allows plasma echo to occur. }
    \label{fig:compare_M30_M60}
\end{figure} 

To further confirm the absence of echo, we perform two additional runs, {\tt K2b} and {\tt K4b}, between which the only difference is the number of Hermite moments ($M=30$ for {\tt K4b} and $M=60$ for {\tt K2b}).
The Hermite spectrum of {\tt K2b} also exhibits a $m^{-1/2}$ scaling (Fig.~\ref{fig:mspec}, top panel).
In Fig.~\ref{fig:compare_M30_M60} we compare the dissipation rate of total free energy between the {\tt K2b} ($M=60$) and {\tt K4b} ($M=30$) runs (dashed and solid blue curves).
The good overlap indicates that the features of energy dissipation remain unchanged when including more Hermite moments and thus extending the range of spatial scales over which a potential plasma echo could arise.
In the same figure we follow Eq.~\eqref{eq:mcr} and plot the dependence of the critical Hermite mode number $m_{\rm cr}$ on $k_\perp \rho_i$ (red curve).
Based on the comparison between advection and phase-mixing rates (Eq.~\eqref{eq:mk_balance}), the plasma echo is allowed in the region above the red curve and below the cutoff $M$, which is a wide region for both runs.
Therefore, the plasma echo is absent not because it is excluded in our simulations due to limited resolution in position or velocity space; instead, its absence must be due to the turbulent dynamics.

\begin{figure}[htp]
    \centering
    \includegraphics[width=0.5\textwidth]{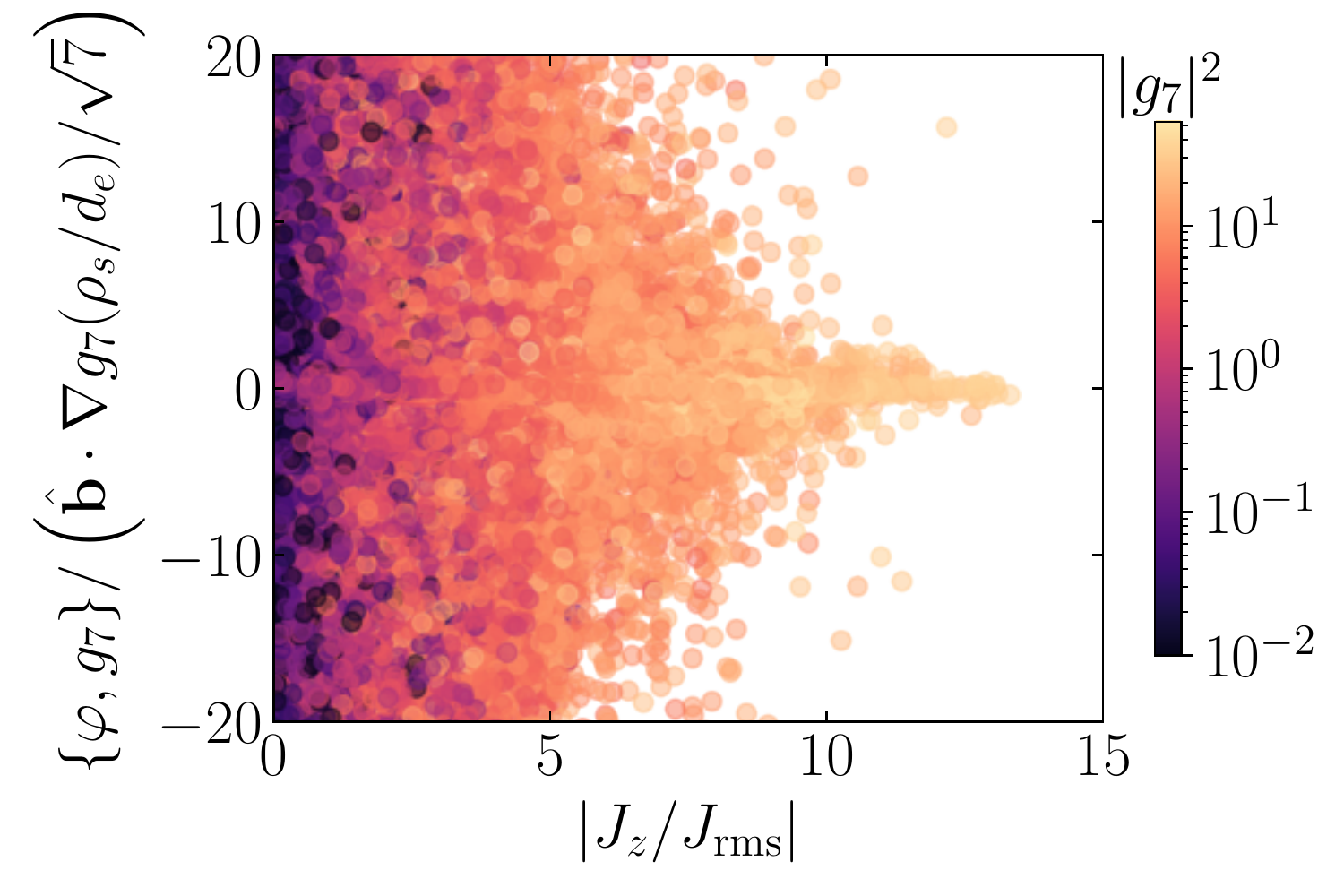}
    \caption{The ratio of the nonlinear-advection rate to phase-mixing rate of $g_7$ as a function of normalized current density $|J_z/J_{\rm rms}|$. The color corresponds to the local energy density of $g_7$. }
    \label{fig:small_nonlinearity}
\end{figure}

We proposed in Sec.~\ref{sec:theory_mcr} that the absence of echo is due to the local reduction of the advective nonlinearity around current sheets. 
The visuals of $J_z$ (Fig.~\ref{fig:compare_Hruns}, top panels) show that indeed, at scales larger than $d_e$, the regions with strong current density are elongated  ``sheet'' structures.
In Fig.~\ref{fig:small_nonlinearity} we show the ratio of the nonlinear-advection rate to the phase-mixing rate of $g_7$ versus the normalized current density (the results using other Hermite moments are qualitatively similar); the color map of the data points represents the local energy density of $g_7$.
This plot clearly shows that nonlinear advection is subdominant with respect to phase mixing at positions with large current density. 
We conjecture that it is the specific configuration of $A_z$, $\varphi$, and $g_m$ around the sheet-like structure of current that leads to the local suppression of echo effects; this interpretation is inspired by observations of strong phase-mixing in supposedly similar geometries that arise in reconnection studies~\cite{loureiro2013fast,numata2015,mccubbin2022characterizing}.
(Note that simply invoking the quasi-1D (in the perpendicular plane) morphology of current sheets is not sufficient to explain the weakening of the advection-to-phase-mixing rates, because such weakening would affect equally the advective nonlinearity and the parallel streaming.)
As a result of enhanced phase-mixing, current sheets are also where the $g_m$ fluctuations are concentrated (see Eq.~\eqref{eq:gm_solution}), as shown by the correlation between $|g_7|^2$ and $|J_z/J_{\rm rms}|$ in Fig.~\ref{fig:small_nonlinearity}.
Therefore, current sheets are energetically important for dissipation.
The overlap in position space between the suppression of plasma echo and the concentration of free energy in $g_e$ results in the efficient energy transfer to small scales in velocity space and, eventually, strong electron heating at the kinetic range.

To further visualize this correlation between the current density and electron heating, in the color map of current density (Fig.~\ref{fig:compare_Hruns} top panels), we show the regions with strong electron heating using the contour (magenta lines) of the electron heating rate (measured by the energy dissipation through hyper-collisions at small scales in velocity space) that is twice its volume-averaged value.  
Indeed we find that, for scales such that $\rho_i \simeq k_\perp^{-1} > d_e$ (left and middle panels), significant electron heating mostly occurs around current sheets, consistent with the above arguments.  
Such correlation between heating and current sheets has been reported in numerical simulations of kinetic turbulence~\cite{tenbarge2013current,navarro2016structure} and observations~\cite{carbone2022high}.
Here, we add to the previous body of work by showing how such heating is possible: the reduction of the advective nonlinearity around current sheets enables phase-mixing to proceed unimpeded in those locations.

\section{Conclusions}
\label{sec:conclusion}
In this work, we conduct an analytical and numerical study of sub-ion-gyroscale (sub-$\rho_i$) turbulence in the low-$\beta$ limit using an analytical formalism known as the \textit{Kinetic Reduced Electron Heating Model} (KREHM)~\cite{zocco2011reduced}.
Two main issues addressed by this work concern the energy spectra of sub-$\rho_i$ turbulence, which is of a kinetic Alfv\'en wave (KAW) nature in this low-$\beta$ regime, and the energy dissipation (and electron heating) in such turbulence. 

In the first part of the paper, we study the energy spectrum, spectral anisotropy, and intermittency of the turbulent fluctuations in the limit of isothermal electrons. 
In our numerical results, the unchanged $k_\perp^{-8/3}$ magnetic (and density) spectrum obtained from simulations with different flux-unfreezing mechanisms (electron inertia versus (hyper-)resistivity) suggests that the spectrum is not set by tearing mediation.
By calculating the 3D, multi-point structure function of the magnetic fields, we obtain a $\propto k_\parallel^{-7/2}$ parallel spectrum, and the scalings $\ell \propto \lambda^{2/3}$ and $\xi \propto \lambda$, which are in good agreement with recent solar wind measurements~\cite{wang2020observational,zhang2022three}.
The measured $\xi \propto \lambda$ scaling indicates isotropy of the fluctuations in the plane perpendicular to the (local) magnetic field, and is consistent with the absence of tearing mediation.
We argue that, instead, the steepening of the spectrum with respect to the straightforward KAW cascade prediction of $k_\perp^{-7/3}$ is likely to be due to intermittency, which we quantify using the higher $q$th-order structure function $S_q(\delta r)$.
The measured $S_q(\delta r) \sim \delta r^{q/3+1}$ scaling reveals that the energy-containing, intermittent structures are quasi-2D, which confirms the main assumption made by~\citet{boldyrev2012spectrum} to derive the $k_\perp^{-8/3}$ spectrum. 

In the second part of the paper, we take into account electron kinetic effects and study the phase-space dynamics and channels for energy dissipation. 
In the $m$-$k_\perp$ phase space, where $m$ is the velocity-space mode number, the energy dissipation is found to mostly occur at high $m$ through collisions, implying that phase mixing dominates nonlinear advection of the free energy and, thus, that the ``kinetic channel'' (as opposed to the ``fluid channel'') is the main route for energy dissipation and electron heating.
We verify that this argument is quantitatively supported, as follows: if we subtract from the magnetic spectrum pertaining to an isothermal run the energy that would be linearly Landau-damped within one eddy turn-over-time, scale-by-scale, we obtain a spectrum that agrees remarkably well with that from a corresponding run with kinetic electrons.
A direct consequence of this result is that the application of a Landau-fluid-type closure~\cite{Hammett1990} to KAW turbulence in the low-$\beta$ limit is justified. 

In Sec.~\ref{sec:theory_zeroth_solution}, we derive a lowest-order solution for the coefficients of the Hermite expansion of the electron distribution function, $g_m$, under the assumption that phase-mixing dominates over nonlinear advection: $g_{m+1}=-\sqrt{m/(m+1)}g_{m-1}$, and $g_3\propto J_z$. 
This solution, as well as the corresponding $m^{-1/2}$ Hermite spectrum, is confirmed by our numerical results.
Lastly, we find that the absence of stochastic plasma echo, which would impede phase mixing, is due to the ubiquitous current sheets that develop self-consistently in our system.
The fluctuations of the Hermite moments are concentrated in the vicinity of current sheets, while at these specific positions, the advective nonlinearity is weakened because of the specific configurations of the magnetic field and the flow around the sheets.
This observed anti-correlation between advective nonlinearity and current density (and thus the density of free-energy) undermines plasma echo and allows the free energy to transfer to higher $m$ by phase mixing.

It is important to note that for all our simulations, the long-wavelength MHD range is either not included or too narrow for the progressive alignment of the fluctuations~\cite{boldyrev2006spectrum} to establish itself. 
In reality, fluctuations cascading from large scales down to the $\rho_i$ scale (i.e., in the MHD range) are expected to become progressively more anisotropic in the field-perpendicular plane, and eventually transition to a tearing-mediated range at a critical scale $\lambda_{\rm cr}>\rho_i$~\cite{loureiro2017collisionless}. 
Below $\lambda_{\rm cr}$, the tearing mediation is expected to decrease the alignment and isotropize the eddies (see Eq.~(22) in~\cite{boldyrev2017magnetohydrodynamic} and Sec. 6 in~\cite{mallet2017disruption}).
Therefore, the fluctuations entering the sub-$\rho_i$ range can be isotropic only when there is a sufficiently wide range between $\lambda_{\rm cr}$ and $\rho_i$.
In weakly collisional turbulence, $\lambda_{\rm cr}$ is expected to be close to $\rho_i$ (see Eqs. (16) and (18) in~\cite{loureiro2017collisionless} and Sec. 4 in~\cite{mallet2017disruptionb}), and so the fluctuations in the sub-$\rho_i$ range can be anisotropic.
However, our numerical results suggests that there is no dynamic alignment effect at sub-$\rho_i$ scales, i.e., the residual anisotropy from the MHD range would not increase in this range.
Therefore, it is possible that the fluctuations around the $\rho_i$ scale are sufficiently anisotropic for tearing mediation; but the tearing mode should become stable as $k_\perp \rho_i$ increases. 
With greater computational resources than currently available, simulations might be possible that cover both the MHD and the kinetic ranges with sufficient resolution such as to enable a more self-consistent study of turbulence, including the microscale feedback to the long-wavelength range (via tearing), and a more accurate  ``large-scale'' condition for the sub-$\rho_i$ turbulence. Having said this, the agreement between our results and several pieces of observational evidence is reassuring.

The efficient electron heating we found in the low-$\beta$ limit is qualitatively consistent with previous numerical studies on electron versus ion energization in kinetic turbulence, in which the electron to ion heating-rate-ratio is a decreasing function of plasma~$\beta$ (e.g.,~\cite{howes2010prescription,kawazura2019thermal,zhdankin2019,arzamasskiy2019hybrid}).
We believe that the weakening of the advective nonlinearity around current sheets --- which enables phase mixing and heating --- remains valid beyond the low-$\beta$ limit that we explore here.
In a gyrokinetic study of reconnection~\cite{numata2015}, it is found that ion heating (via nonlinear phase-mixing) becomes progressively more important as $\beta$ increases. Extrapolating these ideas to kinetic turbulence suggests that, as beta increases (but still below unity), current sheets remain critical as energy dissipation sites, with the balance shifting from heating electrons at low beta (via linear phase mixing) to ions (via nonlinear phase mixing).
Further studies are required to test this hypothesis.

In addition to critical plasma parameters such as the plasma $\beta$ and temperature ratio, the effects of the ``large-scale'' conditions of the turbulence on the energy dissipation are also to be investigated. 
For example, the imbalance of turbulence relevant to the fast-wind streams has recently been found to have profound effects on the heating mechanisms for ions and electrons~\cite{meyrand2021violation,squire2022high}, while the electron kinetic effects are omitted.  
Kinetic studies taking into account both electron and ion physics with relevant large-scale conditions are necessary to explain the extensive measurements taken both at corona and the solar wind by currently operating and future spacecrafts.

In summary, this paper presents a self-consistent explanation for the electromagnetic spectra and electron heating in low-$\beta$ collisionless turbulence by accounting for entwined physical processes in position and velocity spaces. 
On the one hand, we show that intermittency and electron Landau damping are the two physical mechanisms that underlie the steepening of the spectrum with respect to the prediction from a Kolmogorov-like KAW cascade.
On the other, it is due to the presence of these spontaneously formed sheet-like structures that the kinetic effects of phase mixing and Landau damping are locally enhanced and thus play an important role in energy dissipation.
By explaining the above underling physics that connects the position-space and velocity-space dynamics, this study clarifies the long-standing discussion of the relation between current sheets, magnetic reconnection, and electron heating in the low-$\beta$ kinetic turbulence.

\paragraph{Acknowledgments.}
The authors thank L.~Arzamasskiy, S.~Boldyrev, G.~Howes, N.~Mandell, A.~A.~Schekochihin, V.~Zhdankin for insightful discussions.
This work was supported by the National Science Foundation (NSF) under CAREER award No.~1654168 (NFL and MZ), by the National Aeronautics and Space Administration (NASA) under award NNH19ZA001N-FINESST (MZ), and by the 
NSF-DOE Partnership in Basic Plasma Science and Engineering Award No. PHY-2010136 (ZL).
This research used resources of the MIT-PSFC partition of the
Engaging cluster at the MGHPCC facility, funded by DOE award No. DE-FG02-91-ER54109 and the National Energy Research Scientific Computing Center, a DOE Office of Science User Facility supported by the Office of Science of the U.S. Department of Energy under Contract No. DE-AC02-05CH11231 using NERSC award FES-ERCAP0020063.

\twocolumngrid
\bibliographystyle{apsrev4-1}
\bibliography{references}

\include{supplement}

\end{document}

%% file: supplement.tex
\appendix

\setcounter{equation}{0}
\setcounter{figure}{0}
\setcounter{table}{0}
\newcounter{SIfig}

\renewcommand{\theequation}{S\arabic{equation}}
\renewcommand{\thefigure}{S\arabic{figure}}
\renewcommand{\theSIfig}{S\arabic{SIfig}}

\section*{Supplemental Materials}

\subsection{Linear tearing mode with (hyper-)resistivity}
\label{sec:tearing_calcualtion}
In Sec.~\ref{sec:theory_spectra} we invoked scalings pertaining to the (kinetic) tearing instability driven by (hyper-)resistivity. We derive them here.
We perform a linear calculation of the tearing mode with an initial equilibrium described by $\bb{B}_{\perp 0}(x) = \bar{B}_{\perp0}f(x/a) \bb{\hat{y}}$ where $f(x/a)$ is an odd function satisfying $f(0)=0$ and $a$ is the characteristic length scale of the gradient of the equilibrium magnetic field.
We consider the case where $\rho_i \gg a$ and so the GK Poisson's law reduces to Eq.~\eqref{eq:gk_poisson}.
We assume there is no equilibrium flow.
In this setup, the linearized KREHM equations in the isothermal limit (i.e., $g_e=0$) become
\begin{equation}
    \gamma n_e = ik_y B_{\perp 0} (\partial_x^2-k_y^2)A_\parallel - i k_y A_\parallel \bar{B}_{\perp 0}f''(x/a),
\end{equation}
\begin{equation}
    \gamma (1+k_\perp^2 d_e^2) A_\parallel -i k_y \varphi B_{\perp 0} = -\rho_s^2 ik_y n_e B_{\perp 0} + \eta_H k_\perp^2 A_\parallel.
    \label{eq:tearing_inner}
\end{equation}
The tearing instability parameter $\Delta' \equiv \left[ dA_\parallel/dx \right]_{0-}^{0+}$ is obtained by solving the outer layer (MHD) equations: for $k_y a\ll 1$, it is $\Delta'a\propto (k_y a)^{-n}$, where $n=1$ corresponds to a Harris sheet-like configuration for the equilibrium, and $n=2$ corresponds to a sinusoidal profile.

In the inner layer, where $x \ll a$ and $\partial_x^2 \gg k_y^2$, the equilibrium magnetic field can be approximated as $B_{\perp 0} \approx \bar{B}_{\perp 0} x/a$. 
In the large $\Delta'$ limit (the ``Coppi'' modes~\cite{coppi1976resistive}), $\partial_x^2 A_\parallel \approx A_\parallel/\delta^2$, where $\delta$ is the thickness of the inner boundary layer.
In the opposite, small $\Delta'$ limit (the ``FKR'' modes~\cite{furth1963finite}), $\partial_x^2 A_\parallel \approx \Delta' A_\parallel/\delta$. 
Let us first consider the latter case. The balance of terms in the inner layer equations becomes:
\begin{align}
\label{eq:inner_ne}
    \gamma n_e \sim \bar{B_0} \frac{\delta}{a} k_y \frac{\Delta' A_\parallel}{\delta} \sim \frac{\bar{B}_{\perp 0}}{a^2} k_y A_\parallel (k_y a)^{-n},
\end{align}
\begin{equation}
    \gamma(1+\frac{\Delta'}{\delta}d_e^2)A_\parallel \sim k_y \rho_\tau^2 \bar{B}_{\perp 0} \frac{\delta}{a} n_e -\eta_H \frac{\Delta'}{\delta^{\alpha-1}}A_\parallel,
    \label{eq:inner_ohm}
\end{equation}
where $\rho_\tau = \rho_s^2+\rho_i^2$.
In the case of $\eta_H$=0, frozen flux is broken by the electron inertia. 
Balancing the two terms on the {\it l.h.s.} of Ohm's law (Eq.~\eqref{eq:inner_ohm}) yields an expression for the width of the inner layer:
\begin{align}
    \delta \sim \Delta' d_e^2 \sim (k_y a)^{-n} d_e^2/a.  
\end{align}
Using this expression as well as Eq.~\eqref{eq:inner_ne} and Eq.~\eqref{eq:inner_ohm}, we obtain the scaling of the growth rate for FKR-type modes:
\begin{equation}
    \gamma \sim k_y \rho_\tau(k_y a)^{-n} \frac{d_e}{a} \frac{\bar{B}_{\perp 0}}{a}.
\end{equation}

Let us now consider the large $\Delta'$ limit, where \blue{$\Delta'a \sim (k_y a)^{-n} \rightarrow a/\delta$}, and the growth rate and inner layer thickness become \begin{equation}
    \gamma \sim k_y \frac{\rho_\tau}{a}\frac{d_e}{a} \bar{B}_{\perp 0} \frac{d_e}{a} \sim k_y \frac{\rho_\tau}{a} \bar{B}_{\perp 0},
\end{equation}
\begin{equation}
    \delta^2 \sim a \frac{d_e^2}{a} \rightarrow \delta \sim d_e.
\end{equation}

The growth rate of the most unstable mode is obtained by matching the growth rate in the large and the small $\Delta '$ limits:
\begin{equation}
    k_y \frac{\rho_\tau}{a} \bar{B}_{\perp 0} \frac{d_e}{a} (k_y a)^{-n} \sim k_y \frac{\rho_\tau}{a} \bar{B}_{\perp 0},
\end{equation}
from which follows that the most unstable wavenumber scales as
\begin{equation}
    (k_{\rm max}a)^{n} \sim d_e/a \rightarrow k_{\rm max} a \sim (d_e/a)^{1/n}.    
\end{equation}
We thus obtain 
\begin{equation}
\label{eq:gamma_max}
    \gamma_{\rm max} \sim \frac{\bar{B}_{\perp 0}}{a} \frac{\rho_\tau}{a} \left( \frac{d_e}{a} \right)^{1/n}.    
\end{equation}

In the case where the frozen flux is broken by the (hyper or Laplacian) resistivity, we can formally replace the electron inertia term with the resistivity term in the above derivation:
\begin{equation}
    \gamma \frac{\Delta'}{\delta} d_e^2 \rightarrow \eta_H \frac{\Delta '}{\delta^{\alpha-1}}, \quad \text{i.e.}, \quad d_e^2 \rightarrow \eta_H \gamma^{-1} \delta^{2-\alpha}, 
\end{equation}
where $\eta_H$ is the value of the resistivity and $\alpha$ is the order of the spatial derivative in the resistive term.
This then yields:
\begin{equation}
    \gamma_{\rm max}^{(2n+1)/2n} \sim \frac{\bar{B}_{\perp 0}}{a} \frac{\rho_\tau}{a} a^{-1/n} \eta_H^{1/2n} \delta^{(2-\alpha)/2n}.
    \label{eq:gamma_max_step1}
\end{equation}
The thickness of the inner layer should be
\begin{equation}
    \delta \sim d_e \sim (\eta_H/\gamma_{\rm max})^{1/2} \delta^{(2-\alpha)/2} \rightarrow \delta \sim (\eta_H/\gamma_{\rm max})^{1/\alpha}.
    \label{eq:delta_step1}
\end{equation}
Combining Eqs.~\eqref{eq:gamma_max_step1} and \eqref{eq:delta_step1}, we obtain the final expressions for $\gamma_{\rm max}$ and $\delta$ as a function of the parameter for the equilibrium profile $n$, and that for the order of the (hyper-)dissipation $\alpha$:
\begin{equation}
    \gamma_{\rm max} \sim \left(\bar{B}_{\perp 0} \rho_\tau\right)^{n\alpha/(n\alpha+1)} a^{-\alpha(2n+1)/(n\alpha+1)} \eta_H^{1/(n\alpha+1)},
    \label{eq:gamma_max_hyper}
\end{equation}
and 
\begin{equation}
    \delta \sim \left(\bar{B}_{\perp 0}\rho_\tau \right)^{-n/(n\alpha+1)} a^{(2n+1)/(n\alpha+1)} \eta_H^{n/(n\alpha+1)}.
    \label{eq:delta}
\end{equation}

In the model of tearing-mediated turbulence, the nonlinear timescale is set by the growth rate of tearing instability, expressed by Eq.~\eqref{eq:gamma_max_hyper} (or Eq.~\eqref{eq:gamma_max} in the case of electron inertia).
We use the expression for $\delta$ (Eq.\eqref{eq:delta}) to estimate the resolution in the numerical simulation required to resolve the inner layer thickness of the tearing mode set by the electron inertia or the (hyper-) resistivity.

We now derive the magnetic (or density) energy spectrum at sub-$\rho_i$ scales following the tearing-mediated turbulence model (described in Sec.~\ref{sec:theory_spectra}) for the case where the frozen flux is broken by the resistivity (instead of the electron inertia).
In that model, the eddy-turnover rate, which satisfies the dimensional relation $\gamma_{nl} \sim \varepsilon/(\delta B_{\perp \lambda}^2/8\pi)$, is set by the growth rate of tearing mode (Eq.~\eqref{eq:gamma_max_hyper}), where the characteristic length scale and amplitude of the equilibrium magnetic fields correspond to those of eddies: $a \rightarrow \lambda$ and $\bar{B}_{\perp 0} \rightarrow \delta B_{\perp \lambda}$. 
From $\gamma_{nl} \sim \gamma_{\rm max}$ we obtain
\begin{equation}
 \delta B_{\perp \lambda}^{2} \propto \delta B_{\perp \lambda}^{-n\alpha/(n\alpha+1)} \lambda^{\alpha(2n+1)/(n\alpha+1)},
    \label{eq:gammanl_gammamax}
\end{equation}
and so the magnetic energy spectrum should scale as
\begin{equation}
    E_B(k_\perp)dk_\perp \propto k_\perp^{-(7n\alpha+2\alpha+2)/(3n\alpha+2)} dk_\perp.
    \label{eq:resistive_tearing_spectrum}
\end{equation}
In the case of Laplacian resistivity, i.e., $\alpha=2$, the $n=1$ profile gives the $k_\perp^{-5/2}=k_\perp^{-2.5}$ spectrum, and the $n=2$ profile gives $k_\perp^{-17/7}\approx k_\perp^{-2.4}$.
In the case of hyper-resistivity with $\alpha=6$ (the order used in our isothermal simulations), the $n=1$ profile gives the $k_\perp^{-14/5}=k_\perp^{-2.8}$ spectrum, and the $n=2$ profile gives $k_\perp^{-49/19}\approx k_\perp^{-2.6}$.
In Sec.~\ref{sec:theory_spectra} we obtained the corresponding scalings for the case where it is instead electron inertia breaking the frozen flux constraint: the $n=1$ profile then yields a $k_\perp^{-3}$ spectrum whereas $k_\perp^{-8/3}$ is obtained for the $n=2$ case.
Comparing the above derived spectra, we reach the conclusion that in numerical simulations where the flux unfreezing is caused by different mechanisms, if the spectra were indeed set by the tearing modes, the difference of the measured spectra should be sufficient to identify.

\subsection{Kinetic Alfv\'en waves}
\label{sec:kaw_dispersion}
The linear dispersion relation for kinetic Alfv\'en waves (KAWs) in the  KREHM framework is~\cite{zocco2011reduced}:
\begin{equation}
    \left[\zeta^2 - \frac{\tau}{z_c}\frac{k_\perp^2d_e^2/2}{1-\Gamma_0(k_\perp^2\rho_i^2/2)} \right][1+\zeta Z(\zeta)]=\frac{1}{2}k_\perp^2d_e^2,
    \label{eq:KAW_DR}
\end{equation}
where $\zeta \equiv \Omega_{\rm KAW}/|k_\parallel| v_{{\rm th}e}$ ($\Omega_{\rm KAW}=\omega_{\rm KAW}+i\gamma_{\rm KAW}$), $\tau \equiv T_{0i}/T_{0e}$ is the temperature ratio, $z_c$ is the charge ratio, $\Gamma_0(\alpha)=I_0(\alpha)e^{-\alpha}$ with $I_0$ being the zeroth-order modified Bessel function, and $Z(\zeta) \equiv \pi^{-1/2}\int_{-\infty}^{\infty}e^{-t^2}/(t-\zeta)dt$ is the plasma dispersion function. 
In what follows, we specify $z_c=1$ and $\tau=1$.

\begin{figure}[htp]
    \centering
    \includegraphics[width=0.5\textwidth]{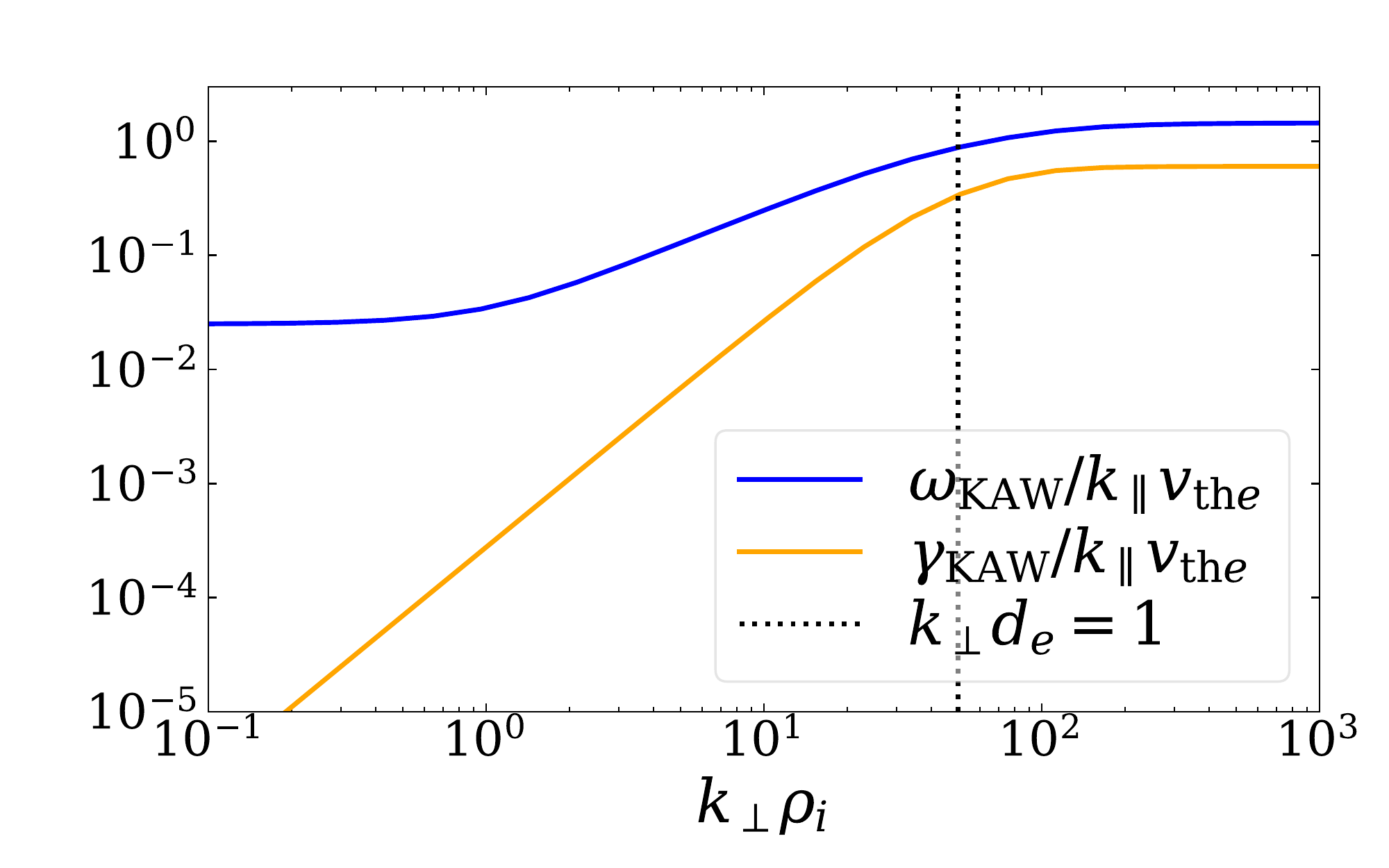}
    \caption{Oscillation frequency ($\omega_{\rm KAW}$) and damping rate ($\gamma_{\rm KAW}$) of KAWs (normalized to $k_\parallel v_{{\rm th}e}$), obtained from Eq.~\eqref{eq:KAW_DR} for $\rho_i/d_e=40$ (corresponding to run H2a ). Vertical dotted line indicates the $d_e$ scale.}
    \label{fig:KAW_rhoide40}
\end{figure}

Fig.~\ref{fig:KAW_rhoide40} shows the scale ($k_\perp \rho_i$) dependence of the oscillation frequency, $\omega_{\rm KAW}$, and the (Landau) damping rate, $\gamma_{\rm KAW}$, of KAWs obtained from Eq.~\eqref{eq:KAW_DR} for $\rho_i/d_e=40$. 
At around $k_\perp d_e \sim 1$, $\omega_{\rm KAW}/k_\parallel v_{{\rm th}e} \sim 1$, while at $k_\perp d_e \ll 1$, $\omega_{\rm KAW} \ll k_\parallel v_{{\rm th}e}$.
In this case, at $k_\perp d_e \ll 1$, the nonlinear eddy turn over rate of the KAW turbulence, which is comparable to the KAW frequency $\omega_{\rm KAW}$, is much smaller than the rate of linear phase mixing, $k_\parallel v_{{\rm th}e}$. 
This timescale ordering leads to the zeroth-order solution we describe in Sec.~\ref{sec:theory_zeroth_solution}.

As argued in Sec.~\ref{sec:theory_zeroth_solution}, the validity of our argument of strong electron heating is predicated on the existence of a large enough dynamical range $\rho_i \gg 1/k_\perp \gg d_e$ such that both condition for the zeroth-order solution,
\begin{align}
    \frac{\omega_{\rm KAW}}{k_\parallel v_{{\rm th}e}} \sim \frac{k_\parallel V_A k_\perp \rho_s}{k_\parallel v_{{\rm th}e}}\sim \frac{V_A}{v_{{\rm th}e}} k_\perp \rho_s \sim \frac{d_e}{\rho_i} k_\perp \rho_s \sim k_\perp d_e \ll 1,
\end{align}
and the considerable damping rate,
\begin{align}
    -\frac{\gamma_{\rm KAW}}{\omega_{\rm KAW}} > \frac{1}{2\pi},
\end{align}
are satisfied.

\begin{figure}[htp]
    \centering
    \includegraphics[width=0.5\textwidth]{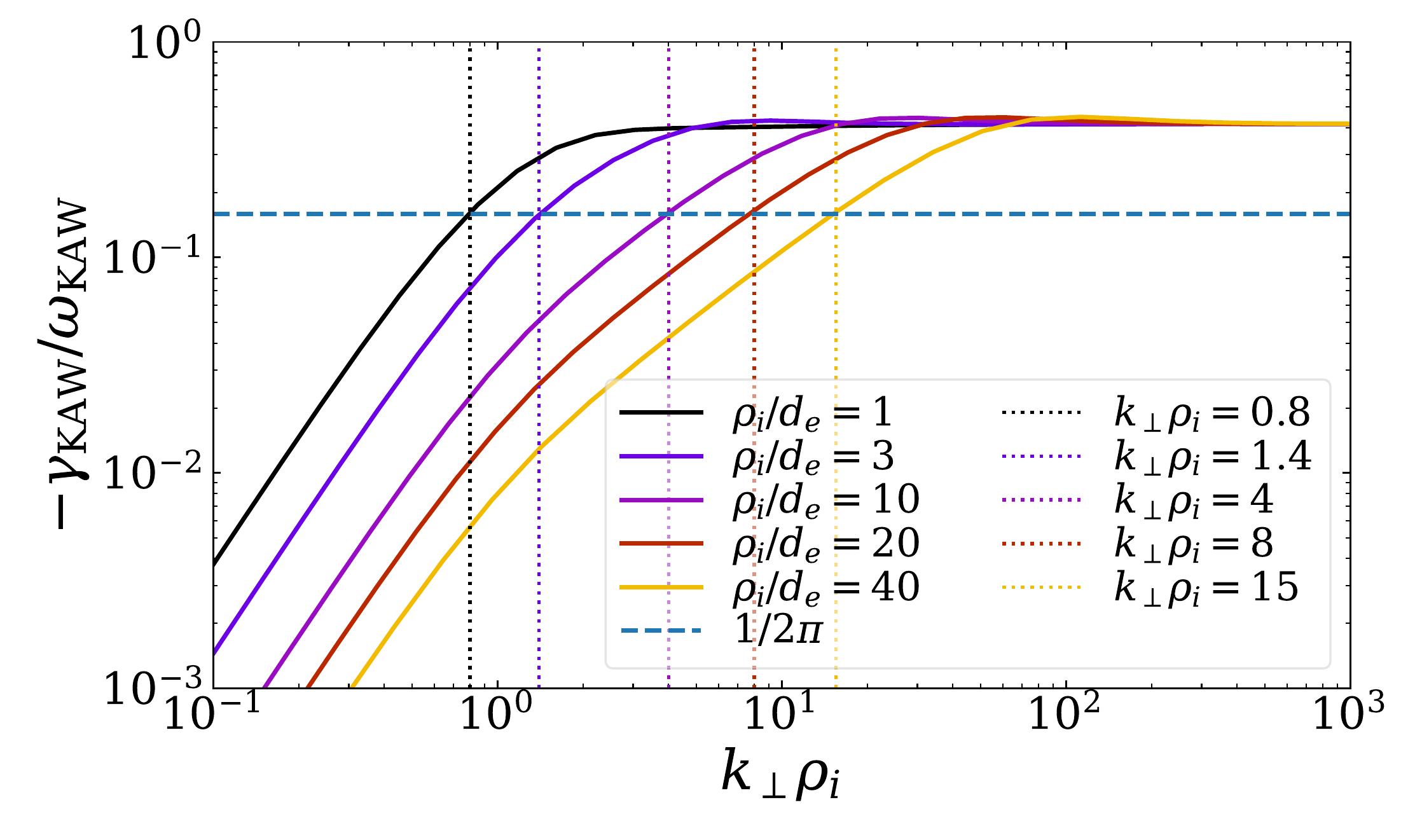}
    \caption{Ratio of damping rate ($\gamma_{\rm KAW}$) to frequency ($\omega_{\rm KAW}$) of KAWs for varying $\rho_i/d_e$, obtained by solving Eq.~\eqref{eq:KAW_DR}.}
    \label{fig:KAW_vary_rhoide}
\end{figure}

In Fig.~\ref{fig:KAW_vary_rhoide} we plot the ratio of damping rate to frequency $\gamma_{\rm KAW}/\omega_{\rm KAW}$ at different values of $\rho_i/d_e$. 
The horizontal dashed line indicates $\gamma_{\rm KAW}/\omega_{\rm KAW}=1/2\pi$, i.e., the value at which Landau damping of KAWs becomes significant; and the vertical dotted lines show the corresponding values of $k_\perp \rho_i$ at which this threshold is reached for each value of $\rho_i/d_e$.  
As $\rho_i/d_e$ increases, there is a progressively wider range of scales where damping of KAWs cannot be ignored.

\subsection{Multi-point, three-dimensional structure function}
\label{sec:structure_function_method}
In order to test the prediction of the parallel spectrum and spectral anisotropy (given by~\citet{boldyrev2012spectrum} and~\citet{loureiro2017collisionless}), and, furthermore, to quantify the intermittency of the system, we calculate the five-point, three-dimensional structure function of the total magnetic fields $\bb{B}(\bb{r},t)$.
The reason to use a multi-point structure function (instead of a more common two-point version~\cite{cho2000,mallet2016}) is to capture the correct scaling for fluctuations whose energy spectra are close to or steeper than $\sim k^{-3}$~\cite{cho2009simulations,cerri2019kinetic} and to filter out more effectively the large-scale fluctuations with wavenumbers below $k
\sim \pi/\delta r$, where $\delta r$ is the incremental scale. 
Following~\citet{cho2009simulations}, we define the five-point, $q$th-order structure function for magnetic-field fluctuations as
\begin{equation}
    S_q (\delta \bb{r}) \equiv \langle |\Delta \bb{B} (\bb{r},\delta \bb{r})|^q \rangle_{\bb{r}}. 
\end{equation}
Here $\delta \bb{r}$ is the increment vector, $\langle ... \rangle_{\bb{r}}$ is the ensemble average over the domain, and the (normalized) local field increment is calculated with the mentioned five-point stencil:
\begin{equation}
\begin{aligned}
    \Delta \bb{B}(\bb{r},\bb{\delta r}) &=[\bb{B}(\bb{r}+2\delta \bb{r})-4\bb{B}(\bb{r}+\delta \bb{r})+6\bb{B}(\bb{r})-\\ & 4\bb{B}(\bb{r}-\delta \bb{r})+\bb{B}(\bb{r}-2\delta \bb{r})]/\sqrt{35}.
\end{aligned}
\end{equation}

The structure function $S_q(\delta \bb{r})$ with respect to $\delta \bb{r}$ can be projected into local coordinates with respect to a local, scale-dependent coordinate system $(\hat{\ell}, \hat{\xi},\hat{\lambda})$, where $\hat{\ell}$ is parallel to the local mean magnetic fields defined as
\begin{equation}
\begin{aligned}
   \bb{B}_{\rm loc} &=[\bb{B}(\bb{r}+2\delta \bb{r})+4\bb{B}(\bb{r}+\delta \bb{r})+6\bb{B}(\bb{r})+\\ & 4\bb{B}(\bb{r}+\delta \bb{r})+\bb{B}(\bb{r}-2\delta \bb{r})]/16,
\end{aligned}
\end{equation}
$\hat{\xi}$ is the direction of the local perpendicular fluctuation $\Delta \bb{B}_\perp=\bb{B}_{\rm loc} \times (\Delta \bb{B} \times \bb{B}_{\rm loc})$, and $\hat{\lambda}=\hat{\ell} \times \hat{\xi}$.
We denote $\theta_B$ as the angle between $\delta \bb{r}$ and $\bb{B}_{\rm loc}$, and $\theta_{\Delta B_\perp}$ as that between $\hat{\xi}$ and the projection of $\delta \bb{r}$ on the plane perpendicular to $\bb{B}_{\rm loc}$.
The structure function in the three orthogonal directions can then be defined as 
\begin{equation}
    S_q(\ell) \equiv S_q(\ell; 0^{\circ}<\theta_B<\delta \theta, 0^{\circ}<\theta_{\Delta B_\perp}<90^{\circ}),
\end{equation}
\begin{equation}
    S_q(\xi) \equiv S_q(\xi; 90^{\circ}-\delta \theta<\theta_B<90^{\circ}, 0^{\circ}<\theta_{\Delta B_\perp}<\delta \theta),
\end{equation}
\begin{equation}
    S_q(\lambda) \equiv S_q(\lambda; 90^{\circ}-\delta \theta<\theta_B<90^{\circ}, 90^{\circ}-\delta \theta<\theta_{\Delta B_\perp}<90^{\circ}),
\end{equation}
where $\delta \theta$ is a finite angular threshold.
When calculating the directional structure functions, we reduce $\delta \theta$ until their scalings appeared converged.
Using the second-order structure function ($q=2$) and equating the value between pairs of $S_2(\ell)$, $S_2(\xi)$, and $S_2(\lambda)$, we obtain the (spectral anisotropy) scalings between $\ell$, $\xi$, and $\lambda$.

%% file: main.bbl
\begin{thebibliography}{107}%
\makeatletter
\providecommand \@ifxundefined [1]{%
 \@ifx{#1\undefined}
}%
\providecommand \@ifnum [1]{%
 \ifnum #1\expandafter \@firstoftwo
 \else \expandafter \@secondoftwo
 \fi
}%
\providecommand \@ifx [1]{%
 \ifx #1\expandafter \@firstoftwo
 \else \expandafter \@secondoftwo
 \fi
}%
\providecommand \natexlab [1]{#1}%
\providecommand \enquote  [1]{``#1''}%
\providecommand \bibnamefont  [1]{#1}%
\providecommand \bibfnamefont [1]{#1}%
\providecommand \citenamefont [1]{#1}%
\providecommand \href@noop [0]{\@secondoftwo}%
\providecommand \href [0]{\begingroup \@sanitize@url \@href}%
\providecommand \@href[1]{\@@startlink{#1}\@@href}%
\providecommand \@@href[1]{\endgroup#1\@@endlink}%
\providecommand \@sanitize@url [0]{\catcode `\\12\catcode `\$12\catcode
  `\&12\catcode `\#12\catcode `\^12\catcode `\_12\catcode `\%12\relax}%
\providecommand \@@startlink[1]{}%
\providecommand \@@endlink[0]{}%
\providecommand \url  [0]{\begingroup\@sanitize@url \@url }%
\providecommand \@url [1]{\endgroup\@href {#1}{\urlprefix }}%
\providecommand \urlprefix  [0]{URL }%
\providecommand \Eprint [0]{\href }%
\providecommand \doibase [0]{http://dx.doi.org/}%
\providecommand \selectlanguage [0]{\@gobble}%
\providecommand \bibinfo  [0]{\@secondoftwo}%
\providecommand \bibfield  [0]{\@secondoftwo}%
\providecommand \translation [1]{[#1]}%
\providecommand \BibitemOpen [0]{}%
\providecommand \bibitemStop [0]{}%
\providecommand \bibitemNoStop [0]{.\EOS\space}%
\providecommand \EOS [0]{\spacefactor3000\relax}%
\providecommand \BibitemShut  [1]{\csname bibitem#1\endcsname}%
\let\auto@bib@innerbib\@empty
\bibitem [{\citenamefont {Borovsky}\ and\ \citenamefont
  {Funsten}(2003)}]{borovsky2003mhd}%
  \BibitemOpen
  \bibfield  {author} {\bibinfo {author} {\bibfnamefont {J.~E.}\ \bibnamefont
  {Borovsky}}\ and\ \bibinfo {author} {\bibfnamefont {H.~O.}\ \bibnamefont
  {Funsten}},\ }\href@noop {} {\bibfield  {journal} {\bibinfo  {journal}
  {Journal of Geophysical Research: Space Physics}\ }\textbf {\bibinfo {volume}
  {108}} (\bibinfo {year} {2003})}\BibitemShut {NoStop}%
\bibitem [{\citenamefont {{Goldreich}}\ and\ \citenamefont
  {{Sridhar}}(1995)}]{Goldreich1995}%
  \BibitemOpen
  \bibfield  {author} {\bibinfo {author} {\bibfnamefont {P.}~\bibnamefont
  {{Goldreich}}}\ and\ \bibinfo {author} {\bibfnamefont {S.}~\bibnamefont
  {{Sridhar}}},\ }\href {\doibase 10.1086/175121} {\bibfield  {journal}
  {\bibinfo  {journal} {The Astrophysical Journal}\ }\textbf {\bibinfo {volume}
  {438}},\ \bibinfo {pages} {763} (\bibinfo {year} {1995})}\BibitemShut
  {NoStop}%
\bibitem [{\citenamefont {Tu}\ and\ \citenamefont {Marsch}(1995)}]{tu1995mhd}%
  \BibitemOpen
  \bibfield  {author} {\bibinfo {author} {\bibfnamefont {C.-Y.}\ \bibnamefont
  {Tu}}\ and\ \bibinfo {author} {\bibfnamefont {E.}~\bibnamefont {Marsch}},\
  }\href@noop {} {\bibfield  {journal} {\bibinfo  {journal} {Space Science
  Reviews}\ }\textbf {\bibinfo {volume} {73}},\ \bibinfo {pages} {1} (\bibinfo
  {year} {1995})}\BibitemShut {NoStop}%
\bibitem [{\citenamefont {{Parker}}(1983)}]{parker1983}%
  \BibitemOpen
  \bibfield  {author} {\bibinfo {author} {\bibfnamefont {E.~N.}\ \bibnamefont
  {{Parker}}},\ }\href@noop {} {\bibfield  {journal} {\bibinfo  {journal} {The
  Astrophysical Journal}\ }\textbf {\bibinfo {volume} {264}},\ \bibinfo {pages}
  {635} (\bibinfo {year} {1983})}\BibitemShut {NoStop}%
\bibitem [{\citenamefont {Balbus}\ and\ \citenamefont
  {Hawley}(1998)}]{balbus1998instability}%
  \BibitemOpen
  \bibfield  {author} {\bibinfo {author} {\bibfnamefont {S.~A.}\ \bibnamefont
  {Balbus}}\ and\ \bibinfo {author} {\bibfnamefont {J.~F.}\ \bibnamefont
  {Hawley}},\ }\href@noop {} {\bibfield  {journal} {\bibinfo  {journal}
  {Reviews of Modern Physics}\ }\textbf {\bibinfo {volume} {70}},\ \bibinfo
  {pages} {1} (\bibinfo {year} {1998})}\BibitemShut {NoStop}%
\bibitem [{\citenamefont {Begelman}\ and\ \citenamefont
  {Fabian}(1990)}]{begelman1990turbulent}%
  \BibitemOpen
  \bibfield  {author} {\bibinfo {author} {\bibfnamefont {M.~C.}\ \bibnamefont
  {Begelman}}\ and\ \bibinfo {author} {\bibfnamefont {A.~C.}\ \bibnamefont
  {Fabian}},\ }\href@noop {} {\bibfield  {journal} {\bibinfo  {journal}
  {Monthly Notices of the Royal Astronomical Society}\ }\textbf {\bibinfo
  {volume} {244}},\ \bibinfo {pages} {26P} (\bibinfo {year}
  {1990})}\BibitemShut {NoStop}%
\bibitem [{\citenamefont {Sahraoui}\ \emph {et~al.}(2009)\citenamefont
  {Sahraoui}, \citenamefont {Goldstein}, \citenamefont {Robert},\ and\
  \citenamefont {Khotyaintsev}}]{sahraoui2009evidence}%
  \BibitemOpen
  \bibfield  {author} {\bibinfo {author} {\bibfnamefont {F.}~\bibnamefont
  {Sahraoui}}, \bibinfo {author} {\bibfnamefont {M.~L.}\ \bibnamefont
  {Goldstein}}, \bibinfo {author} {\bibfnamefont {P.}~\bibnamefont {Robert}}, \
  and\ \bibinfo {author} {\bibfnamefont {Y.~V.}\ \bibnamefont {Khotyaintsev}},\
  }\href@noop {} {\bibfield  {journal} {\bibinfo  {journal} {Physical Review
  Letters}\ }\textbf {\bibinfo {volume} {102}},\ \bibinfo {pages} {231102}
  (\bibinfo {year} {2009})}\BibitemShut {NoStop}%
\bibitem [{\citenamefont {Alexandrova}\ \emph {et~al.}(2009)\citenamefont
  {Alexandrova}, \citenamefont {Saur}, \citenamefont {Lacombe}, \citenamefont
  {Mangeney}, \citenamefont {Mitchell}, \citenamefont {Schwartz},\ and\
  \citenamefont {Robert}}]{alexandrova2009universality}%
  \BibitemOpen
  \bibfield  {author} {\bibinfo {author} {\bibfnamefont {O.}~\bibnamefont
  {Alexandrova}}, \bibinfo {author} {\bibfnamefont {J.}~\bibnamefont {Saur}},
  \bibinfo {author} {\bibfnamefont {C.}~\bibnamefont {Lacombe}}, \bibinfo
  {author} {\bibfnamefont {A.}~\bibnamefont {Mangeney}}, \bibinfo {author}
  {\bibfnamefont {J.}~\bibnamefont {Mitchell}}, \bibinfo {author}
  {\bibfnamefont {S.~J.}\ \bibnamefont {Schwartz}}, \ and\ \bibinfo {author}
  {\bibfnamefont {P.}~\bibnamefont {Robert}},\ }\href@noop {} {\bibfield
  {journal} {\bibinfo  {journal} {Physical Review Letters}\ }\textbf {\bibinfo
  {volume} {103}},\ \bibinfo {pages} {165003} (\bibinfo {year}
  {2009})}\BibitemShut {NoStop}%
\bibitem [{\citenamefont {Alexandrova}\ \emph {et~al.}(2012)\citenamefont
  {Alexandrova}, \citenamefont {Lacombe}, \citenamefont {Mangeney},
  \citenamefont {Grappin},\ and\ \citenamefont
  {Maksimovic}}]{alexandrova2012solar}%
  \BibitemOpen
  \bibfield  {author} {\bibinfo {author} {\bibfnamefont {O.}~\bibnamefont
  {Alexandrova}}, \bibinfo {author} {\bibfnamefont {C.}~\bibnamefont
  {Lacombe}}, \bibinfo {author} {\bibfnamefont {A.}~\bibnamefont {Mangeney}},
  \bibinfo {author} {\bibfnamefont {R.}~\bibnamefont {Grappin}}, \ and\
  \bibinfo {author} {\bibfnamefont {M.}~\bibnamefont {Maksimovic}},\
  }\href@noop {} {\bibfield  {journal} {\bibinfo  {journal} {The Astrophysical
  Journal}\ }\textbf {\bibinfo {volume} {760}},\ \bibinfo {pages} {121}
  (\bibinfo {year} {2012})}\BibitemShut {NoStop}%
\bibitem [{\citenamefont {Alexandrova}\ \emph {et~al.}(2013)\citenamefont
  {Alexandrova}, \citenamefont {Chen}, \citenamefont {Sorriso-Valvo},
  \citenamefont {Horbury},\ and\ \citenamefont {Bale}}]{alexandrova2013solar}%
  \BibitemOpen
  \bibfield  {author} {\bibinfo {author} {\bibfnamefont {O.}~\bibnamefont
  {Alexandrova}}, \bibinfo {author} {\bibfnamefont {C.~H.~K.}\ \bibnamefont
  {Chen}}, \bibinfo {author} {\bibfnamefont {L.}~\bibnamefont {Sorriso-Valvo}},
  \bibinfo {author} {\bibfnamefont {T.~S.}\ \bibnamefont {Horbury}}, \ and\
  \bibinfo {author} {\bibfnamefont {S.~D.}\ \bibnamefont {Bale}},\ }\href@noop
  {} {\bibfield  {journal} {\bibinfo  {journal} {Space Science Reviews}\
  }\textbf {\bibinfo {volume} {178}},\ \bibinfo {pages} {101} (\bibinfo {year}
  {2013})}\BibitemShut {NoStop}%
\bibitem [{\citenamefont {Chen}\ \emph {et~al.}(2019)\citenamefont {Chen},
  \citenamefont {Klein},\ and\ \citenamefont {Howes}}]{chen2019evidence}%
  \BibitemOpen
  \bibfield  {author} {\bibinfo {author} {\bibfnamefont {C.~H.~K.}\
  \bibnamefont {Chen}}, \bibinfo {author} {\bibfnamefont {K.~G.}\ \bibnamefont
  {Klein}}, \ and\ \bibinfo {author} {\bibfnamefont {G.~G.}\ \bibnamefont
  {Howes}},\ }\href@noop {} {\bibfield  {journal} {\bibinfo  {journal} {Nature
  Communications}\ }\textbf {\bibinfo {volume} {10}},\ \bibinfo {pages} {1}
  (\bibinfo {year} {2019})}\BibitemShut {NoStop}%
\bibitem [{\citenamefont {Gary}\ and\ \citenamefont
  {Nishimura}(2004)}]{gary2004kinetic}%
  \BibitemOpen
  \bibfield  {author} {\bibinfo {author} {\bibfnamefont {S.~P.}\ \bibnamefont
  {Gary}}\ and\ \bibinfo {author} {\bibfnamefont {K.}~\bibnamefont
  {Nishimura}},\ }\href@noop {} {\bibfield  {journal} {\bibinfo  {journal}
  {Journal of Geophysical Research: Space Physics}\ }\textbf {\bibinfo {volume}
  {109}} (\bibinfo {year} {2004})}\BibitemShut {NoStop}%
\bibitem [{\citenamefont {Howes}\ \emph
  {et~al.}(2008{\natexlab{a}})\citenamefont {Howes}, \citenamefont {Cowley},
  \citenamefont {Dorland}, \citenamefont {Hammett}, \citenamefont {Quataert},\
  and\ \citenamefont {Schekochihin}}]{howes2008model}%
  \BibitemOpen
  \bibfield  {author} {\bibinfo {author} {\bibfnamefont {G.~G.}\ \bibnamefont
  {Howes}}, \bibinfo {author} {\bibfnamefont {S.~C.}\ \bibnamefont {Cowley}},
  \bibinfo {author} {\bibfnamefont {W.}~\bibnamefont {Dorland}}, \bibinfo
  {author} {\bibfnamefont {G.~W.}\ \bibnamefont {Hammett}}, \bibinfo {author}
  {\bibfnamefont {E.}~\bibnamefont {Quataert}}, \ and\ \bibinfo {author}
  {\bibfnamefont {A.~A.}\ \bibnamefont {Schekochihin}},\ }\href@noop {}
  {\bibfield  {journal} {\bibinfo  {journal} {Journal of Geophysical Research:
  Space Physics}\ }\textbf {\bibinfo {volume} {113}} (\bibinfo {year}
  {2008}{\natexlab{a}})}\BibitemShut {NoStop}%
\bibitem [{\citenamefont {Howes}\ \emph
  {et~al.}(2008{\natexlab{b}})\citenamefont {Howes}, \citenamefont {Dorland},
  \citenamefont {Cowley}, \citenamefont {Hammett}, \citenamefont {Quataert},
  \citenamefont {Schekochihin},\ and\ \citenamefont
  {Tatsuno}}]{howes2008kinetic}%
  \BibitemOpen
  \bibfield  {author} {\bibinfo {author} {\bibfnamefont {G.~G.}\ \bibnamefont
  {Howes}}, \bibinfo {author} {\bibfnamefont {W.}~\bibnamefont {Dorland}},
  \bibinfo {author} {\bibfnamefont {S.~C.}\ \bibnamefont {Cowley}}, \bibinfo
  {author} {\bibfnamefont {G.~W.}\ \bibnamefont {Hammett}}, \bibinfo {author}
  {\bibfnamefont {E.}~\bibnamefont {Quataert}}, \bibinfo {author}
  {\bibfnamefont {A.~A.}\ \bibnamefont {Schekochihin}}, \ and\ \bibinfo
  {author} {\bibfnamefont {T.}~\bibnamefont {Tatsuno}},\ }\href@noop {}
  {\bibfield  {journal} {\bibinfo  {journal} {Physical Review Letters}\
  }\textbf {\bibinfo {volume} {100}},\ \bibinfo {pages} {065004} (\bibinfo
  {year} {2008}{\natexlab{b}})}\BibitemShut {NoStop}%
\bibitem [{\citenamefont {Salem}\ \emph {et~al.}(2012)\citenamefont {Salem},
  \citenamefont {Howes}, \citenamefont {Sundkvist}, \citenamefont {Bale},
  \citenamefont {Chaston}, \citenamefont {Chen},\ and\ \citenamefont
  {Mozer}}]{salem2012identification}%
  \BibitemOpen
  \bibfield  {author} {\bibinfo {author} {\bibfnamefont {C.~S.}\ \bibnamefont
  {Salem}}, \bibinfo {author} {\bibfnamefont {G.~G.}\ \bibnamefont {Howes}},
  \bibinfo {author} {\bibfnamefont {D.}~\bibnamefont {Sundkvist}}, \bibinfo
  {author} {\bibfnamefont {S.~D.}\ \bibnamefont {Bale}}, \bibinfo {author}
  {\bibfnamefont {C.~C.}\ \bibnamefont {Chaston}}, \bibinfo {author}
  {\bibfnamefont {C.~H.~K.}\ \bibnamefont {Chen}}, \ and\ \bibinfo {author}
  {\bibfnamefont {F.~S.}\ \bibnamefont {Mozer}},\ }\href@noop {} {\bibfield
  {journal} {\bibinfo  {journal} {The Astrophysical Journal Letters}\ }\textbf
  {\bibinfo {volume} {745}},\ \bibinfo {pages} {L9} (\bibinfo {year}
  {2012})}\BibitemShut {NoStop}%
\bibitem [{\citenamefont {TenBarge}\ \emph {et~al.}(2012)\citenamefont
  {TenBarge}, \citenamefont {Podesta}, \citenamefont {Klein},\ and\
  \citenamefont {Howes}}]{tenbarge2012interpreting}%
  \BibitemOpen
  \bibfield  {author} {\bibinfo {author} {\bibfnamefont {J.~M.}\ \bibnamefont
  {TenBarge}}, \bibinfo {author} {\bibfnamefont {J.~J.}\ \bibnamefont
  {Podesta}}, \bibinfo {author} {\bibfnamefont {K.~G.}\ \bibnamefont {Klein}},
  \ and\ \bibinfo {author} {\bibfnamefont {G.~G.}\ \bibnamefont {Howes}},\
  }\href@noop {} {\bibfield  {journal} {\bibinfo  {journal} {The Astrophysical
  Journal}\ }\textbf {\bibinfo {volume} {753}},\ \bibinfo {pages} {107}
  (\bibinfo {year} {2012})}\BibitemShut {NoStop}%
\bibitem [{\citenamefont {Franci}\ \emph {et~al.}(2018)\citenamefont {Franci},
  \citenamefont {Landi}, \citenamefont {Verdini}, \citenamefont {Matteini},\
  and\ \citenamefont {Hellinger}}]{franci2018solar}%
  \BibitemOpen
  \bibfield  {author} {\bibinfo {author} {\bibfnamefont {L.}~\bibnamefont
  {Franci}}, \bibinfo {author} {\bibfnamefont {S.}~\bibnamefont {Landi}},
  \bibinfo {author} {\bibfnamefont {A.}~\bibnamefont {Verdini}}, \bibinfo
  {author} {\bibfnamefont {L.}~\bibnamefont {Matteini}}, \ and\ \bibinfo
  {author} {\bibfnamefont {P.}~\bibnamefont {Hellinger}},\ }\href@noop {}
  {\bibfield  {journal} {\bibinfo  {journal} {The Astrophysical Journal}\
  }\textbf {\bibinfo {volume} {853}},\ \bibinfo {pages} {26} (\bibinfo {year}
  {2018})}\BibitemShut {NoStop}%
\bibitem [{\citenamefont {Cerri}\ \emph {et~al.}(2019)\citenamefont {Cerri},
  \citenamefont {Gro{\v{s}}elj},\ and\ \citenamefont
  {Franci}}]{cerri2019kinetic}%
  \BibitemOpen
  \bibfield  {author} {\bibinfo {author} {\bibfnamefont {S.~S.}\ \bibnamefont
  {Cerri}}, \bibinfo {author} {\bibfnamefont {D.}~\bibnamefont
  {Gro{\v{s}}elj}}, \ and\ \bibinfo {author} {\bibfnamefont {L.}~\bibnamefont
  {Franci}},\ }\href@noop {} {\bibfield  {journal} {\bibinfo  {journal}
  {Frontiers in Astronomy and Space Sciences}\ ,\ \bibinfo {pages} {64}}
  (\bibinfo {year} {2019})}\BibitemShut {NoStop}%
\bibitem [{\citenamefont {Gro{\v{s}}elj}\ \emph {et~al.}(2019)\citenamefont
  {Gro{\v{s}}elj}, \citenamefont {Chen}, \citenamefont {Mallet}, \citenamefont
  {Samtaney}, \citenamefont {Schneider},\ and\ \citenamefont
  {Jenko}}]{grovselj2019kinetic}%
  \BibitemOpen
  \bibfield  {author} {\bibinfo {author} {\bibfnamefont {D.}~\bibnamefont
  {Gro{\v{s}}elj}}, \bibinfo {author} {\bibfnamefont {C.~H.~K.}\ \bibnamefont
  {Chen}}, \bibinfo {author} {\bibfnamefont {A.}~\bibnamefont {Mallet}},
  \bibinfo {author} {\bibfnamefont {R.}~\bibnamefont {Samtaney}}, \bibinfo
  {author} {\bibfnamefont {K.}~\bibnamefont {Schneider}}, \ and\ \bibinfo
  {author} {\bibfnamefont {F.}~\bibnamefont {Jenko}},\ }\href@noop {}
  {\bibfield  {journal} {\bibinfo  {journal} {Physical Review X}\ }\textbf
  {\bibinfo {volume} {9}},\ \bibinfo {pages} {031037} (\bibinfo {year}
  {2019})}\BibitemShut {NoStop}%
\bibitem [{\citenamefont {Leamon}\ \emph {et~al.}(1998)\citenamefont {Leamon},
  \citenamefont {Smith}, \citenamefont {Ness}, \citenamefont {Matthaeus},\ and\
  \citenamefont {Wong}}]{leamon1998observational}%
  \BibitemOpen
  \bibfield  {author} {\bibinfo {author} {\bibfnamefont {R.~J.}\ \bibnamefont
  {Leamon}}, \bibinfo {author} {\bibfnamefont {C.~W.}\ \bibnamefont {Smith}},
  \bibinfo {author} {\bibfnamefont {N.~F.}\ \bibnamefont {Ness}}, \bibinfo
  {author} {\bibfnamefont {W.~H.}\ \bibnamefont {Matthaeus}}, \ and\ \bibinfo
  {author} {\bibfnamefont {H.~K.}\ \bibnamefont {Wong}},\ }\href@noop {}
  {\bibfield  {journal} {\bibinfo  {journal} {Journal of Geophysical Research:
  Space Physics}\ }\textbf {\bibinfo {volume} {103}},\ \bibinfo {pages} {4775}
  (\bibinfo {year} {1998})}\BibitemShut {NoStop}%
\bibitem [{\citenamefont {Podesta}\ \emph {et~al.}(2010)\citenamefont
  {Podesta}, \citenamefont {Borovsky},\ and\ \citenamefont
  {Gary}}]{podesta2010kinetic}%
  \BibitemOpen
  \bibfield  {author} {\bibinfo {author} {\bibfnamefont {J.~J.}\ \bibnamefont
  {Podesta}}, \bibinfo {author} {\bibfnamefont {J.~E.}\ \bibnamefont
  {Borovsky}}, \ and\ \bibinfo {author} {\bibfnamefont {S.~P.}\ \bibnamefont
  {Gary}},\ }\href@noop {} {\bibfield  {journal} {\bibinfo  {journal} {The
  Astrophysical Journal}\ }\textbf {\bibinfo {volume} {712}},\ \bibinfo {pages}
  {685} (\bibinfo {year} {2010})}\BibitemShut {NoStop}%
\bibitem [{\citenamefont {Chen}\ \emph {et~al.}(2013)\citenamefont {Chen},
  \citenamefont {Boldyrev}, \citenamefont {Xia},\ and\ \citenamefont
  {Perez}}]{chen2013nature}%
  \BibitemOpen
  \bibfield  {author} {\bibinfo {author} {\bibfnamefont {C.~H.~K.}\
  \bibnamefont {Chen}}, \bibinfo {author} {\bibfnamefont {S.}~\bibnamefont
  {Boldyrev}}, \bibinfo {author} {\bibfnamefont {Q.}~\bibnamefont {Xia}}, \
  and\ \bibinfo {author} {\bibfnamefont {J.~C.}\ \bibnamefont {Perez}},\
  }\href@noop {} {\bibfield  {journal} {\bibinfo  {journal} {Physical Review
  Letters}\ }\textbf {\bibinfo {volume} {110}},\ \bibinfo {pages} {225002}
  (\bibinfo {year} {2013})}\BibitemShut {NoStop}%
\bibitem [{\citenamefont {Kiyani}\ \emph {et~al.}(2012)\citenamefont {Kiyani},
  \citenamefont {Chapman}, \citenamefont {Sahraoui}, \citenamefont {Hnat},
  \citenamefont {Fauvarque},\ and\ \citenamefont
  {Khotyaintsev}}]{kiyani2012enhanced}%
  \BibitemOpen
  \bibfield  {author} {\bibinfo {author} {\bibfnamefont {K.~H.}\ \bibnamefont
  {Kiyani}}, \bibinfo {author} {\bibfnamefont {S.~C.}\ \bibnamefont {Chapman}},
  \bibinfo {author} {\bibfnamefont {F.}~\bibnamefont {Sahraoui}}, \bibinfo
  {author} {\bibfnamefont {B.}~\bibnamefont {Hnat}}, \bibinfo {author}
  {\bibfnamefont {O.}~\bibnamefont {Fauvarque}}, \ and\ \bibinfo {author}
  {\bibfnamefont {Y.~V.}\ \bibnamefont {Khotyaintsev}},\ }\href@noop {}
  {\bibfield  {journal} {\bibinfo  {journal} {The Astrophysical Journal}\
  }\textbf {\bibinfo {volume} {763}},\ \bibinfo {pages} {10} (\bibinfo {year}
  {2012})}\BibitemShut {NoStop}%
\bibitem [{\citenamefont {Chen}(2016)}]{chen2016recent}%
  \BibitemOpen
  \bibfield  {author} {\bibinfo {author} {\bibfnamefont {C.~H.~K.}\
  \bibnamefont {Chen}},\ }\href@noop {} {\bibfield  {journal} {\bibinfo
  {journal} {Journal of Plasma Physics}\ }\textbf {\bibinfo {volume} {82}}
  (\bibinfo {year} {2016})}\BibitemShut {NoStop}%
\bibitem [{\citenamefont {Cho}\ and\ \citenamefont
  {Lazarian}(2004)}]{cho2004anisotropy}%
  \BibitemOpen
  \bibfield  {author} {\bibinfo {author} {\bibfnamefont {J.}~\bibnamefont
  {Cho}}\ and\ \bibinfo {author} {\bibfnamefont {A.}~\bibnamefont {Lazarian}},\
  }\href@noop {} {\bibfield  {journal} {\bibinfo  {journal} {The Astrophysical
  Journal Letters}\ }\textbf {\bibinfo {volume} {615}},\ \bibinfo {pages} {L41}
  (\bibinfo {year} {2004})}\BibitemShut {NoStop}%
\bibitem [{\citenamefont {Cho}\ and\ \citenamefont
  {Lazarian}(2009)}]{cho2009simulations}%
  \BibitemOpen
  \bibfield  {author} {\bibinfo {author} {\bibfnamefont {J.}~\bibnamefont
  {Cho}}\ and\ \bibinfo {author} {\bibfnamefont {A.}~\bibnamefont {Lazarian}},\
  }\href@noop {} {\bibfield  {journal} {\bibinfo  {journal} {The Astrophysical
  Journal}\ }\textbf {\bibinfo {volume} {701}},\ \bibinfo {pages} {236}
  (\bibinfo {year} {2009})}\BibitemShut {NoStop}%
\bibitem [{\citenamefont {{Schekochihin}}\ \emph {et~al.}(2009)\citenamefont
  {{Schekochihin}}, \citenamefont {{Cowley}}, \citenamefont {{Dorland}},
  \citenamefont {{Hammett}}, \citenamefont {{Howes}}, \citenamefont
  {{Quataert}},\ and\ \citenamefont
  {{Tatsuno}}}]{schekochihin2009astrophysical}%
  \BibitemOpen
  \bibfield  {author} {\bibinfo {author} {\bibfnamefont {A.~A.}\ \bibnamefont
  {{Schekochihin}}}, \bibinfo {author} {\bibfnamefont {S.~C.}\ \bibnamefont
  {{Cowley}}}, \bibinfo {author} {\bibfnamefont {W.}~\bibnamefont {{Dorland}}},
  \bibinfo {author} {\bibfnamefont {G.~W.}\ \bibnamefont {{Hammett}}}, \bibinfo
  {author} {\bibfnamefont {G.~G.}\ \bibnamefont {{Howes}}}, \bibinfo {author}
  {\bibfnamefont {E.}~\bibnamefont {{Quataert}}}, \ and\ \bibinfo {author}
  {\bibfnamefont {T.}~\bibnamefont {{Tatsuno}}},\ }\href@noop {} {\bibfield
  {journal} {\bibinfo  {journal} {The Astrophysical Journal Supplement Series}\
  }\textbf {\bibinfo {volume} {182}},\ \bibinfo {pages} {310} (\bibinfo {year}
  {2009})}\BibitemShut {NoStop}%
\bibitem [{\citenamefont {Bale}\ \emph {et~al.}(2005)\citenamefont {Bale},
  \citenamefont {Kellogg}, \citenamefont {Mozer}, \citenamefont {Horbury},\
  and\ \citenamefont {Reme}}]{bale2005measurement}%
  \BibitemOpen
  \bibfield  {author} {\bibinfo {author} {\bibfnamefont {S.~D.}\ \bibnamefont
  {Bale}}, \bibinfo {author} {\bibfnamefont {P.~J.}\ \bibnamefont {Kellogg}},
  \bibinfo {author} {\bibfnamefont {F.~S.}\ \bibnamefont {Mozer}}, \bibinfo
  {author} {\bibfnamefont {T.~S.}\ \bibnamefont {Horbury}}, \ and\ \bibinfo
  {author} {\bibfnamefont {H.}~\bibnamefont {Reme}},\ }\href@noop {} {\bibfield
   {journal} {\bibinfo  {journal} {Physical Review Letters}\ }\textbf {\bibinfo
  {volume} {94}},\ \bibinfo {pages} {215002} (\bibinfo {year}
  {2005})}\BibitemShut {NoStop}%
\bibitem [{\citenamefont {Servidio}\ \emph {et~al.}(2015)\citenamefont
  {Servidio}, \citenamefont {Valentini}, \citenamefont {Perrone}, \citenamefont
  {Greco}, \citenamefont {Califano}, \citenamefont {Matthaeus},\ and\
  \citenamefont {Veltri}}]{servidio2015kinetic}%
  \BibitemOpen
  \bibfield  {author} {\bibinfo {author} {\bibfnamefont {S.}~\bibnamefont
  {Servidio}}, \bibinfo {author} {\bibfnamefont {F.}~\bibnamefont {Valentini}},
  \bibinfo {author} {\bibfnamefont {D.}~\bibnamefont {Perrone}}, \bibinfo
  {author} {\bibfnamefont {A.}~\bibnamefont {Greco}}, \bibinfo {author}
  {\bibfnamefont {F.}~\bibnamefont {Califano}}, \bibinfo {author}
  {\bibfnamefont {W.~H.}\ \bibnamefont {Matthaeus}}, \ and\ \bibinfo {author}
  {\bibfnamefont {P.}~\bibnamefont {Veltri}},\ }\href@noop {} {\bibfield
  {journal} {\bibinfo  {journal} {Journal of Plasma Physics}\ }\textbf
  {\bibinfo {volume} {81}} (\bibinfo {year} {2015})}\BibitemShut {NoStop}%
\bibitem [{\citenamefont {Wan}\ \emph {et~al.}(2015{\natexlab{a}})\citenamefont
  {Wan}, \citenamefont {Matthaeus}, \citenamefont {Roytershteyn}, \citenamefont
  {Karimabadi}, \citenamefont {Parashar}, \citenamefont {Wu},\ and\
  \citenamefont {Shay}}]{wan2015intermittent}%
  \BibitemOpen
  \bibfield  {author} {\bibinfo {author} {\bibfnamefont {M.}~\bibnamefont
  {Wan}}, \bibinfo {author} {\bibfnamefont {W.~H.}\ \bibnamefont {Matthaeus}},
  \bibinfo {author} {\bibfnamefont {V.}~\bibnamefont {Roytershteyn}}, \bibinfo
  {author} {\bibfnamefont {H.}~\bibnamefont {Karimabadi}}, \bibinfo {author}
  {\bibfnamefont {T.}~\bibnamefont {Parashar}}, \bibinfo {author}
  {\bibfnamefont {P.}~\bibnamefont {Wu}}, \ and\ \bibinfo {author}
  {\bibfnamefont {M.}~\bibnamefont {Shay}},\ }\href@noop {} {\bibfield
  {journal} {\bibinfo  {journal} {Physical Review Letters}\ }\textbf {\bibinfo
  {volume} {114}},\ \bibinfo {pages} {175002} (\bibinfo {year}
  {2015}{\natexlab{a}})}\BibitemShut {NoStop}%
\bibitem [{\citenamefont {Told}\ \emph {et~al.}(2015)\citenamefont {Told},
  \citenamefont {Jenko}, \citenamefont {TenBarge}, \citenamefont {Howes},\ and\
  \citenamefont {Hammett}}]{told2015multiscale}%
  \BibitemOpen
  \bibfield  {author} {\bibinfo {author} {\bibfnamefont {D.}~\bibnamefont
  {Told}}, \bibinfo {author} {\bibfnamefont {F.}~\bibnamefont {Jenko}},
  \bibinfo {author} {\bibfnamefont {J.~M.}\ \bibnamefont {TenBarge}}, \bibinfo
  {author} {\bibfnamefont {G.~G.}\ \bibnamefont {Howes}}, \ and\ \bibinfo
  {author} {\bibfnamefont {G.}~\bibnamefont {Hammett}},\ }\href@noop {}
  {\bibfield  {journal} {\bibinfo  {journal} {Physical Review Letters}\
  }\textbf {\bibinfo {volume} {115}},\ \bibinfo {pages} {025003} (\bibinfo
  {year} {2015})}\BibitemShut {NoStop}%
\bibitem [{\citenamefont {Cerri}\ \emph {et~al.}(2018)\citenamefont {Cerri},
  \citenamefont {Kunz},\ and\ \citenamefont {Califano}}]{cerri2018dual}%
  \BibitemOpen
  \bibfield  {author} {\bibinfo {author} {\bibfnamefont {S.~S.}\ \bibnamefont
  {Cerri}}, \bibinfo {author} {\bibfnamefont {M.~W.}\ \bibnamefont {Kunz}}, \
  and\ \bibinfo {author} {\bibfnamefont {F.}~\bibnamefont {Califano}},\
  }\href@noop {} {\bibfield  {journal} {\bibinfo  {journal} {The Astrophysical
  Journal Letters}\ }\textbf {\bibinfo {volume} {856}},\ \bibinfo {pages} {L13}
  (\bibinfo {year} {2018})}\BibitemShut {NoStop}%
\bibitem [{\citenamefont {Arzamasskiy}\ \emph {et~al.}(2019)\citenamefont
  {Arzamasskiy}, \citenamefont {Kunz}, \citenamefont {Chandran},\ and\
  \citenamefont {Quataert}}]{arzamasskiy2019hybrid}%
  \BibitemOpen
  \bibfield  {author} {\bibinfo {author} {\bibfnamefont {L.}~\bibnamefont
  {Arzamasskiy}}, \bibinfo {author} {\bibfnamefont {M.~W.}\ \bibnamefont
  {Kunz}}, \bibinfo {author} {\bibfnamefont {B.~D.~G.}\ \bibnamefont
  {Chandran}}, \ and\ \bibinfo {author} {\bibfnamefont {E.}~\bibnamefont
  {Quataert}},\ }\href@noop {} {\bibfield  {journal} {\bibinfo  {journal} {The
  Astrophysical Journal}\ }\textbf {\bibinfo {volume} {879}},\ \bibinfo {pages}
  {53} (\bibinfo {year} {2019})}\BibitemShut {NoStop}%
\bibitem [{\citenamefont {Gro{\v{s}}elj}\ \emph {et~al.}(2018)\citenamefont
  {Gro{\v{s}}elj}, \citenamefont {Mallet}, \citenamefont {Loureiro},\ and\
  \citenamefont {Jenko}}]{grovselj2018fully}%
  \BibitemOpen
  \bibfield  {author} {\bibinfo {author} {\bibfnamefont {D.}~\bibnamefont
  {Gro{\v{s}}elj}}, \bibinfo {author} {\bibfnamefont {A.}~\bibnamefont
  {Mallet}}, \bibinfo {author} {\bibfnamefont {N.~F.}\ \bibnamefont
  {Loureiro}}, \ and\ \bibinfo {author} {\bibfnamefont {F.}~\bibnamefont
  {Jenko}},\ }\href@noop {} {\bibfield  {journal} {\bibinfo  {journal}
  {Physical Review Letters}\ }\textbf {\bibinfo {volume} {120}},\ \bibinfo
  {pages} {105101} (\bibinfo {year} {2018})}\BibitemShut {NoStop}%
\bibitem [{\citenamefont {Kiyani}\ \emph {et~al.}(2009)\citenamefont {Kiyani},
  \citenamefont {Chapman}, \citenamefont {Khotyaintsev}, \citenamefont
  {Dunlop},\ and\ \citenamefont {Sahraoui}}]{kiyani2009global}%
  \BibitemOpen
  \bibfield  {author} {\bibinfo {author} {\bibfnamefont {K.~H.}\ \bibnamefont
  {Kiyani}}, \bibinfo {author} {\bibfnamefont {S.~C.}\ \bibnamefont {Chapman}},
  \bibinfo {author} {\bibfnamefont {Y.~V.}\ \bibnamefont {Khotyaintsev}},
  \bibinfo {author} {\bibfnamefont {M.~W.}\ \bibnamefont {Dunlop}}, \ and\
  \bibinfo {author} {\bibfnamefont {F.}~\bibnamefont {Sahraoui}},\ }\href@noop
  {} {\bibfield  {journal} {\bibinfo  {journal} {Physical Review Letters}\
  }\textbf {\bibinfo {volume} {103}},\ \bibinfo {pages} {075006} (\bibinfo
  {year} {2009})}\BibitemShut {NoStop}%
\bibitem [{\citenamefont {Sahraoui}\ \emph {et~al.}(2013)\citenamefont
  {Sahraoui}, \citenamefont {Huang}, \citenamefont {Belmont}, \citenamefont
  {Goldstein}, \citenamefont {Retin{\`o}}, \citenamefont {Robert},\ and\
  \citenamefont {De~Patoul}}]{sahraoui2013scaling}%
  \BibitemOpen
  \bibfield  {author} {\bibinfo {author} {\bibfnamefont {F.}~\bibnamefont
  {Sahraoui}}, \bibinfo {author} {\bibfnamefont {S.~Y.}\ \bibnamefont {Huang}},
  \bibinfo {author} {\bibfnamefont {G.}~\bibnamefont {Belmont}}, \bibinfo
  {author} {\bibfnamefont {M.~L.}\ \bibnamefont {Goldstein}}, \bibinfo {author}
  {\bibfnamefont {A.}~\bibnamefont {Retin{\`o}}}, \bibinfo {author}
  {\bibfnamefont {P.}~\bibnamefont {Robert}}, \ and\ \bibinfo {author}
  {\bibfnamefont {J.}~\bibnamefont {De~Patoul}},\ }\href@noop {} {\bibfield
  {journal} {\bibinfo  {journal} {The Astrophysical Journal}\ }\textbf
  {\bibinfo {volume} {777}},\ \bibinfo {pages} {15} (\bibinfo {year}
  {2013})}\BibitemShut {NoStop}%
\bibitem [{\citenamefont {Sundkvist}\ \emph {et~al.}(2007)\citenamefont
  {Sundkvist}, \citenamefont {Retin\`o}, \citenamefont {Vaivads},\ and\
  \citenamefont {Bale}}]{Sundkvist2007}%
  \BibitemOpen
  \bibfield  {author} {\bibinfo {author} {\bibfnamefont {D.}~\bibnamefont
  {Sundkvist}}, \bibinfo {author} {\bibfnamefont {A.}~\bibnamefont {Retin\`o}},
  \bibinfo {author} {\bibfnamefont {A.}~\bibnamefont {Vaivads}}, \ and\
  \bibinfo {author} {\bibfnamefont {S.~D.}\ \bibnamefont {Bale}},\ }\href@noop
  {} {\bibfield  {journal} {\bibinfo  {journal} {Physical Review Letters}\
  }\textbf {\bibinfo {volume} {99}},\ \bibinfo {pages} {025004} (\bibinfo
  {year} {2007})}\BibitemShut {NoStop}%
\bibitem [{\citenamefont {{Karimabadi}}\ \emph {et~al.}(2011)\citenamefont
  {{Karimabadi}}, \citenamefont {{Dorelli}}, \citenamefont {{Roytershteyn}},
  \citenamefont {{Daughton}},\ and\ \citenamefont
  {{Chac{\'o}n}}}]{karimabadi2011flux}%
  \BibitemOpen
  \bibfield  {author} {\bibinfo {author} {\bibfnamefont {H.}~\bibnamefont
  {{Karimabadi}}}, \bibinfo {author} {\bibfnamefont {J.}~\bibnamefont
  {{Dorelli}}}, \bibinfo {author} {\bibfnamefont {V.}~\bibnamefont
  {{Roytershteyn}}}, \bibinfo {author} {\bibfnamefont {W.}~\bibnamefont
  {{Daughton}}}, \ and\ \bibinfo {author} {\bibfnamefont {L.}~\bibnamefont
  {{Chac{\'o}n}}},\ }\href {\doibase 10.1103/PhysRevLett.107.025002} {\bibfield
   {journal} {\bibinfo  {journal} {Physical Review Letters}\ }\textbf {\bibinfo
  {volume} {107}},\ \bibinfo {eid} {025002} (\bibinfo {year}
  {2011})}\BibitemShut {NoStop}%
\bibitem [{\citenamefont {Osman}\ \emph {et~al.}(2014)\citenamefont {Osman},
  \citenamefont {Matthaeus}, \citenamefont {Gosling}, \citenamefont {Greco},
  \citenamefont {Servidio}, \citenamefont {Hnat}, \citenamefont {Chapman},\
  and\ \citenamefont {Phan}}]{Osman2014}%
  \BibitemOpen
  \bibfield  {author} {\bibinfo {author} {\bibfnamefont {K.~T.}\ \bibnamefont
  {Osman}}, \bibinfo {author} {\bibfnamefont {W.~H.}\ \bibnamefont
  {Matthaeus}}, \bibinfo {author} {\bibfnamefont {J.~T.}\ \bibnamefont
  {Gosling}}, \bibinfo {author} {\bibfnamefont {A.}~\bibnamefont {Greco}},
  \bibinfo {author} {\bibfnamefont {S.}~\bibnamefont {Servidio}}, \bibinfo
  {author} {\bibfnamefont {B.}~\bibnamefont {Hnat}}, \bibinfo {author}
  {\bibfnamefont {S.~C.}\ \bibnamefont {Chapman}}, \ and\ \bibinfo {author}
  {\bibfnamefont {T.~D.}\ \bibnamefont {Phan}},\ }\href@noop {} {\bibfield
  {journal} {\bibinfo  {journal} {Physical Review Letters}\ }\textbf {\bibinfo
  {volume} {112}},\ \bibinfo {pages} {215002} (\bibinfo {year}
  {2014})}\BibitemShut {NoStop}%
\bibitem [{\citenamefont {Wan}\ \emph {et~al.}(2015{\natexlab{b}})\citenamefont
  {Wan}, \citenamefont {Matthaeus}, \citenamefont {Roytershteyn}, \citenamefont
  {Karimabadi}, \citenamefont {Parashar}, \citenamefont {Wu},\ and\
  \citenamefont {Shay}}]{wan2015}%
  \BibitemOpen
  \bibfield  {author} {\bibinfo {author} {\bibfnamefont {M.}~\bibnamefont
  {Wan}}, \bibinfo {author} {\bibfnamefont {W.~H.}\ \bibnamefont {Matthaeus}},
  \bibinfo {author} {\bibfnamefont {V.}~\bibnamefont {Roytershteyn}}, \bibinfo
  {author} {\bibfnamefont {H.}~\bibnamefont {Karimabadi}}, \bibinfo {author}
  {\bibfnamefont {T.}~\bibnamefont {Parashar}}, \bibinfo {author}
  {\bibfnamefont {P.}~\bibnamefont {Wu}}, \ and\ \bibinfo {author}
  {\bibfnamefont {M.}~\bibnamefont {Shay}},\ }\href@noop {} {\bibfield
  {journal} {\bibinfo  {journal} {Physical Review Letters}\ }\textbf {\bibinfo
  {volume} {114}},\ \bibinfo {pages} {175002} (\bibinfo {year}
  {2015}{\natexlab{b}})}\BibitemShut {NoStop}%
\bibitem [{\citenamefont {Chasapis}\ \emph {et~al.}(2015)\citenamefont
  {Chasapis}, \citenamefont {Retin{\`o}}, \citenamefont {Sahraoui},
  \citenamefont {Vaivads}, \citenamefont {Khotyaintsev}, \citenamefont
  {Sundkvist}, \citenamefont {Greco}, \citenamefont {Sorriso-Valvo},\ and\
  \citenamefont {Canu}}]{chasapis2015thin}%
  \BibitemOpen
  \bibfield  {author} {\bibinfo {author} {\bibfnamefont {A.}~\bibnamefont
  {Chasapis}}, \bibinfo {author} {\bibfnamefont {A.}~\bibnamefont
  {Retin{\`o}}}, \bibinfo {author} {\bibfnamefont {F.}~\bibnamefont
  {Sahraoui}}, \bibinfo {author} {\bibfnamefont {A.}~\bibnamefont {Vaivads}},
  \bibinfo {author} {\bibfnamefont {Y.~V.}\ \bibnamefont {Khotyaintsev}},
  \bibinfo {author} {\bibfnamefont {D.}~\bibnamefont {Sundkvist}}, \bibinfo
  {author} {\bibfnamefont {A.}~\bibnamefont {Greco}}, \bibinfo {author}
  {\bibfnamefont {L.}~\bibnamefont {Sorriso-Valvo}}, \ and\ \bibinfo {author}
  {\bibfnamefont {P.}~\bibnamefont {Canu}},\ }\href@noop {} {\bibfield
  {journal} {\bibinfo  {journal} {The Astrophysical Journal Letters}\ }\textbf
  {\bibinfo {volume} {804}},\ \bibinfo {pages} {L1} (\bibinfo {year}
  {2015})}\BibitemShut {NoStop}%
\bibitem [{\citenamefont {Camporeale}\ \emph {et~al.}(2018)\citenamefont
  {Camporeale}, \citenamefont {Sorriso-Valvo}, \citenamefont {Califano},\ and\
  \citenamefont {Retin\`o}}]{Camporeale2018}%
  \BibitemOpen
  \bibfield  {author} {\bibinfo {author} {\bibfnamefont {E.}~\bibnamefont
  {Camporeale}}, \bibinfo {author} {\bibfnamefont {L.}~\bibnamefont
  {Sorriso-Valvo}}, \bibinfo {author} {\bibfnamefont {F.}~\bibnamefont
  {Califano}}, \ and\ \bibinfo {author} {\bibfnamefont {A.}~\bibnamefont
  {Retin\`o}},\ }\href@noop {} {\bibfield  {journal} {\bibinfo  {journal}
  {Physical Review Letters}\ }\textbf {\bibinfo {volume} {120}},\ \bibinfo
  {pages} {125101} (\bibinfo {year} {2018})}\BibitemShut {NoStop}%
\bibitem [{\citenamefont {Boldyrev}\ and\ \citenamefont
  {Perez}(2012)}]{boldyrev2012spectrum}%
  \BibitemOpen
  \bibfield  {author} {\bibinfo {author} {\bibfnamefont {S.}~\bibnamefont
  {Boldyrev}}\ and\ \bibinfo {author} {\bibfnamefont {J.~C.}\ \bibnamefont
  {Perez}},\ }\href@noop {} {\bibfield  {journal} {\bibinfo  {journal} {The
  Astrophysical Journal Letters}\ }\textbf {\bibinfo {volume} {758}},\ \bibinfo
  {pages} {L44} (\bibinfo {year} {2012})}\BibitemShut {NoStop}%
\bibitem [{\citenamefont {Boldyrev}(2006)}]{boldyrev2006spectrum}%
  \BibitemOpen
  \bibfield  {author} {\bibinfo {author} {\bibfnamefont {S.}~\bibnamefont
  {Boldyrev}},\ }\href@noop {} {\bibfield  {journal} {\bibinfo  {journal}
  {Physics Review Letters}\ }\textbf {\bibinfo {volume} {96}},\ \bibinfo
  {pages} {115002} (\bibinfo {year} {2006})}\BibitemShut {NoStop}%
\bibitem [{\citenamefont {Loureiro}\ and\ \citenamefont
  {Boldyrev}(2017{\natexlab{a}})}]{loureiro2017role}%
  \BibitemOpen
  \bibfield  {author} {\bibinfo {author} {\bibfnamefont {N.~F.}\ \bibnamefont
  {Loureiro}}\ and\ \bibinfo {author} {\bibfnamefont {S.}~\bibnamefont
  {Boldyrev}},\ }\href@noop {} {\bibfield  {journal} {\bibinfo  {journal}
  {Physical Review Letters}\ }\textbf {\bibinfo {volume} {118}},\ \bibinfo
  {pages} {245101} (\bibinfo {year} {2017}{\natexlab{a}})}\BibitemShut
  {NoStop}%
\bibitem [{\citenamefont {Mallet}\ and\ \citenamefont
  {Schekochihin}(2017)}]{mallet2017statistical}%
  \BibitemOpen
  \bibfield  {author} {\bibinfo {author} {\bibfnamefont {A.}~\bibnamefont
  {Mallet}}\ and\ \bibinfo {author} {\bibfnamefont {A.~A.}\ \bibnamefont
  {Schekochihin}},\ }\href@noop {} {\bibfield  {journal} {\bibinfo  {journal}
  {Monthly Notices of the Royal Astronomical Society}\ }\textbf {\bibinfo
  {volume} {466}},\ \bibinfo {pages} {3918} (\bibinfo {year}
  {2017})}\BibitemShut {NoStop}%
\bibitem [{\citenamefont {Boldyrev}\ and\ \citenamefont
  {Loureiro}(2017)}]{boldyrev2017magnetohydrodynamic}%
  \BibitemOpen
  \bibfield  {author} {\bibinfo {author} {\bibfnamefont {S.}~\bibnamefont
  {Boldyrev}}\ and\ \bibinfo {author} {\bibfnamefont {N.~F.}\ \bibnamefont
  {Loureiro}},\ }\href@noop {} {\bibfield  {journal} {\bibinfo  {journal} {The
  Astrophysical Journal}\ }\textbf {\bibinfo {volume} {844}},\ \bibinfo {pages}
  {125} (\bibinfo {year} {2017})}\BibitemShut {NoStop}%
\bibitem [{\citenamefont {Dong}\ \emph {et~al.}(2018)\citenamefont {Dong},
  \citenamefont {Wang}, \citenamefont {Huang}, \citenamefont {Comisso},\ and\
  \citenamefont {Bhattacharjee}}]{dong2018role}%
  \BibitemOpen
  \bibfield  {author} {\bibinfo {author} {\bibfnamefont {C.}~\bibnamefont
  {Dong}}, \bibinfo {author} {\bibfnamefont {L.}~\bibnamefont {Wang}}, \bibinfo
  {author} {\bibfnamefont {Y.-M.}\ \bibnamefont {Huang}}, \bibinfo {author}
  {\bibfnamefont {L.}~\bibnamefont {Comisso}}, \ and\ \bibinfo {author}
  {\bibfnamefont {A.}~\bibnamefont {Bhattacharjee}},\ }\href@noop {} {\bibfield
   {journal} {\bibinfo  {journal} {Physical Review Letters}\ }\textbf {\bibinfo
  {volume} {121}},\ \bibinfo {pages} {165101} (\bibinfo {year}
  {2018})}\BibitemShut {NoStop}%
\bibitem [{\citenamefont {Walker}\ \emph {et~al.}(2018)\citenamefont {Walker},
  \citenamefont {Boldyrev},\ and\ \citenamefont
  {Loureiro}}]{walker2018influence}%
  \BibitemOpen
  \bibfield  {author} {\bibinfo {author} {\bibfnamefont {J.}~\bibnamefont
  {Walker}}, \bibinfo {author} {\bibfnamefont {S.}~\bibnamefont {Boldyrev}}, \
  and\ \bibinfo {author} {\bibfnamefont {N.~F.}\ \bibnamefont {Loureiro}},\
  }\href@noop {} {\bibfield  {journal} {\bibinfo  {journal} {Physical Review
  E}\ }\textbf {\bibinfo {volume} {98}},\ \bibinfo {pages} {033209} (\bibinfo
  {year} {2018})}\BibitemShut {NoStop}%
\bibitem [{\citenamefont {Loureiro}\ and\ \citenamefont
  {Boldyrev}(2020)}]{loureiro2020nonlinear}%
  \BibitemOpen
  \bibfield  {author} {\bibinfo {author} {\bibfnamefont {N.~F.}\ \bibnamefont
  {Loureiro}}\ and\ \bibinfo {author} {\bibfnamefont {S.}~\bibnamefont
  {Boldyrev}},\ }\href@noop {} {\bibfield  {journal} {\bibinfo  {journal} {The
  Astrophysical Journal}\ }\textbf {\bibinfo {volume} {890}},\ \bibinfo {pages}
  {55} (\bibinfo {year} {2020})}\BibitemShut {NoStop}%
\bibitem [{\citenamefont {Loureiro}\ and\ \citenamefont
  {Boldyrev}(2017{\natexlab{b}})}]{loureiro2017collisionless}%
  \BibitemOpen
  \bibfield  {author} {\bibinfo {author} {\bibfnamefont {N.~F.}\ \bibnamefont
  {Loureiro}}\ and\ \bibinfo {author} {\bibfnamefont {S.}~\bibnamefont
  {Boldyrev}},\ }\href@noop {} {\bibfield  {journal} {\bibinfo  {journal} {The
  Astrophysical Journal}\ }\textbf {\bibinfo {volume} {850}},\ \bibinfo {pages}
  {182} (\bibinfo {year} {2017}{\natexlab{b}})}\BibitemShut {NoStop}%
\bibitem [{\citenamefont {Parashar}\ \emph {et~al.}(2015)\citenamefont
  {Parashar}, \citenamefont {Salem}, \citenamefont {Wicks}, \citenamefont
  {Karimabadi}, \citenamefont {Gary},\ and\ \citenamefont
  {Matthaeus}}]{parashar2015turbulent}%
  \BibitemOpen
  \bibfield  {author} {\bibinfo {author} {\bibfnamefont {T.~N.}\ \bibnamefont
  {Parashar}}, \bibinfo {author} {\bibfnamefont {C.}~\bibnamefont {Salem}},
  \bibinfo {author} {\bibfnamefont {R.~T.}\ \bibnamefont {Wicks}}, \bibinfo
  {author} {\bibfnamefont {H.}~\bibnamefont {Karimabadi}}, \bibinfo {author}
  {\bibfnamefont {S.~P.}\ \bibnamefont {Gary}}, \ and\ \bibinfo {author}
  {\bibfnamefont {W.~H.}\ \bibnamefont {Matthaeus}},\ }\href@noop {} {\bibfield
   {journal} {\bibinfo  {journal} {Journal of Plasma Physics}\ }\textbf
  {\bibinfo {volume} {81}} (\bibinfo {year} {2015})}\BibitemShut {NoStop}%
\bibitem [{\citenamefont {Dmitruk}\ \emph {et~al.}(2004)\citenamefont
  {Dmitruk}, \citenamefont {Matthaeus},\ and\ \citenamefont
  {Seenu}}]{dmitruk2004test}%
  \BibitemOpen
  \bibfield  {author} {\bibinfo {author} {\bibfnamefont {P.}~\bibnamefont
  {Dmitruk}}, \bibinfo {author} {\bibfnamefont {W.~H.}\ \bibnamefont
  {Matthaeus}}, \ and\ \bibinfo {author} {\bibfnamefont {N.}~\bibnamefont
  {Seenu}},\ }\href@noop {} {\bibfield  {journal} {\bibinfo  {journal} {The
  Astrophysical Journal}\ }\textbf {\bibinfo {volume} {617}},\ \bibinfo {pages}
  {667} (\bibinfo {year} {2004})}\BibitemShut {NoStop}%
\bibitem [{\citenamefont {{Matthaeus}}\ and\ \citenamefont
  {{Velli}}(2011)}]{Matthaeus2011who}%
  \BibitemOpen
  \bibfield  {author} {\bibinfo {author} {\bibfnamefont {W.~H.}\ \bibnamefont
  {{Matthaeus}}}\ and\ \bibinfo {author} {\bibfnamefont {M.}~\bibnamefont
  {{Velli}}},\ }\href@noop {} {\bibfield  {journal} {\bibinfo  {journal} {Space
  Science Reviews}\ }\textbf {\bibinfo {volume} {160}},\ \bibinfo {pages} {145}
  (\bibinfo {year} {2011})}\BibitemShut {NoStop}%
\bibitem [{\citenamefont {{Servidio}}\ \emph {et~al.}(2011)\citenamefont
  {{Servidio}}, \citenamefont {{Dmitruk}}, \citenamefont {{Greco}},
  \citenamefont {{Wan}}, \citenamefont {{Donato}}, \citenamefont {{Cassak}},
  \citenamefont {{Shay}}, \citenamefont {{Carbone}},\ and\ \citenamefont
  {{Matthaeus}}}]{servidio2011}%
  \BibitemOpen
  \bibfield  {author} {\bibinfo {author} {\bibfnamefont {S.}~\bibnamefont
  {{Servidio}}}, \bibinfo {author} {\bibfnamefont {P.}~\bibnamefont
  {{Dmitruk}}}, \bibinfo {author} {\bibfnamefont {A.}~\bibnamefont {{Greco}}},
  \bibinfo {author} {\bibfnamefont {M.}~\bibnamefont {{Wan}}}, \bibinfo
  {author} {\bibfnamefont {S.}~\bibnamefont {{Donato}}}, \bibinfo {author}
  {\bibfnamefont {P.~A.}\ \bibnamefont {{Cassak}}}, \bibinfo {author}
  {\bibfnamefont {M.~A.}\ \bibnamefont {{Shay}}}, \bibinfo {author}
  {\bibfnamefont {V.}~\bibnamefont {{Carbone}}}, \ and\ \bibinfo {author}
  {\bibfnamefont {W.~H.}\ \bibnamefont {{Matthaeus}}},\ }\href@noop {}
  {\bibfield  {journal} {\bibinfo  {journal} {Nonlinear Processes in
  Geophysics}\ }\textbf {\bibinfo {volume} {18}},\ \bibinfo {pages} {675}
  (\bibinfo {year} {2011})}\BibitemShut {NoStop}%
\bibitem [{\citenamefont {{Osman}}\ \emph {et~al.}(2011)\citenamefont
  {{Osman}}, \citenamefont {{Matthaeus}}, \citenamefont {{Greco}},\ and\
  \citenamefont {{Servidio}}}]{osman2011}%
  \BibitemOpen
  \bibfield  {author} {\bibinfo {author} {\bibfnamefont {K.~T.}\ \bibnamefont
  {{Osman}}}, \bibinfo {author} {\bibfnamefont {W.~H.}\ \bibnamefont
  {{Matthaeus}}}, \bibinfo {author} {\bibfnamefont {A.}~\bibnamefont
  {{Greco}}}, \ and\ \bibinfo {author} {\bibfnamefont {S.}~\bibnamefont
  {{Servidio}}},\ }\href@noop {} {\bibfield  {journal} {\bibinfo  {journal}
  {\apjl}\ ,\ \bibinfo {eid} {L11}} (\bibinfo {year} {2011})}\BibitemShut
  {NoStop}%
\bibitem [{\citenamefont {Zhdankin}\ \emph {et~al.}(2013)\citenamefont
  {Zhdankin}, \citenamefont {Uzdensky}, \citenamefont {Perez},\ and\
  \citenamefont {Boldyrev}}]{zhdankin2013statistical}%
  \BibitemOpen
  \bibfield  {author} {\bibinfo {author} {\bibfnamefont {V.}~\bibnamefont
  {Zhdankin}}, \bibinfo {author} {\bibfnamefont {D.~A.}\ \bibnamefont
  {Uzdensky}}, \bibinfo {author} {\bibfnamefont {J.~C.}\ \bibnamefont {Perez}},
  \ and\ \bibinfo {author} {\bibfnamefont {S.}~\bibnamefont {Boldyrev}},\
  }\href@noop {} {\bibfield  {journal} {\bibinfo  {journal} {The Astrophysical
  Journal}\ }\textbf {\bibinfo {volume} {771}},\ \bibinfo {pages} {124}
  (\bibinfo {year} {2013})}\BibitemShut {NoStop}%
\bibitem [{\citenamefont {Chapman}\ and\ \citenamefont
  {Cowling}(1990)}]{chapman1990mathematical}%
  \BibitemOpen
  \bibfield  {author} {\bibinfo {author} {\bibfnamefont {S.}~\bibnamefont
  {Chapman}}\ and\ \bibinfo {author} {\bibfnamefont {T.~G.}\ \bibnamefont
  {Cowling}},\ }\href@noop {} {\emph {\bibinfo {title} {The mathematical theory
  of non-uniform gases: an account of the kinetic theory of viscosity, thermal
  conduction and diffusion in gases}}}\ (\bibinfo  {publisher} {Cambridge
  university press},\ \bibinfo {year} {1990})\BibitemShut {NoStop}%
\bibitem [{\citenamefont {Schekochihin}\ \emph {et~al.}(2008)\citenamefont
  {Schekochihin}, \citenamefont {Cowley}, \citenamefont {Dorland},
  \citenamefont {Hammett}, \citenamefont {Howes}, \citenamefont {Plunk},
  \citenamefont {Quataert},\ and\ \citenamefont
  {Tatsuno}}]{schekochihin2008gyrokinetic}%
  \BibitemOpen
  \bibfield  {author} {\bibinfo {author} {\bibfnamefont {A.~A.}\ \bibnamefont
  {Schekochihin}}, \bibinfo {author} {\bibfnamefont {S.~C.}\ \bibnamefont
  {Cowley}}, \bibinfo {author} {\bibfnamefont {W.}~\bibnamefont {Dorland}},
  \bibinfo {author} {\bibfnamefont {G.~W.}\ \bibnamefont {Hammett}}, \bibinfo
  {author} {\bibfnamefont {G.~G.}\ \bibnamefont {Howes}}, \bibinfo {author}
  {\bibfnamefont {G.~G.}\ \bibnamefont {Plunk}}, \bibinfo {author}
  {\bibfnamefont {E.}~\bibnamefont {Quataert}}, \ and\ \bibinfo {author}
  {\bibfnamefont {T.}~\bibnamefont {Tatsuno}},\ }\href@noop {} {\bibfield
  {journal} {\bibinfo  {journal} {Plasma Physics and Controlled Fusion}\
  }\textbf {\bibinfo {volume} {50}},\ \bibinfo {pages} {124024} (\bibinfo
  {year} {2008})}\BibitemShut {NoStop}%
\bibitem [{\citenamefont {Navarro}\ \emph {et~al.}(2016)\citenamefont
  {Navarro}, \citenamefont {Teaca}, \citenamefont {Told}, \citenamefont
  {Groselj}, \citenamefont {Crandall},\ and\ \citenamefont
  {Jenko}}]{navarro2016structure}%
  \BibitemOpen
  \bibfield  {author} {\bibinfo {author} {\bibfnamefont {A.~B.}\ \bibnamefont
  {Navarro}}, \bibinfo {author} {\bibfnamefont {B.}~\bibnamefont {Teaca}},
  \bibinfo {author} {\bibfnamefont {D.}~\bibnamefont {Told}}, \bibinfo {author}
  {\bibfnamefont {D.}~\bibnamefont {Groselj}}, \bibinfo {author} {\bibfnamefont
  {P.}~\bibnamefont {Crandall}}, \ and\ \bibinfo {author} {\bibfnamefont
  {F.}~\bibnamefont {Jenko}},\ }\href@noop {} {\bibfield  {journal} {\bibinfo
  {journal} {Physical Review Letters}\ }\textbf {\bibinfo {volume} {117}},\
  \bibinfo {pages} {245101} (\bibinfo {year} {2016})}\BibitemShut {NoStop}%
\bibitem [{\citenamefont {Gro{\v{s}}elj}\ \emph {et~al.}(2017)\citenamefont
  {Gro{\v{s}}elj}, \citenamefont {Cerri}, \citenamefont {Navarro},
  \citenamefont {Willmott}, \citenamefont {Told}, \citenamefont {Loureiro},
  \citenamefont {Califano},\ and\ \citenamefont {Jenko}}]{grovselj2017fully}%
  \BibitemOpen
  \bibfield  {author} {\bibinfo {author} {\bibfnamefont {D.}~\bibnamefont
  {Gro{\v{s}}elj}}, \bibinfo {author} {\bibfnamefont {S.~S.}\ \bibnamefont
  {Cerri}}, \bibinfo {author} {\bibfnamefont {A.~B.}\ \bibnamefont {Navarro}},
  \bibinfo {author} {\bibfnamefont {C.}~\bibnamefont {Willmott}}, \bibinfo
  {author} {\bibfnamefont {D.}~\bibnamefont {Told}}, \bibinfo {author}
  {\bibfnamefont {N.~F.}\ \bibnamefont {Loureiro}}, \bibinfo {author}
  {\bibfnamefont {F.}~\bibnamefont {Califano}}, \ and\ \bibinfo {author}
  {\bibfnamefont {F.}~\bibnamefont {Jenko}},\ }\href@noop {} {\bibfield
  {journal} {\bibinfo  {journal} {The Astrophysical Journal}\ }\textbf
  {\bibinfo {volume} {847}},\ \bibinfo {pages} {28} (\bibinfo {year}
  {2017})}\BibitemShut {NoStop}%
\bibitem [{\citenamefont {Servidio}\ \emph {et~al.}(2017)\citenamefont
  {Servidio}, \citenamefont {Chasapis}, \citenamefont {Matthaeus},
  \citenamefont {Perrone}, \citenamefont {Valentini}, \citenamefont {Parashar},
  \citenamefont {Veltri}, \citenamefont {Gershman}, \citenamefont {Russell},
  \citenamefont {Giles} \emph {et~al.}}]{servidio2017magnetospheric}%
  \BibitemOpen
  \bibfield  {author} {\bibinfo {author} {\bibfnamefont {S.}~\bibnamefont
  {Servidio}}, \bibinfo {author} {\bibfnamefont {A.}~\bibnamefont {Chasapis}},
  \bibinfo {author} {\bibfnamefont {W.}~\bibnamefont {Matthaeus}}, \bibinfo
  {author} {\bibfnamefont {D.}~\bibnamefont {Perrone}}, \bibinfo {author}
  {\bibfnamefont {F.}~\bibnamefont {Valentini}}, \bibinfo {author}
  {\bibfnamefont {T.}~\bibnamefont {Parashar}}, \bibinfo {author}
  {\bibfnamefont {P.}~\bibnamefont {Veltri}}, \bibinfo {author} {\bibfnamefont
  {D.}~\bibnamefont {Gershman}}, \bibinfo {author} {\bibfnamefont
  {C.}~\bibnamefont {Russell}}, \bibinfo {author} {\bibfnamefont
  {B.}~\bibnamefont {Giles}},  \emph {et~al.},\ }\href@noop {} {\bibfield
  {journal} {\bibinfo  {journal} {Physical Review Letters}\ }\textbf {\bibinfo
  {volume} {119}},\ \bibinfo {pages} {205101} (\bibinfo {year}
  {2017})}\BibitemShut {NoStop}%
\bibitem [{\citenamefont {Schekochihin}\ \emph {et~al.}(2016)\citenamefont
  {Schekochihin}, \citenamefont {Parker}, \citenamefont {Highcock},
  \citenamefont {Dellar}, \citenamefont {Dorland},\ and\ \citenamefont
  {Hammett}}]{schekochihin2016phase}%
  \BibitemOpen
  \bibfield  {author} {\bibinfo {author} {\bibfnamefont {A.~A.}\ \bibnamefont
  {Schekochihin}}, \bibinfo {author} {\bibfnamefont {J.~T.}\ \bibnamefont
  {Parker}}, \bibinfo {author} {\bibfnamefont {E.~G.}\ \bibnamefont
  {Highcock}}, \bibinfo {author} {\bibfnamefont {P.~J.}\ \bibnamefont
  {Dellar}}, \bibinfo {author} {\bibfnamefont {W.}~\bibnamefont {Dorland}}, \
  and\ \bibinfo {author} {\bibfnamefont {G.~W.}\ \bibnamefont {Hammett}},\
  }\href@noop {} {\bibfield  {journal} {\bibinfo  {journal} {Journal of Plasma
  Physics}\ }\textbf {\bibinfo {volume} {82}} (\bibinfo {year}
  {2016})}\BibitemShut {NoStop}%
\bibitem [{\citenamefont {Adkins}\ and\ \citenamefont
  {Schekochihin}(2018)}]{adkins2018solvable}%
  \BibitemOpen
  \bibfield  {author} {\bibinfo {author} {\bibfnamefont {T.}~\bibnamefont
  {Adkins}}\ and\ \bibinfo {author} {\bibfnamefont {A.~A.}\ \bibnamefont
  {Schekochihin}},\ }\href@noop {} {\bibfield  {journal} {\bibinfo  {journal}
  {Journal of Plasma Physics}\ }\textbf {\bibinfo {volume} {84}} (\bibinfo
  {year} {2018})}\BibitemShut {NoStop}%
\bibitem [{\citenamefont {Eyink}(2018)}]{eyink2018cascades}%
  \BibitemOpen
  \bibfield  {author} {\bibinfo {author} {\bibfnamefont {G.~L.}\ \bibnamefont
  {Eyink}},\ }\href@noop {} {\bibfield  {journal} {\bibinfo  {journal}
  {Physical Review X}\ }\textbf {\bibinfo {volume} {8}},\ \bibinfo {pages}
  {041020} (\bibinfo {year} {2018})}\BibitemShut {NoStop}%
\bibitem [{\citenamefont {Gould}\ \emph {et~al.}(1967)\citenamefont {Gould},
  \citenamefont {O'Neil},\ and\ \citenamefont {Malmberg}}]{Gould1967echo}%
  \BibitemOpen
  \bibfield  {author} {\bibinfo {author} {\bibfnamefont {R.~W.}\ \bibnamefont
  {Gould}}, \bibinfo {author} {\bibfnamefont {T.~M.}\ \bibnamefont {O'Neil}}, \
  and\ \bibinfo {author} {\bibfnamefont {J.~H.}\ \bibnamefont {Malmberg}},\
  }\href@noop {} {\bibfield  {journal} {\bibinfo  {journal} {Physical Review
  Letters}\ }\textbf {\bibinfo {volume} {19}},\ \bibinfo {pages} {219}
  (\bibinfo {year} {1967})}\BibitemShut {NoStop}%
\bibitem [{\citenamefont {Malmberg}\ \emph {et~al.}(1968)\citenamefont
  {Malmberg}, \citenamefont {Wharton}, \citenamefont {Gould},\ and\
  \citenamefont {O'Neil}}]{malmberg1968plasma}%
  \BibitemOpen
  \bibfield  {author} {\bibinfo {author} {\bibfnamefont {J.~H.}\ \bibnamefont
  {Malmberg}}, \bibinfo {author} {\bibfnamefont {C.~B.}\ \bibnamefont
  {Wharton}}, \bibinfo {author} {\bibfnamefont {R.~W.}\ \bibnamefont {Gould}},
  \ and\ \bibinfo {author} {\bibfnamefont {T.~M.}\ \bibnamefont {O'Neil}},\
  }\href@noop {} {\bibfield  {journal} {\bibinfo  {journal} {Physical Review
  Letters}\ }\textbf {\bibinfo {volume} {20}},\ \bibinfo {pages} {95} (\bibinfo
  {year} {1968})}\BibitemShut {NoStop}%
\bibitem [{\citenamefont {Meyrand}\ \emph {et~al.}(2019)\citenamefont
  {Meyrand}, \citenamefont {Kanekar}, \citenamefont {Dorland},\ and\
  \citenamefont {Schekochihin}}]{meyrand2019fluidization}%
  \BibitemOpen
  \bibfield  {author} {\bibinfo {author} {\bibfnamefont {R.}~\bibnamefont
  {Meyrand}}, \bibinfo {author} {\bibfnamefont {A.}~\bibnamefont {Kanekar}},
  \bibinfo {author} {\bibfnamefont {W.}~\bibnamefont {Dorland}}, \ and\
  \bibinfo {author} {\bibfnamefont {A.~A.}\ \bibnamefont {Schekochihin}},\
  }\href@noop {} {\bibfield  {journal} {\bibinfo  {journal} {Proceedings of the
  National Academy of Sciences}\ }\textbf {\bibinfo {volume} {116}},\ \bibinfo
  {pages} {1185} (\bibinfo {year} {2019})}\BibitemShut {NoStop}%
\bibitem [{\citenamefont {Parker}\ \emph {et~al.}(2016)\citenamefont {Parker},
  \citenamefont {Highcock}, \citenamefont {Schekochihin},\ and\ \citenamefont
  {Dellar}}]{parker2016suppression}%
  \BibitemOpen
  \bibfield  {author} {\bibinfo {author} {\bibfnamefont {J.~T.}\ \bibnamefont
  {Parker}}, \bibinfo {author} {\bibfnamefont {E.}~\bibnamefont {Highcock}},
  \bibinfo {author} {\bibfnamefont {A.~A.}\ \bibnamefont {Schekochihin}}, \
  and\ \bibinfo {author} {\bibfnamefont {P.}~\bibnamefont {Dellar}},\
  }\href@noop {} {\bibfield  {journal} {\bibinfo  {journal} {Physics of
  Plasmas}\ }\textbf {\bibinfo {volume} {23}},\ \bibinfo {pages} {070703}
  (\bibinfo {year} {2016})}\BibitemShut {NoStop}%
\bibitem [{\citenamefont {Zocco}\ and\ \citenamefont
  {Schekochihin}(2011)}]{zocco2011reduced}%
  \BibitemOpen
  \bibfield  {author} {\bibinfo {author} {\bibfnamefont {A.}~\bibnamefont
  {Zocco}}\ and\ \bibinfo {author} {\bibfnamefont {A.~A.}\ \bibnamefont
  {Schekochihin}},\ }\href@noop {} {\bibfield  {journal} {\bibinfo  {journal}
  {Physics of Plasmas}\ }\textbf {\bibinfo {volume} {18}},\ \bibinfo {pages}
  {102309} (\bibinfo {year} {2011})}\BibitemShut {NoStop}%
\bibitem [{\citenamefont {Kadomtsev}\ and\ \citenamefont
  {Pogutse}(1973)}]{kadomtsev1973nonlinear}%
  \BibitemOpen
  \bibfield  {author} {\bibinfo {author} {\bibfnamefont {B.~B.}\ \bibnamefont
  {Kadomtsev}}\ and\ \bibinfo {author} {\bibfnamefont {O.~P.}\ \bibnamefont
  {Pogutse}},\ }\href@noop {} {\bibfield  {journal} {\bibinfo  {journal}
  {Soviet physics, Journal of Experimental and Theoretical Physics}\ }\textbf
  {\bibinfo {volume} {5}},\ \bibinfo {pages} {575} (\bibinfo {year}
  {1973})}\BibitemShut {NoStop}%
\bibitem [{\citenamefont {Strauss}(1976)}]{strauss1976nonlinear}%
  \BibitemOpen
  \bibfield  {author} {\bibinfo {author} {\bibfnamefont {H.~R.}\ \bibnamefont
  {Strauss}},\ }\href@noop {} {\bibfield  {journal} {\bibinfo  {journal} {The
  Physics of Fluids}\ }\textbf {\bibinfo {volume} {19}},\ \bibinfo {pages}
  {134} (\bibinfo {year} {1976})}\BibitemShut {NoStop}%
\bibitem [{\citenamefont {{Zank}}\ and\ \citenamefont
  {{Matthaeus}}(1992)}]{zank1992}%
  \BibitemOpen
  \bibfield  {author} {\bibinfo {author} {\bibfnamefont {G.~P.}\ \bibnamefont
  {{Zank}}}\ and\ \bibinfo {author} {\bibfnamefont {W.~H.}\ \bibnamefont
  {{Matthaeus}}},\ }\href {\doibase 10.1017/S002237780001638X} {\bibfield
  {journal} {\bibinfo  {journal} {Journal of Plasma Physics}\ }\textbf
  {\bibinfo {volume} {48}},\ \bibinfo {pages} {85} (\bibinfo {year}
  {1992})}\BibitemShut {NoStop}%
\bibitem [{Note1()}]{Note1}%
  \BibitemOpen
  \bibinfo {note} {In the conventional $\beta $-model~\cite
  {frisch1995turbulence} for intermittency widely used in hydrodynamics, the
  dimension of intermittent structures is a fractal dimension and is not
  necessarily the same as their spatial dimension. However, in~\protect \citet
  {boldyrev2012spectrum} and other turbulence simulations, it is found that
  intermittent structures do appear to be 2D in position space.}\BibitemShut
  {Stop}%
\bibitem [{\citenamefont {Frisch}(1995)}]{frisch1995turbulence}%
  \BibitemOpen
  \bibfield  {author} {\bibinfo {author} {\bibfnamefont {U.}~\bibnamefont
  {Frisch}},\ }\href@noop {} {\emph {\bibinfo {title} {Turbulence: the legacy
  of A. N. Kolmogorov}}}\ (\bibinfo  {publisher} {Cambridge university press},\
  \bibinfo {year} {1995})\BibitemShut {NoStop}%
\bibitem [{\citenamefont {Mallet}\ \emph
  {et~al.}(2017{\natexlab{a}})\citenamefont {Mallet}, \citenamefont
  {Schekochihin},\ and\ \citenamefont {Chandran}}]{mallet2017disruption}%
  \BibitemOpen
  \bibfield  {author} {\bibinfo {author} {\bibfnamefont {A.}~\bibnamefont
  {Mallet}}, \bibinfo {author} {\bibfnamefont {A.~A.}\ \bibnamefont
  {Schekochihin}}, \ and\ \bibinfo {author} {\bibfnamefont {B.~D.~G.}\
  \bibnamefont {Chandran}},\ }\href@noop {} {\bibfield  {journal} {\bibinfo
  {journal} {Monthly Notices of the Royal Astronomical Society}\ }\textbf
  {\bibinfo {volume} {468}},\ \bibinfo {pages} {4862} (\bibinfo {year}
  {2017}{\natexlab{a}})}\BibitemShut {NoStop}%
\bibitem [{Note2()}]{Note2}%
  \BibitemOpen
  \bibinfo {note} {For the Harris sheet-type configuration, the growth rate of
  the fastest tearing mode becomes $\gamma _t \sim v_{A\lambda } d_e \rho
  _s/\lambda ^3$~\cite {loureiro2017collisionless}, leading to the scale
  dependence of magnetic fluctuation $v_{A\lambda } \sim \varepsilon
  ^{1/3}\lambda (d_e \rho _s)^{-1/3}$, the magnetic spectrum $E_B (k_\perp )
  \sim 4 \pi \rho _0 \varepsilon ^{2/3} (d_e \rho _s)^{-2/3} k_\perp ^{-3}
  dk_\perp $, and the spectral anisotropy $k_\parallel \sim \varepsilon ^{1/3}
  (d_e^2/\rho _s)^{1/3} V_A^{-1} k_\perp $.}\BibitemShut {Stop}%
\bibitem [{\citenamefont {Howes}\ \emph {et~al.}(2018)\citenamefont {Howes},
  \citenamefont {McCubbin},\ and\ \citenamefont {Klein}}]{howes2018spatially}%
  \BibitemOpen
  \bibfield  {author} {\bibinfo {author} {\bibfnamefont {G.~G.}\ \bibnamefont
  {Howes}}, \bibinfo {author} {\bibfnamefont {A.~J.}\ \bibnamefont {McCubbin}},
  \ and\ \bibinfo {author} {\bibfnamefont {K.~G.}\ \bibnamefont {Klein}},\
  }\href@noop {} {\bibfield  {journal} {\bibinfo  {journal} {Journal of Plasma
  Physics}\ }\textbf {\bibinfo {volume} {84}} (\bibinfo {year}
  {2018})}\BibitemShut {NoStop}%
\bibitem [{\citenamefont {Klein}\ and\ \citenamefont
  {Howes}(2016)}]{klein2016measuring}%
  \BibitemOpen
  \bibfield  {author} {\bibinfo {author} {\bibfnamefont {K.~G.}\ \bibnamefont
  {Klein}}\ and\ \bibinfo {author} {\bibfnamefont {G.~G.}\ \bibnamefont
  {Howes}},\ }\href@noop {} {\bibfield  {journal} {\bibinfo  {journal} {The
  Astrophysical Journal Letters}\ }\textbf {\bibinfo {volume} {826}},\ \bibinfo
  {pages} {L30} (\bibinfo {year} {2016})}\BibitemShut {NoStop}%
\bibitem [{\citenamefont {Howes}\ \emph {et~al.}(2017)\citenamefont {Howes},
  \citenamefont {Klein},\ and\ \citenamefont {Li}}]{howes2017diagnosing}%
  \BibitemOpen
  \bibfield  {author} {\bibinfo {author} {\bibfnamefont {G.~G.}\ \bibnamefont
  {Howes}}, \bibinfo {author} {\bibfnamefont {K.~G.}\ \bibnamefont {Klein}}, \
  and\ \bibinfo {author} {\bibfnamefont {T.~C.}\ \bibnamefont {Li}},\
  }\href@noop {} {\bibfield  {journal} {\bibinfo  {journal} {Journal of Plasma
  Physics}\ }\textbf {\bibinfo {volume} {83}} (\bibinfo {year}
  {2017})}\BibitemShut {NoStop}%
\bibitem [{\citenamefont {{Loureiro}}\ \emph {et~al.}(2013)\citenamefont
  {{Loureiro}}, \citenamefont {{Schekochihin}},\ and\ \citenamefont
  {{Zocco}}}]{loureiro2013fast}%
  \BibitemOpen
  \bibfield  {author} {\bibinfo {author} {\bibfnamefont {N.~F.}\ \bibnamefont
  {{Loureiro}}}, \bibinfo {author} {\bibfnamefont {A.~A.}\ \bibnamefont
  {{Schekochihin}}}, \ and\ \bibinfo {author} {\bibfnamefont {A.}~\bibnamefont
  {{Zocco}}},\ }\href@noop {} {\bibfield  {journal} {\bibinfo  {journal}
  {Physical Review Letters}\ }\textbf {\bibinfo {volume} {111}},\ \bibinfo
  {pages} {025002} (\bibinfo {year} {2013})}\BibitemShut {NoStop}%
\bibitem [{\citenamefont {{Numata}}\ and\ \citenamefont
  {{Loureiro}}(2015)}]{numata2015}%
  \BibitemOpen
  \bibfield  {author} {\bibinfo {author} {\bibfnamefont {R.}~\bibnamefont
  {{Numata}}}\ and\ \bibinfo {author} {\bibfnamefont {N.~F.}\ \bibnamefont
  {{Loureiro}}},\ }\href {\doibase 10.1017/S002237781400107X} {\bibfield
  {journal} {\bibinfo  {journal} {Journal of Plasma Physics}\ }\textbf
  {\bibinfo {volume} {81}},\ \bibinfo {eid} {305810201} (\bibinfo {year}
  {2015})}\BibitemShut {NoStop}%
\bibitem [{\citenamefont {Kanekar}\ \emph {et~al.}(2015)\citenamefont
  {Kanekar}, \citenamefont {Schekochihin}, \citenamefont {Dorland},\ and\
  \citenamefont {Loureiro}}]{kanekar2015fluctuation}%
  \BibitemOpen
  \bibfield  {author} {\bibinfo {author} {\bibfnamefont {A.}~\bibnamefont
  {Kanekar}}, \bibinfo {author} {\bibfnamefont {A.~A.}\ \bibnamefont
  {Schekochihin}}, \bibinfo {author} {\bibfnamefont {W.}~\bibnamefont
  {Dorland}}, \ and\ \bibinfo {author} {\bibfnamefont {N.~F.}\ \bibnamefont
  {Loureiro}},\ }\href@noop {} {\bibfield  {journal} {\bibinfo  {journal}
  {Journal of Plasma Physics}\ }\textbf {\bibinfo {volume} {81}} (\bibinfo
  {year} {2015})}\BibitemShut {NoStop}%
\bibitem [{\citenamefont {Loureiro}\ \emph {et~al.}(2016)\citenamefont
  {Loureiro}, \citenamefont {Dorland}, \citenamefont {Fazendeiro},
  \citenamefont {Kanekar}, \citenamefont {Mallet}, \citenamefont {Vilelas},\
  and\ \citenamefont {Zocco}}]{loureiro2016viriato}%
  \BibitemOpen
  \bibfield  {author} {\bibinfo {author} {\bibfnamefont {N.~F.}\ \bibnamefont
  {Loureiro}}, \bibinfo {author} {\bibfnamefont {W.}~\bibnamefont {Dorland}},
  \bibinfo {author} {\bibfnamefont {L.}~\bibnamefont {Fazendeiro}}, \bibinfo
  {author} {\bibfnamefont {A.}~\bibnamefont {Kanekar}}, \bibinfo {author}
  {\bibfnamefont {A.}~\bibnamefont {Mallet}}, \bibinfo {author} {\bibfnamefont
  {M.~S.}\ \bibnamefont {Vilelas}}, \ and\ \bibinfo {author} {\bibfnamefont
  {A.}~\bibnamefont {Zocco}},\ }\href@noop {} {\bibfield  {journal} {\bibinfo
  {journal} {Computer Physics Communications}\ }\textbf {\bibinfo {volume}
  {206}},\ \bibinfo {pages} {45} (\bibinfo {year} {2016})}\BibitemShut
  {NoStop}%
\bibitem [{\citenamefont {{Mallet}}\ \emph {et~al.}(2016)\citenamefont
  {{Mallet}}, \citenamefont {{Schekochihin}}, \citenamefont {{Chandran}},
  \citenamefont {{Chen}}, \citenamefont {{Horbury}}, \citenamefont {{Wicks}},\
  and\ \citenamefont {{Greenan}}}]{mallet2016}%
  \BibitemOpen
  \bibfield  {author} {\bibinfo {author} {\bibfnamefont {A.}~\bibnamefont
  {{Mallet}}}, \bibinfo {author} {\bibfnamefont {A.~A.}\ \bibnamefont
  {{Schekochihin}}}, \bibinfo {author} {\bibfnamefont {B.~D.~G.}\ \bibnamefont
  {{Chandran}}}, \bibinfo {author} {\bibfnamefont {C.~H.~K.}\ \bibnamefont
  {{Chen}}}, \bibinfo {author} {\bibfnamefont {T.~S.}\ \bibnamefont
  {{Horbury}}}, \bibinfo {author} {\bibfnamefont {R.~T.}\ \bibnamefont
  {{Wicks}}}, \ and\ \bibinfo {author} {\bibfnamefont {C.~C.}\ \bibnamefont
  {{Greenan}}},\ }\href {\doibase 10.1093/mnras/stw802} {\bibfield  {journal}
  {\bibinfo  {journal} {Monthly Notices of the Royal Astronomical Society}\
  }\textbf {\bibinfo {volume} {459}},\ \bibinfo {pages} {2130} (\bibinfo {year}
  {2016})}\BibitemShut {NoStop}%
\bibitem [{\citenamefont {Schekochihin}(2020)}]{schekochihin2020mhd}%
  \BibitemOpen
  \bibfield  {author} {\bibinfo {author} {\bibfnamefont {A.~A.}\ \bibnamefont
  {Schekochihin}},\ }\href@noop {} {\bibfield  {journal} {\bibinfo  {journal}
  {arXiv preprint arXiv:2010.00699}\ } (\bibinfo {year} {2020})}\BibitemShut
  {NoStop}%
\bibitem [{\citenamefont {Wang}\ \emph {et~al.}(2020)\citenamefont {Wang},
  \citenamefont {He}, \citenamefont {Alexandrova}, \citenamefont {Dunlop},\
  and\ \citenamefont {Perrone}}]{wang2020observational}%
  \BibitemOpen
  \bibfield  {author} {\bibinfo {author} {\bibfnamefont {T.}~\bibnamefont
  {Wang}}, \bibinfo {author} {\bibfnamefont {J.}~\bibnamefont {He}}, \bibinfo
  {author} {\bibfnamefont {O.}~\bibnamefont {Alexandrova}}, \bibinfo {author}
  {\bibfnamefont {M.}~\bibnamefont {Dunlop}}, \ and\ \bibinfo {author}
  {\bibfnamefont {D.}~\bibnamefont {Perrone}},\ }\href@noop {} {\bibfield
  {journal} {\bibinfo  {journal} {The Astrophysical Journal}\ }\textbf
  {\bibinfo {volume} {898}},\ \bibinfo {pages} {91} (\bibinfo {year}
  {2020})}\BibitemShut {NoStop}%
\bibitem [{\citenamefont {Zhang}\ \emph {et~al.}(2022)\citenamefont {Zhang},
  \citenamefont {Huang}, \citenamefont {He}, \citenamefont {Wang},
  \citenamefont {Yuan}, \citenamefont {Deng}, \citenamefont {Jiang},
  \citenamefont {Xu}, \citenamefont {Xiong} \emph {et~al.}}]{zhang2022three}%
  \BibitemOpen
  \bibfield  {author} {\bibinfo {author} {\bibfnamefont {J.}~\bibnamefont
  {Zhang}}, \bibinfo {author} {\bibfnamefont {S.~Y.}\ \bibnamefont {Huang}},
  \bibinfo {author} {\bibfnamefont {J.~S.}\ \bibnamefont {He}}, \bibinfo
  {author} {\bibfnamefont {T.~Y.}\ \bibnamefont {Wang}}, \bibinfo {author}
  {\bibfnamefont {Z.~G.}\ \bibnamefont {Yuan}}, \bibinfo {author}
  {\bibfnamefont {X.~H.}\ \bibnamefont {Deng}}, \bibinfo {author}
  {\bibfnamefont {Y.~Y.}\ \bibnamefont {Jiang}, \bibfnamefont {K~.and~Wei}},
  \bibinfo {author} {\bibfnamefont {S.~B.}\ \bibnamefont {Xu}}, \bibinfo
  {author} {\bibfnamefont {Q.~Y.}\ \bibnamefont {Xiong}},  \emph {et~al.},\
  }\href@noop {} {\bibfield  {journal} {\bibinfo  {journal} {The Astrophysical
  Journal Letters}\ }\textbf {\bibinfo {volume} {924}},\ \bibinfo {pages} {L21}
  (\bibinfo {year} {2022})}\BibitemShut {NoStop}%
\bibitem [{\citenamefont {{Loureiro}}\ \emph {et~al.}(2007)\citenamefont
  {{Loureiro}}, \citenamefont {{Schekochihin}},\ and\ \citenamefont
  {{Cowley}}}]{loureiro2007instability}%
  \BibitemOpen
  \bibfield  {author} {\bibinfo {author} {\bibfnamefont {N.~F.}\ \bibnamefont
  {{Loureiro}}}, \bibinfo {author} {\bibfnamefont {A.~A.}\ \bibnamefont
  {{Schekochihin}}}, \ and\ \bibinfo {author} {\bibfnamefont {S.~C.}\
  \bibnamefont {{Cowley}}},\ }\href {\doibase 10.1063/1.2783986} {\bibfield
  {journal} {\bibinfo  {journal} {Physics of Plasmas}\ }\textbf {\bibinfo
  {volume} {14}},\ \bibinfo {pages} {100703} (\bibinfo {year}
  {2007})}\BibitemShut {NoStop}%
\bibitem [{Note3()}]{Note3}%
  \BibitemOpen
  \bibinfo {note} {\protect \leavevmode {\color {blue}The large-$m$ dissipation
  at the system scale is artificial, caused by the forcing.}}\BibitemShut
  {Stop}%
\bibitem [{\citenamefont {{Howes}}\ \emph {et~al.}(2011)\citenamefont
  {{Howes}}, \citenamefont {{Tenbarge}},\ and\ \citenamefont
  {{Dorland}}}]{howes2011weakened}%
  \BibitemOpen
  \bibfield  {author} {\bibinfo {author} {\bibfnamefont {G.~G.}\ \bibnamefont
  {{Howes}}}, \bibinfo {author} {\bibfnamefont {J.~M.}\ \bibnamefont
  {{Tenbarge}}}, \ and\ \bibinfo {author} {\bibfnamefont {W.}~\bibnamefont
  {{Dorland}}},\ }\href {\doibase 10.1063/1.3646400} {\bibfield  {journal}
  {\bibinfo  {journal} {Physics of Plasmas}\ }\textbf {\bibinfo {volume}
  {18}},\ \bibinfo {pages} {102305} (\bibinfo {year} {2011})}\BibitemShut
  {NoStop}%
\bibitem [{\citenamefont {TenBarge}\ \emph {et~al.}(2013)\citenamefont
  {TenBarge}, \citenamefont {Howes},\ and\ \citenamefont
  {Dorland}}]{tenbarge2013collisionless}%
  \BibitemOpen
  \bibfield  {author} {\bibinfo {author} {\bibfnamefont {J.~M.}\ \bibnamefont
  {TenBarge}}, \bibinfo {author} {\bibfnamefont {G.~G.}\ \bibnamefont {Howes}},
  \ and\ \bibinfo {author} {\bibfnamefont {W.}~\bibnamefont {Dorland}},\
  }\href@noop {} {\bibfield  {journal} {\bibinfo  {journal} {The Astrophysical
  Journal}\ }\textbf {\bibinfo {volume} {774}},\ \bibinfo {pages} {139}
  (\bibinfo {year} {2013})}\BibitemShut {NoStop}%
\bibitem [{\citenamefont {TenBarge}\ and\ \citenamefont
  {Howes}(2013)}]{tenbarge2013current}%
  \BibitemOpen
  \bibfield  {author} {\bibinfo {author} {\bibfnamefont {J.~M.}\ \bibnamefont
  {TenBarge}}\ and\ \bibinfo {author} {\bibfnamefont {G.~G.}\ \bibnamefont
  {Howes}},\ }\href@noop {} {\bibfield  {journal} {\bibinfo  {journal} {The
  Astrophysical Journal Letters}\ }\textbf {\bibinfo {volume} {771}},\ \bibinfo
  {pages} {L27} (\bibinfo {year} {2013})}\BibitemShut {NoStop}%
\bibitem [{\citenamefont {Howes}(2010)}]{howes2010prescription}%
  \BibitemOpen
  \bibfield  {author} {\bibinfo {author} {\bibfnamefont {G.~G.}\ \bibnamefont
  {Howes}},\ }\href@noop {} {\bibfield  {journal} {\bibinfo  {journal} {Monthly
  Notices of the Royal Astronomical Society: Letters}\ }\textbf {\bibinfo
  {volume} {409}},\ \bibinfo {pages} {L104} (\bibinfo {year}
  {2010})}\BibitemShut {NoStop}%
\bibitem [{\citenamefont {Passot}\ and\ \citenamefont
  {Sulem}(2015)}]{passot2015model}%
  \BibitemOpen
  \bibfield  {author} {\bibinfo {author} {\bibfnamefont {T.}~\bibnamefont
  {Passot}}\ and\ \bibinfo {author} {\bibfnamefont {P.~L.}\ \bibnamefont
  {Sulem}},\ }\href@noop {} {\bibfield  {journal} {\bibinfo  {journal} {The
  Astrophysical Journal Letters}\ }\textbf {\bibinfo {volume} {812}},\ \bibinfo
  {pages} {L37} (\bibinfo {year} {2015})}\BibitemShut {NoStop}%
\bibitem [{\citenamefont {Kunz}\ \emph {et~al.}(2018)\citenamefont {Kunz},
  \citenamefont {Abel}, \citenamefont {Klein},\ and\ \citenamefont
  {Schekochihin}}]{kunz2018astrophysical}%
  \BibitemOpen
  \bibfield  {author} {\bibinfo {author} {\bibfnamefont {M.~W.}\ \bibnamefont
  {Kunz}}, \bibinfo {author} {\bibfnamefont {I.~G.}\ \bibnamefont {Abel}},
  \bibinfo {author} {\bibfnamefont {K.~G.}\ \bibnamefont {Klein}}, \ and\
  \bibinfo {author} {\bibfnamefont {A.~A.}\ \bibnamefont {Schekochihin}},\
  }\href@noop {} {\bibfield  {journal} {\bibinfo  {journal} {Journal of Plasma
  Physics}\ }\textbf {\bibinfo {volume} {84}} (\bibinfo {year}
  {2018})}\BibitemShut {NoStop}%
\bibitem [{\citenamefont {McCubbin}\ \emph {et~al.}(2022)\citenamefont
  {McCubbin}, \citenamefont {Howes},\ and\ \citenamefont
  {TenBarge}}]{mccubbin2022characterizing}%
  \BibitemOpen
  \bibfield  {author} {\bibinfo {author} {\bibfnamefont {A.~J.}\ \bibnamefont
  {McCubbin}}, \bibinfo {author} {\bibfnamefont {G.~G.}\ \bibnamefont {Howes}},
  \ and\ \bibinfo {author} {\bibfnamefont {J.~M.}\ \bibnamefont {TenBarge}},\
  }\href@noop {} {\bibfield  {journal} {\bibinfo  {journal} {Physics of
  Plasmas}\ }\textbf {\bibinfo {volume} {29}},\ \bibinfo {pages} {052105}
  (\bibinfo {year} {2022})}\BibitemShut {NoStop}%
\bibitem [{\citenamefont {Carbone}\ \emph {et~al.}(2022)\citenamefont
  {Carbone}, \citenamefont {Telloni}, \citenamefont {Lepreti},\ and\
  \citenamefont {Vecchio}}]{carbone2022high}%
  \BibitemOpen
  \bibfield  {author} {\bibinfo {author} {\bibfnamefont {V.}~\bibnamefont
  {Carbone}}, \bibinfo {author} {\bibfnamefont {D.}~\bibnamefont {Telloni}},
  \bibinfo {author} {\bibfnamefont {F.}~\bibnamefont {Lepreti}}, \ and\
  \bibinfo {author} {\bibfnamefont {A.}~\bibnamefont {Vecchio}},\ }\href@noop
  {} {\bibfield  {journal} {\bibinfo  {journal} {The Astrophysical Journal
  Letters}\ }\textbf {\bibinfo {volume} {924}},\ \bibinfo {pages} {L26}
  (\bibinfo {year} {2022})}\BibitemShut {NoStop}%
\bibitem [{\citenamefont {Hammett}\ and\ \citenamefont
  {Perkins}(1990)}]{Hammett1990}%
  \BibitemOpen
  \bibfield  {author} {\bibinfo {author} {\bibfnamefont {G.~W.}\ \bibnamefont
  {Hammett}}\ and\ \bibinfo {author} {\bibfnamefont {F.~W.}\ \bibnamefont
  {Perkins}},\ }\href {\doibase 10.1103/PhysRevLett.64.3019} {\bibfield
  {journal} {\bibinfo  {journal} {Physical Review Letters}\ }\textbf {\bibinfo
  {volume} {64}},\ \bibinfo {pages} {3019} (\bibinfo {year}
  {1990})}\BibitemShut {NoStop}%
\bibitem [{\citenamefont {Mallet}\ \emph
  {et~al.}(2017{\natexlab{b}})\citenamefont {Mallet}, \citenamefont
  {Schekochihin},\ and\ \citenamefont {Chandran}}]{mallet2017disruptionb}%
  \BibitemOpen
  \bibfield  {author} {\bibinfo {author} {\bibfnamefont {A.}~\bibnamefont
  {Mallet}}, \bibinfo {author} {\bibfnamefont {A.~A.}\ \bibnamefont
  {Schekochihin}}, \ and\ \bibinfo {author} {\bibfnamefont {B.~D.~G.}\
  \bibnamefont {Chandran}},\ }\href@noop {} {\bibfield  {journal} {\bibinfo
  {journal} {Journal of Plasma Physics}\ }\textbf {\bibinfo {volume} {83}}
  (\bibinfo {year} {2017}{\natexlab{b}})}\BibitemShut {NoStop}%
\bibitem [{\citenamefont {Kawazura}\ \emph {et~al.}(2019)\citenamefont
  {Kawazura}, \citenamefont {Barnes},\ and\ \citenamefont
  {Schekochihin}}]{kawazura2019thermal}%
  \BibitemOpen
  \bibfield  {author} {\bibinfo {author} {\bibfnamefont {Y.}~\bibnamefont
  {Kawazura}}, \bibinfo {author} {\bibfnamefont {M.}~\bibnamefont {Barnes}}, \
  and\ \bibinfo {author} {\bibfnamefont {A.~A.}\ \bibnamefont {Schekochihin}},\
  }\href@noop {} {\bibfield  {journal} {\bibinfo  {journal} {Proceedings of the
  National Academy of Sciences}\ }\textbf {\bibinfo {volume} {116}},\ \bibinfo
  {pages} {771} (\bibinfo {year} {2019})}\BibitemShut {NoStop}%
\bibitem [{\citenamefont {Zhdankin}\ \emph {et~al.}(2019)\citenamefont
  {Zhdankin}, \citenamefont {Uzdensky}, \citenamefont {Werner},\ and\
  \citenamefont {Begelman}}]{zhdankin2019}%
  \BibitemOpen
  \bibfield  {author} {\bibinfo {author} {\bibfnamefont {V.}~\bibnamefont
  {Zhdankin}}, \bibinfo {author} {\bibfnamefont {D.~A.}\ \bibnamefont
  {Uzdensky}}, \bibinfo {author} {\bibfnamefont {G.~R.}\ \bibnamefont
  {Werner}}, \ and\ \bibinfo {author} {\bibfnamefont {M.~C.}\ \bibnamefont
  {Begelman}},\ }\href {\doibase 10.1103/PhysRevLett.122.055101} {\bibfield
  {journal} {\bibinfo  {journal} {Phys. Rev. Lett.}\ }\textbf {\bibinfo
  {volume} {122}},\ \bibinfo {pages} {055101} (\bibinfo {year}
  {2019})}\BibitemShut {NoStop}%
\bibitem [{\citenamefont {Meyrand}\ \emph {et~al.}(2021)\citenamefont
  {Meyrand}, \citenamefont {Squire}, \citenamefont {Schekochihin},\ and\
  \citenamefont {Dorland}}]{meyrand2021violation}%
  \BibitemOpen
  \bibfield  {author} {\bibinfo {author} {\bibfnamefont {R.}~\bibnamefont
  {Meyrand}}, \bibinfo {author} {\bibfnamefont {J.}~\bibnamefont {Squire}},
  \bibinfo {author} {\bibfnamefont {A.~A.}\ \bibnamefont {Schekochihin}}, \
  and\ \bibinfo {author} {\bibfnamefont {W.}~\bibnamefont {Dorland}},\
  }\href@noop {} {\bibfield  {journal} {\bibinfo  {journal} {Journal of Plasma
  Physics}\ }\textbf {\bibinfo {volume} {87}} (\bibinfo {year}
  {2021})}\BibitemShut {NoStop}%
\bibitem [{\citenamefont {Squire}\ \emph {et~al.}(2022)\citenamefont {Squire},
  \citenamefont {Meyrand}, \citenamefont {Kunz}, \citenamefont {Arzamasskiy},
  \citenamefont {Schekochihin},\ and\ \citenamefont
  {Quataert}}]{squire2022high}%
  \BibitemOpen
  \bibfield  {author} {\bibinfo {author} {\bibfnamefont {J.}~\bibnamefont
  {Squire}}, \bibinfo {author} {\bibfnamefont {R.}~\bibnamefont {Meyrand}},
  \bibinfo {author} {\bibfnamefont {M.~W.}\ \bibnamefont {Kunz}}, \bibinfo
  {author} {\bibfnamefont {L.}~\bibnamefont {Arzamasskiy}}, \bibinfo {author}
  {\bibfnamefont {A.~A.}\ \bibnamefont {Schekochihin}}, \ and\ \bibinfo
  {author} {\bibfnamefont {E.}~\bibnamefont {Quataert}},\ }\href@noop {}
  {\bibfield  {journal} {\bibinfo  {journal} {Nature Astronomy}\ ,\ \bibinfo
  {pages} {1}} (\bibinfo {year} {2022})}\BibitemShut {NoStop}%
\bibitem [{\citenamefont {Coppi}\ \emph {et~al.}(1976)\citenamefont {Coppi},
  \citenamefont {Galvao}, \citenamefont {Pellat}, \citenamefont {Rosenbluth},\
  and\ \citenamefont {Rutherford}}]{coppi1976resistive}%
  \BibitemOpen
  \bibfield  {author} {\bibinfo {author} {\bibfnamefont {B.}~\bibnamefont
  {Coppi}}, \bibinfo {author} {\bibfnamefont {R.}~\bibnamefont {Galvao}},
  \bibinfo {author} {\bibfnamefont {R.}~\bibnamefont {Pellat}}, \bibinfo
  {author} {\bibfnamefont {M.~N.}\ \bibnamefont {Rosenbluth}}, \ and\ \bibinfo
  {author} {\bibfnamefont {P.}~\bibnamefont {Rutherford}},\ }\href@noop {}
  {\bibfield  {journal} {\bibinfo  {journal} {Soviet Journal of Plasma
  Physics}\ }\textbf {\bibinfo {volume} {2}},\ \bibinfo {pages} {533} (\bibinfo
  {year} {1976})}\BibitemShut {NoStop}%
\bibitem [{\citenamefont {Furth}\ \emph {et~al.}(1963)\citenamefont {Furth},
  \citenamefont {Killeen},\ and\ \citenamefont {Rosenbluth}}]{furth1963finite}%
  \BibitemOpen
  \bibfield  {author} {\bibinfo {author} {\bibfnamefont {H.~P.}\ \bibnamefont
  {Furth}}, \bibinfo {author} {\bibfnamefont {J.}~\bibnamefont {Killeen}}, \
  and\ \bibinfo {author} {\bibfnamefont {M.~N.}\ \bibnamefont {Rosenbluth}},\
  }\href@noop {} {\bibfield  {journal} {\bibinfo  {journal} {The physics of
  Fluids}\ }\textbf {\bibinfo {volume} {6}},\ \bibinfo {pages} {459} (\bibinfo
  {year} {1963})}\BibitemShut {NoStop}%
\bibitem [{\citenamefont {{Cho}}\ and\ \citenamefont
  {{Vishniac}}(2000)}]{cho2000}%
  \BibitemOpen
  \bibfield  {author} {\bibinfo {author} {\bibfnamefont {J.}~\bibnamefont
  {{Cho}}}\ and\ \bibinfo {author} {\bibfnamefont {E.~T.}\ \bibnamefont
  {{Vishniac}}},\ }\href {\doibase 10.1086/309213} {\bibfield  {journal}
  {\bibinfo  {journal} {The Astrophysical Journal}\ }\textbf {\bibinfo {volume}
  {539}},\ \bibinfo {pages} {273} (\bibinfo {year} {2000})}\BibitemShut
  {NoStop}%
\end{thebibliography}%
